\documentclass[apj]{emulateapj}
\pdfoutput=1
\usepackage{graphics,graphicx}
\usepackage{epstopdf}
\usepackage{amsmath}
\usepackage{natbib}
\usepackage{color}

\DeclareGraphicsExtensions{.jpg,.pdf,.png,.eps,.ps}

\newcommand{\msun}{$M_{\sun}$}

\def\lsim{\hbox{\rlap{\raise 0.425ex\hbox{$<$}}\lower 0.65ex\hbox{$\sim$}}}
\def\gsim{\hbox{\rlap{\raise 0.425ex\hbox{$>$}}\lower 0.65ex\hbox{$\sim$}}}
\def\arcmin{\hbox{$^\prime$}}
\def\arcsec{\hbox{$^{\prime\prime}$}}

\newcommand{\halpha}{H-$\alpha$}
\newcommand{\hbeta}{H-$\beta$}
\newcommand{\hgamma}{H-$\gamma$}
\newcommand{\hdelta}{H-$\delta$}
\newcommand{\oii}{[O {\small II}] $\lambda$3727}

\shorttitle{The SPT-GMOS Spectroscopy Survey}
\shortauthors{Bayliss+2016}

\begin{document}

\def\Harvard{1}
\def\CfA{2}
\def\Colby{3}
\def\StanfordKPAC{4}
\def\StanfordPhys{5}
\def\SLAC{6}
\def\AIfA{7}
\def\MIT{8}
\def\AAUChicago{9}
\def\KICPChicago{10}
\def\FNAL{11}
\def\PhysicsUChicago{12}
\def\Argonne{13}
\def\Munich{14}
\def\ExcellenceCluster{15}
\def\Miss{16}
\def\EFIChicago{17}
\def\NIST{18}
\def\PUC{19}
\def\Caltech{20}
\def\Berkeley{21}
\def\McGill{22}
\def\CIFAR{23}
\def\Melbourne{24}
\def\UCSC{25}
\def\illast{26}
\def\illphy{27}
\def\Huntingdon{28}
\def\MPE{29}
\def\UFlorida{30}
\def\Colorado{31}
\def\UMon{32}
\def\KIPAC{33}
\def\LeidenObservatory{34}
\def\UChicago{35}
\def\Davis{36}
\def\LBNL{37}
\def\StonyBrook{38}
\def\DARK{39}
\def\Arizona{40}
\def\Michigan{41}
\def\Minnesota{42}
\def\STScI{43}
\def\CaseWestern{44}
\def\ArtInstChicago{45}
\def\KASI{46}
\def\LLNL{47}
\def\Dunlap{48}
\def\Toronto{49}
\def\CTIO{50}

\altaffiltext{\Harvard}{Department of Physics, Harvard University, 17 Oxford Street, Cambridge, MA 02138}
\altaffiltext{\CfA}{Harvard-Smithsonian Center for Astrophysics, 60 Garden Street, Cambridge, MA 02138}
\altaffiltext{\Colby}{Department of Physics and Astronomy, Colby College, 5100 Mayflower Hill Dr, Waterville, ME 04901}
\altaffiltext{\Berkeley}{Department of Physics, University of California, Berkeley, CA, USA 94720}
\altaffiltext{\McGill}{Department of Physics, McGill University, Montreal, Quebec H3A 2T8, Canada}
\altaffiltext{\AAUChicago}{Department of Astronomy and Astrophysics, University of Chicago, Chicago, IL, USA 60637}
\altaffiltext{\KICPChicago}{Kavli Institute for Cosmological Physics, University of Chicago, Chicago, IL, USA 60637}
\altaffiltext{\FNAL}{Fermi National Accelerator Laboratory, Batavia, IL 60510-0500, USA}
\altaffiltext{\PhysicsUChicago}{Department of Physics, University of Chicago, Chicago, IL, USA 60637}
\altaffiltext{\Argonne}{Argonne National Laboratory, Argonne, IL, USA 60439}
\altaffiltext{\StanfordKPAC}{Kavli Institute for Particle Astrophysics and Cosmology, Stanford University, 452 Lomita Mall, Stanford, CA 94305}
\altaffiltext{\StanfordPhys}{Department of Physics, Stanford University, 382 Via Pueblo Mall, Stanford, CA 94305}
\altaffiltext{\SLAC}{SLAC National Accelerator Laboratory, 2575 Sand Hill Road, Menlo Park, CA 94025}
\altaffiltext{\AIfA}{Argelander-Institut f\"ur Astronomie, Auf dem H\"ugel 71, D-53121 Bonn, Germany}
\altaffiltext{\MIT}{Kavli Institute for Astrophysics and Space Research, Massachusetts Institute of Technology, 77 Massachusetts Avenue, Cambridge, MA 02139}
\altaffiltext{\Munich}{Faculty of Physics, Ludwig-Maximilians-Universit\"{a}t, Scheinerstr.\ 1, 81679 Munich, Germany}
\altaffiltext{\ExcellenceCluster}{Excellence Cluster Universe, Boltzmannstr.\ 2, 85748 Garching, Germany}
\altaffiltext{\Miss}{Department of Physics and Astronomy, University of Missouri, 5110 Rockhill Road, Kansas City, MO 64110}
\altaffiltext{\EFIChicago}{Enrico Fermi Institute, University of Chicago, Chicago, IL, USA 60637}
\altaffiltext{\NIST}{NIST Quantum Devices Group, Boulder, CO, USA 80305}
\altaffiltext{\PUC}{Departamento de Astronomia y Astrosifica, Pontificia Universidad Catolica,Chile}
\altaffiltext{\Caltech}{California Institute of Technology, Pasadena, CA, USA 91125}
\altaffiltext{\CIFAR}{Canadian Institute for Advanced Research, CIFAR Program in Cosmology and Gravity, Toronto, ON, M5G 1Z8, Canada}
\altaffiltext{\Melbourne}{School of Physics, University of Melbourne, Parkville, VIC 3010, Australia}
\altaffiltext{\UCSC}{Department of Astronomy and Astrophysics, University of California, Santa Cruz, CA 95064, USA}
\altaffiltext{\illast}{Astronomy Department, University of Illinois at Urbana-Champaign, 1002 W.\ Green Street, Urbana, IL 61801, USA}
\altaffiltext{\illphy}{Department of Physics, University of Illinois Urbana-Champaign, 1110 W.\ Green Street, Urbana, IL 61801, USA}
\altaffiltext{\Huntingdon}{Huntingdon Institute for X-ray Astronomy, LLC}
\altaffiltext{\MPE}{Max Planck Institute for Extraterrestrial Physics, Giessenbachstr.\ 1, 85748 Garching, Germany}
\altaffiltext{\UFlorida}{Department of Astronomy, University of Florida, Gainesville, FL 32611}
\altaffiltext{\Colorado}{Department of Astrophysical and Planetary Sciences and Department of Physics, University of Colorado, Boulder, CO, USA 80309}
\altaffiltext{\UMon}{Department of Physics, Universit\'e de Montr\'eal, Montreal, Quebec H3T 1J4, Canada}
\altaffiltext{\KIPAC}{Kavli Institute for Particle Astrophysics and Cosmology, Stanford University, 452 Lomita Mall, Stanford, CA 94305-4085, USA}
\altaffiltext{\LeidenObservatory}{Leiden Observatory, Leiden University, Niels Bohrweg 2, 2333 CA, Leiden, the Netherlands}
\altaffiltext{\UChicago}{University of Chicago, Chicago, IL, USA 60637}
\altaffiltext{\Davis}{Department of Physics, University of California, Davis, CA, USA 95616}
\altaffiltext{\LBNL}{Physics Division, Lawrence Berkeley National Laboratory, Berkeley, CA, USA 94720}
\altaffiltext{\StonyBrook}{Department of Physics and Astronomy, Stony Brook University, Stony Brook, NY 11794, USA}
\altaffiltext{\DARK}{Dark Cosmology Centre, Niels Bohr Institute, University of Copenhagen Juliane Maries Vej 30, 2100 Copenhagen, Denmark}
\altaffiltext{\Arizona}{Steward Observatory, University of Arizona, 933 North Cherry Avenue, Tucson, AZ 85721}
\altaffiltext{\Michigan}{Department of Physics, University of Michigan, Ann  Arbor, MI, USA 48109}
\altaffiltext{\Minnesota}{Department of Physics, University of Minnesota, Minneapolis, MN, USA 55455}
\altaffiltext{\STScI}{Space Telescope Science Institute, 3700 San Martin Dr., Baltimore, MD 21218}
\altaffiltext{\CaseWestern}{Physics Department, Center for Education and Research in Cosmology and Astrophysics, Case Western Reserve University,Cleveland, OH, USA 44106}
\altaffiltext{\ArtInstChicago}{Liberal Arts Department, School of the Art Institute of Chicago, Chicago, IL, USA 60603}
\altaffiltext{\KASI}{Korea Astronomy and Space Science Institute, Daejeon 305-348, Republic of Korea}
\altaffiltext{\LLNL}{Institute of Geophysics and Planetary Physics, Lawrence Livermore National Laboratory, Livermore, CA 94551}
\altaffiltext{\Dunlap}{Dunlap Institute for Astronomy \& Astrophysics, University of Toronto, 50 St George St, Toronto, ON, M5S 3H4, Canada}
\altaffiltext{\Toronto}{Department of Astronomy \& Astrophysics, University of Toronto, 50 St George St, Toronto, ON, M5S 3H4, Canada}
\altaffiltext{\CTIO}{Cerro Tololo Inter-American Observatory, Casilla 603, La Serena, Chile}

\title{SPT-GMOS: A Gemini/GMOS-South Spectroscopic Survey of Galaxy Clusters in the SPT-SZ Survey}

\author{M.~B.~Bayliss\altaffilmark{\Harvard,\CfA,\Colby},
J.~Ruel\altaffilmark{\Harvard},
C.~W.~Stubbs\altaffilmark{\Harvard,\CfA}, 
  S.~W.~Allen\altaffilmark{\StanfordKPAC,\StanfordPhys,\SLAC},
  D.~E.~Applegate\altaffilmark{\AIfA},
  M.~L.~N.~Ashby\altaffilmark{\CfA},
  M.~Bautz\altaffilmark{\MIT},
  B.~A.~Benson\altaffilmark{\AAUChicago,\KICPChicago,\FNAL},
  L.~E.~Bleem\altaffilmark{\KICPChicago,\PhysicsUChicago,\Argonne},
  S.~Bocquet\altaffilmark{\KICPChicago,\Argonne,\Munich,\ExcellenceCluster},
  M.~Brodwin\altaffilmark{\Miss},
  R.~Capasso\altaffilmark{\Munich,\ExcellenceCluster}
  J.~E.~Carlstrom\altaffilmark{\AAUChicago,\KICPChicago,\PhysicsUChicago,\Argonne,\EFIChicago},
  C.~L.~Chang\altaffilmark{\AAUChicago,\KICPChicago,\Argonne},
  I.~Chiu\altaffilmark{\Munich,\ExcellenceCluster},
  H-M.~Cho\altaffilmark{\NIST},
  A.~Clocchiatti\altaffilmark{\PUC},
  T.~M.~Crawford\altaffilmark{\AAUChicago,\KICPChicago}
  A.~T.~Crites\altaffilmark{\Caltech},
  T.~de~Haan\altaffilmark{\Berkeley, \McGill},
  S.~Desai\altaffilmark{\Munich,\ExcellenceCluster},
  J.~P.~Dietrich\altaffilmark{\Munich,\ExcellenceCluster},
  M.~A.~Dobbs\altaffilmark{\McGill,\CIFAR},
  A.~N.~Doucouliagos\altaffilmark{\Melbourne},
  R.~J.~Foley\altaffilmark{\UCSC,\illast,\illphy},
  W.~R.~Forman\altaffilmark{\CfA},
  G.~P.~Garmire\altaffilmark{\Huntingdon},
  E.~M.~George\altaffilmark{\Berkeley,\MPE},
  M.~D.~Gladders\altaffilmark{\AAUChicago,\KICPChicago},
  A.~H.~Gonzalez\altaffilmark{\UFlorida},
  N.~Gupta\altaffilmark{\Munich,\ExcellenceCluster},
  N.~W.~Halverson\altaffilmark{\Colorado},
  J.~Hlavacek-Larrondo\altaffilmark{\UMon,\KIPAC,\StanfordPhys},
  H.~Hoekstra\altaffilmark{\LeidenObservatory},
  G.~P.~Holder\altaffilmark{\McGill},
  W.~L.~Holzapfel\altaffilmark{\Berkeley},
  Z.~Hou\altaffilmark{\KICPChicago,\PhysicsUChicago},
  J.~D.~Hrubes\altaffilmark{\UChicago},
  N.~Huang\altaffilmark{\Berkeley},
  C.~Jones\altaffilmark{\CfA},
  R.~Keisler\altaffilmark{\KICPChicago,\StanfordKPAC,\StanfordPhys,\PhysicsUChicago},
  L.~Knox\altaffilmark{\Davis},
  A.~T.~Lee\altaffilmark{\Berkeley,\LBNL},
  E.~M.~Leitch\altaffilmark{\AAUChicago,\KICPChicago},
  A.~von~der~Linden\altaffilmark{\StonyBrook,\DARK,\StanfordKPAC,\StanfordPhys},
  D.~Luong-Van\altaffilmark{\UChicago},
  A.~Mantz\altaffilmark{\KICPChicago,\StanfordKPAC,\StanfordPhys},
  D.~P.~Marrone\altaffilmark{\Arizona},
  M.~McDonald\altaffilmark{\MIT},
  J.~J.~McMahon\altaffilmark{\Michigan},
  S.~S.~Meyer\altaffilmark{\AAUChicago,\KICPChicago,\PhysicsUChicago,\EFIChicago},
  L.~M.~Mocanu\altaffilmark{\AAUChicago,\KICPChicago},
  J.~J.~Mohr\altaffilmark{\MPE,\Munich,\ExcellenceCluster},
  S.~S.~Murray\altaffilmark{\CfA},
  S.~Padin\altaffilmark{\AAUChicago,\KICPChicago,\Caltech},
  C.~Pryke\altaffilmark{\Minnesota},
  D.~Rapetti\altaffilmark{\Munich, \ExcellenceCluster},
  C.~L.~Reichardt\altaffilmark{\Melbourne},
  A.~Rest\altaffilmark{\STScI},
  J.~E.~Ruhl\altaffilmark{\CaseWestern},
  B.~R.~Saliwanchik\altaffilmark{\CaseWestern},
  A.~Saro\altaffilmark{\Munich, \ExcellenceCluster},
  J.~T.~Sayre\altaffilmark{\Colorado},
  K.~K.~Schaffer\altaffilmark{\KICPChicago,\EFIChicago,\ArtInstChicago},
  T.~Schrabback\altaffilmark{\AIfA},
  E.~Shirokoff\altaffilmark{\AAUChicago,\KICPChicago},
  J.~Song\altaffilmark{\Michigan,\KASI},
  H.~G.~Spieler\altaffilmark{\LBNL},
  B.~Stalder\altaffilmark{\CfA},
  S.~A.~Stanford\altaffilmark{\Davis,\LLNL},
  Z.~Staniszewski\altaffilmark{\CaseWestern},
  A.~A.~Stark\altaffilmark{\CfA},
  K.~T.~Story\altaffilmark{\StanfordKPAC,\StanfordPhys},
  K.~Vanderlinde\altaffilmark{\Dunlap,\Toronto},
  J.~D.~Vieira\altaffilmark{\illast,\illphy},
  A. Vikhlinin\altaffilmark{\CfA},
  R.~Williamson\altaffilmark{\AAUChicago,\KICPChicago,\Caltech},
  and
  A.~Zenteno\altaffilmark{\CTIO}}

\email{mbbayliss@cfa.harvard.edu}

\begin{abstract}

We present the results of SPT-GMOS, a spectroscopic survey with the Gemini Multi-Object Spectrograph 
(GMOS) on Gemini South. The targets of SPT-GMOS are galaxy clusters identified in the SPT-SZ survey, 
a millimeter-wave survey of 2500 deg$^{2}$ of the southern sky using the South Pole Telescope (SPT). 
Multi-object spectroscopic observations of 62 SPT-selected galaxy clusters were performed between 
January 2011 and December 2015, yielding spectra with radial velocity measurements for 2595 sources. 
We identify 2243 of these sources as galaxies, and 352 as stars. Of the galaxies, we identify 1579 as 
members of SPT-SZ galaxy clusters. The primary goal of these observations was to obtain spectra of 
cluster member galaxies to estimate cluster redshifts and velocity dispersions. We describe the full 
spectroscopic dataset and resulting data products, including galaxy redshifts, cluster redshifts and 
velocity dispersions, and measurements of several well-known spectral indices for each galaxy: the 
equivalent width, $W$, of [O {\small II}] $\lambda$$\lambda$3727,3729 and \hdelta, and the 
4000\AA\ break strength, D4000. We use the spectral indices to classify galaxies by spectral type 
(i.e., passive, post-starburst, star-forming), and we match the spectra against photometric catalogs 
to characterize spectroscopically-observed cluster members as a function of brightness (relative to 
m$^{\star}$). Finally, we report several new measurements of redshifts for ten bright, strongly-lensed 
background galaxies in the cores of eight galaxy clusters. Combining the SPT-GMOS dataset with 
previous spectroscopic follow-up of SPT-SZ galaxy clusters results in spectroscopic measurements 
for $> 100$ clusters, or $\sim$20\% of the full SPT-SZ sample.
%
\end{abstract}

\vspace*{0.5cm}


\section{Introduction}

Precise spectroscopic measurements of the recession velocities of distant galaxies 
are among the most important cosmological observables available for studying large 
scale structure in the universe 
\citep{geller89,colless01,colless03,eisenstein05,geller05,drinkwater10,eisenstein11,geller14}. 
On cosmological scales, galaxy 
line-of-sight recession velocities increase monotonically, on average, with their distance; 
this bulk recession velocity is known as the Hubble Flow \citep{hubble31}. The 
line-of-sight velocities of individual galaxies are perturbed off of the Hubble Flow via two 
distinct kinds of gravitational interactions: gravitational redshifts, as described by general 
relativity \citep[e.g.,][]{chant30}, and peculiar velocities induced by local gradients in the matter 
density \citep[e.g.,][]{jackson72,kaiser87}. The former effect is typically very small 
\citep[$\sim 11$km s$^{-1}$][]{wojtak11,sadeh15} and rarely observed, but 
the latter is a standard tool for constraining the statistical properties of density fluctuations 
on large scales \citep[redshift space distortions, e.g.,][]{percival09} and for measuring the 
depths of the gravitational potential wells of individual large fluctuations, namely clusters 
of galaxies 
\citep{dressler99,rines03,white10,rines13,geller13,saro13,sifon13,ruel14,bocquet15,kirk15,sifon16}. 

The first large samples of galaxy clusters were identified as over-densities of galaxies \citep{abell58}, 
and have more recently been identified out to high redshift using optical and near-infrared 
observations \citep[e.g.;][]{gladders00,koester07a,eisenhardt08,wen12,rykoff14}. Galaxy clusters 
are also identifiable from the observational signatures associated with the hot, diffuse intracluster gas 
that accounts for the vast majority of their baryonic content, and there is a long history in the literature 
of galaxy cluster samples based on the characteristic extended X-ray emission that results from that 
hot intracluster gas \citep[e.g.;][]{edge90,ebeling98,rosati98,bohringer00,bohringer01,burenin07,pacaud16}.

In recent years astronomers have been able to produce dedicated surveys at millimeter 
wavelengths that identify massive galaxy clusters via the Sunyaev Zel'dovich (SZ) 
effect \citep{sunyaev72,sunyaev80}. 
The {\it Planck} satellite \citep{planck13-XX}, the Atacama Cosmology Telescope 
\citep[ACT;][]{marriage11b,hasselfield2013}, and the South Pole Telescope 
\citep{staniszewski09,vanderlinde10,williamson11,reichardt13,bleem15} 
have all published SZ-based galaxy cluster catalogs. Galaxy cluster surveys that select clusters 
based on the SZ effect and have sufficient angular resolution to resolve galaxy clusters at all 
redshifts (e.g., SPT and ACT with $\sim$ 1\arcmin\ beams) benefit from an approximately flat 
selection in mass beyond $z \gtrsim 0.25$ \citep{carlstrom02}, which results in 
clean, mass-selected samples extending well beyond a redshift of $z = 1$. These 
SZ-selected galaxy cluster samples present us with new opportunities to characterize 
the properties of galaxy clusters in well-defined bins of mass and redshift. Such samples 
can be powerful tools for testing cosmological models via the growth of structure 
\citep[e.g.,][]{planck13-XX,dehaan16}, and for understanding the 
astrophysical processes that govern how galaxies evolve in the most overdense environments 
\citep[e.g.,][]{zenteno11,bayliss14c,chiu16,hennig16,mcdonald16,sifon16,zenteno16}

In this work we present spectroscopic observations from SPT-GMOS --- a large NOAO survey 
program (11A-0034, PI: C. Stubbs) using the Gemini Multi-Object Spectrograph \citep[GMOS;][]{hook04} 
on Gemini-South. The objective of this program was to measure cosmological redshifts of cluster 
member galaxies and other galaxies along the line of sight toward galaxy clusters that were 
identified in the SPT-SZ survey \citep{bleem15}. In this work we describe observations of 
62 galaxy clusters carried out between September 2011 and May 2015. This program can be 
combined with numerous smaller programs to obtain spectroscopic observations of SPT 
clusters \citep{brodwin10,foley11,stalder13,bayliss14c,ruel14} to produce a sample 
of $\sim100$ SPT clusters that have been followed up with multi-object spectroscopy (MOS).

Throughout the paper, we assume a standard Lambda Cold Dark Matter ($\Lambda$CDM) 
cosmology with $\Omega_{M} = 0.3$, $\Omega_{\Lambda} = 0.7$, $\sigma_{8} = 0.8$, 
$H_{0} = 70$ km s$^{-1}$ Mpc$^{-1}$, and $h=H_{0}/100= 0.7$. All quoted magnitudes are 
in the AB system.

\begin{figure}[t]
\includegraphics[scale=0.56]{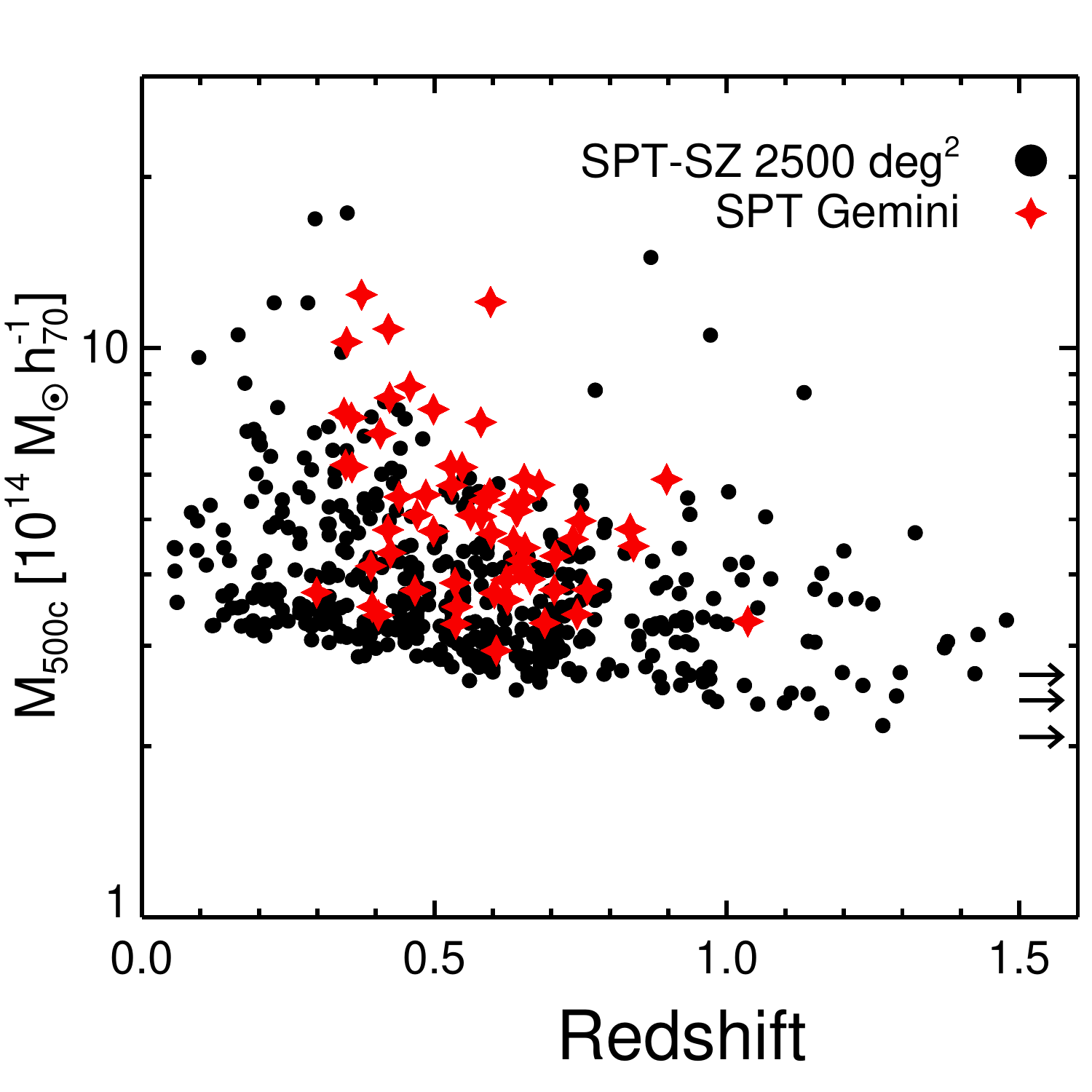}
\caption{\scriptsize{
The full 2500 deg$^{2}$ SPT-SZ sample of 516 confirmed galaxy clusters (black dots) with the 
62 SPT-GMOS clusters marked with red stars. Redshifts and masses for the full SPT-SZ sample 
are those described in \citet{bleem15}, where three clusters only have approximate redshift lower 
limits based on {\it Spitzer} infrared imaging.}}
\label{fig:sample}
\end{figure}

\section{The SPT-GMOS Survey and Observations} 
\label{sec:spectroscopy}

\subsection{Motivation and Design}
\label{sec:motivation}

SPT-GMOS was motivated by the challenge of calibrating mass-observable relations for galaxy clusters, and the reality 
that current cosmological constraints from galaxy cluster counts are systematically limited by uncertainty in estimating 
cluster masses \citep{majumdar03,majumdar04,rozo10,williamson11,benson13,planck13-XX,vonderLinden14,bocquet15,dehaan16}.
The primary goal of the SPT-GMOS survey is to measure line-of-sight velocity dispersions for a large fraction 
of the SPT-SZ galaxy cluster sample. These dispersions can be combined with other mass-proxies (X-ray, weak 
lensing) to more accurately calibrate the SZ-mass scaling relation so that precise dark energy constraints can be 
obtained using the SPT cluster sample. To this end, the SPT-GMOS program represents a survey-level investment 
of resources to expand the sample of velocity dispersion measurements that we have for SPT-SZ galaxy 
clusters. The survey results presented here greatly expand upon previously published spectroscopic 
measurements of SPT clusters that were obtained through numerous observing programs \citep{ruel14}, 
including the results of the first semester of SPT-GMOS spectroscopy.

The data presented in this paper follow the same observational design described by \citet{ruel14}. Specifically, 
we pursue a relatively ``low-N'' strategy to measure velocity dispersions for a large number of clusters using 
typically N$\lesssim 40$ cluster members. This strategy allows us to efficiently observe a large number of 
galaxy clusters; we design two multi-object spectroscopic masks for each cluster, generally placing slits on 
approximately $60-70$ galaxies within a $\sim$3\arcmin\ radius of the center of each targeted galaxy cluster. 
The efficiency advantage of this approach is twofold. Firstly, by pursuing $\lesssim 40$ cluster member galaxies 
we avoid reliance on measuring redshifts for extremely faint cluster members, which means that we require 
significantly less integration time for each spectroscopic mask. In practice this means that all of the masks 
observed in the SPT-GMOS program are exposed for less than 1.9hrs, and the vast majority for less than 
1.5hrs. Secondly, we require only two spectroscopic masks per cluster, which results in a total integration 
time investment that is always less than $<3.8$hrs per cluster, and less than $<2.5$hrs per cluster for the 
vast majority ($\sim80$\%) of observed clusters (see Table~\ref{tab:obs}). The final Gemini-S observing 
allocation for SPT-GMOS concluded at the end of the 2015B semester. Over the course of the entire survey 
we observed 121 individual spectroscopic masks targeting 62  SPT-SZ galaxy clusters.

All final data products from SPT-GMOS are publicly released via the Harvard Dataverse 
Network\footnote{https://dataverse.harvard.edu/dataverse/SPT\_Clusters}, which has hosted all 
partial SPT-GMOS data releases to date.

\subsection{The South Pole Telescope Galaxy Cluster Sample}

The galaxy clusters observed in the SPT-GMOS are all drawn from the SPT-SZ survey, completed in 
November 2011 \citep{carlstrom11}. The full SPT-SZ survey covered approximately 2500 deg$^2$ of 
the southern sky at 95, 150, and 220 GHz with an angular resolution of $\sim$1\arcmin. Noise levels 
in the SPT maps are $\sim$ 40, 18, and 70 $\mu$K-arcmin in the 95, 150, and 220 GHz bands, 
respectively. Galaxy cluster candidates were identified in the SPT-SZ survey via the signal imprinted 
by the inverse Compton scattering of cosmic microwave background 
(CMB) photons off of hot intra-cluster gas, i.e., via the thermal SZ effect. 

The full SPT-SZ galaxy cluster sample contains 409 (677) cluster 
candidates with SZ detection significance, $\xi_\mathrm{SPT} \geq 5 (4.5)$, with the 
$\xi_\mathrm{SPT} \geq 5$ 
candidates having a measured purity of 95\% \citep{song12,bleem15}. The SPT cluster 
selection extends to high redshift \citep[e.g., $z\sim1.5$][]{bayliss14c} and is approximately flat 
in mass beyond $z \sim 0.25$, with a mass 
threshold of M$_{500c} \gtrsim 5 \times 10^{14}$ \msun~h$_{70}^{-1}$ at $z = 0.25$, and M$_{500c} 
\gtrsim 3 \times 10^{14}$ \msun~h$_{70}^{-1}$ at $z > 1.0$ 
\citep[Figure~\ref{fig:sample};][]{benson13,bocquet15,dehaan16}, where M$_{500c}$ refers to the mass contained 
within the radius for which the mean enclosed density is 500 times the critical density of the universe. For more 
information regarding the survey strategy and data analysis we direct the reader to the publications describing the 
SPT-SZ survey and resulting cluster catalogs in detail \citep{staniszewski09,vanderlinde10,schaffer11,reichardt13,bleem15}. 

\subsection{Gemini/GMOS-South Spectroscopy}

\subsubsection{Selecting Cluster Targets}

Individual target selection for GMOS spectroscopy was determined by three main factors:

\begin{itemize}

	\item First consideration was given to clusters that are being targeted as part of a broad program to 
		support multi-wavelength mass calibration of SPT-SZ galaxy clusters. Where possible we 
		obtained SPT-GMOS spectroscopy for systems that already had weak lensing and/or X-ray data. 

	\item The pool of available SPT-SZ clusters changed over the four year lifetime of the SPT-GMOS 
		survey because the full SPT-SZ galaxy cluster catalog was not finalized until approximately two 
		years after SPT-GMOS spectroscopic observations began. 

	\item SPT-GMOS targets were restricted to a redshift range of $0.3 < z < 0.8$ for the first four years of 
		survey observations due to the limitations inherent to the \textit{e2v} detectors that were used in  
		GMOS-South prior to the 2014B semester.

\end{itemize}

The ultimate goal was to obtain comprehensive multi-wavelength follow-up for as many SPT-SZ clusters 
as possible, which will optimize the potential for scaling relation analyses using SZ, X-ray, 
lensing, and dynamical observables. Other mature SPT-SZ cluster follow-up programs include 
a large {\it Chandra-XVP} \citep[PI: B. Benson; see][]{mcdonald13,mcdonald14}, and weak lensing programs 
 \citep{high12}. These programs are converging toward a sample of $\sim$100 SPT-SZ clusters that 
 have spectroscopic/velocity dispersions, weak lensing measurements, and X-ray observations.

In practice, the SPT-GMOS cluster targets were chosen preferentially from the higher significance --- 
and higher mass/purity --- SPT galaxy cluster candidates (generally $\xi_{SPT} > 5$), 
though some lower-significance clusters were observed because target selection was a rolling process. 
Specifically, the spectroscopic observations began in 2011A, prior to the completion of the full 2500 deg$^{2}$ 
SPT-SZ survey, resulting in spectroscopic targets for the first two years of the program being drawn 
from only a fraction of the ultimate 2500 deg$^{2}$ survey area --- primarily from the first 720 deg$^{2}$ 
\citep{reichardt13}.

Target selection for the SPT-GMOS survey program was further constrained to focus on low and 
moderate redshift clusters from the SPT sample, specifically those within the redshift range 
$0.3 < z < 0.8$, as estimated from red-sequence based photometric redshifts \citep{song12,bleem15}. 
We chose this range because it was a good match to the capabilities of the original GMOS-South  
instrument. Prior to August 2014 GMOS-South used thinned \textit{e2v} detectors designed to optimize 
throughput in the blue ($\lambda \lesssim 6000$\AA) while sacrificing quantum efficiency in the red; 
these detectors also exhibit severe fringing at redder ($\lambda \gtrsim 7300$\AA) wavelengths. 
The poor performance in the red made GMOS-South a suboptimal choice for pursuing galaxy redshifts 
beyond $z = 0.8$, where most of the strong spectral features that are common in cluster 
member galaxy spectra --- e.g., Ca {\small II} H\&K, \hdelta, G-band, and H$\gamma$ 
--- are redshifted into the fringe-affected wavelength range. 
 
The final SPT-GMOS sample is plotted relative to the full SPT-SZ 2500 deg$^{2}$ cluster sample in 
Figure~\ref{fig:sample}, and the complete SPT-GMOS list of clusters observed through early May 2015 
is given in Table~\ref{tab:obs}. Observations are complete for 121 custom spectroscopic slitmasks targeting 
62 individual galaxy cluster fields.  Additional observations remained active in the 
Gemini-South queue throughout 2015. Reduction of observations taken through the end of the 2015B are in 
progress and will be made publicly available alongside the data presented here. As of 2014B 
the GMOS detector was upgraded to new red-sensitive chips, and we relaxed the $z < 0.8$ redshift 
constraint for selected cluster targets beginning in that semester.

{
\LongTables
\def\arraystretch{1.20}
\begin{deluxetable*}{cccccccccc}
\tablecaption{Gemini/GMOS-South Observations of SPT-SZ Galaxy Clusters\label{tab:obs}}
\tablewidth{0pt}
\tabletypesize{\tiny}
\tablehead{
\colhead{Cluster} &
\colhead{RA} &
\colhead{Dec} &
\colhead{$\xi_{SPT}$} &
\colhead{Program ID} &
\colhead{Mask} &
\colhead{Grating} &
\colhead{Filter} &
\colhead{$\lambda_{c}$ (\AA)} &
\colhead{t$_{exp}$ (s) }}
\startdata  
SPT-CL~J0013-4906 & 00:13:19.0 & -49:06:54 & 11.22 & GS-2012A-Q-37 & 01 & {\tiny B600\_G5323} & {\tiny ---} & 5400, 5500 & 2200 \\
   &   &  &   &  GS-2012A-Q-37 & 02 & {\tiny B600\_G5323} & {\tiny ---} & 5400, 5500 & 2200 \\
SPT-CL~J0033-6326 & 00:33:54.4 & -63:26:46 &  7.50 & GS-2012B-Q-29 & 05 & {\tiny B600\_G5323} & {\tiny ---} & 6000, 6100 & 2600 \\
   &   &  &   & GS-2012B-Q-29 & 06 & {\tiny B600\_G5323} & {\tiny ---} & 5900, 6000 & 2600 \\
SPT-CL~J0040-4407 & 00:40:49.2 & -44:07:58 & 19.34 & GS-2011A-C-03 & 09 & {\tiny B600\_G5323} & {\tiny ---} & 5200, 5250 & 2100  \\
   &   &  &   & GS-2011A-C-03 & 10 & {\tiny B600\_G5323} & {\tiny ---} & 5200, 5250  & 2100 \\
SPT-CL~J0102-4603 & 01:02:40.6 & -46:03:53 &  7.33 & GS-2012B-Q-29 & 13 & {\tiny R400\_G5325} & {\tiny GG455\_G0329} & 7000, 7100 & 5760 \\
   &   &  &   & GS-2012B-Q-29 & 14 & {\tiny R400\_G5325} & {\tiny GG455\_G0329} & 7000, 7100 & 5760 \\
SPT-CL~J0106-5943 & 01:06:27.7 & -59:43:16 &  9.57 & GS-2012B-Q-59 & 13 & {\tiny B600\_G5323} & {\tiny ---} & 5800, 5900 & 2400 \\
   &   &  &   & GS-2012B-Q-59 & 14 & {\tiny B600\_G5323} & {\tiny ---} & 5800, 5900 & 2400 \\
SPT-CL~J0118-5156 & 01:18:23.8 & -51:56:36 &  5.97 & GS-2011B-C-06 & 52 & {\tiny R400\_G5325} & {\tiny GG455\_G0329} & 6600, 6650 & 4560 \\
   &   &  &   &  GS-2011B-C-06 & 53 & {\tiny R400\_G5325} & {\tiny GG455\_G0329} & 6600, 6650 & 4560 \\
SPT-CL~J0123-4821 & 01:23:10.1 & -48:21:31 &  6.92 & GS-2012B-Q-29 & 23 & {\tiny R400\_G5325} & {\tiny GG455\_G0329} & 7000, 7100 & 5760 \\
SPT-CL~J0142-5032 & 01:42:10.8 & -50:32:37 & 10.12 & GS-2012B-Q-29 & 21 & {\tiny R400\_G5325} & {\tiny GG455\_G0329} & 7000, 7100 & 4800 \\
   &   &  &   & GS-2012B-Q-29 & 22 & {\tiny R400\_G5325} & {\tiny GG455\_G0329} & 7000, 7100 & 4800 \\
SPT-CL~J0200-4852 & 02:00:34.5 & -48:52:32 &  7.38 & GS-2012B-Q-29 & 07 & {\tiny B600\_G5323} & {\tiny ---} & 5800, 5900 & 2500 \\
   &   &  &   & GS-2012B-Q-29 & 08 & {\tiny B600\_G5323} & {\tiny ---} & 5800, 5900 & 2500 \\
SPT-CL~J0205-6432 & 02:05:07.1 & -64:32:44 &  5.83 & GS-2011B-C-06 & 56 & {\tiny R400\_G5325} & {\tiny GG455\_G0329} & 6600, 6650 & 4800 \\
   &   &  &   & GS-2011B-C-06 & 57 & {\tiny R400\_G5325} & {\tiny GG455\_G0329} & 6600, 6650 & 4800 \\
SPT-CL~J0212-4657 & 02:12:25.5 & -46:57:00 & 10.05 & GS-2012B-Q-29 & 19 & {\tiny R400\_G5325} & {\tiny GG455\_G0329} & 6600, 6700 & 3600 \\
   &   &  &   &  GS-2012B-Q-29 & 20 & {\tiny R400\_G5325} & {\tiny GG455\_G0329} & 6600, 6700 & 3600 \\
SPT-CL~J0233-5819 & 02:33:01.3 & -58:19:38 &  6.55 & GS-2011B-C-06 & 54 & {\tiny R400\_G5325} & {\tiny GG455\_G0329} & 6600, 6650 & 4800 \\
SPT-CL~J0243-4833 & 02:43:39.3 & -48:33:36 & 13.90 & GS-2012B-Q-29 & 03 & {\tiny R400\_G5325} & {\tiny ---} & 6000, 6100 & 2600 \\
   &   &  &   &  GS-2012B-Q-29 & 04 & {\tiny R400\_G5325} & {\tiny ---} & 6000, 6100 & 3900 \\
SPT-CL~J0243-5930 & 02:43:26.8 & -59:30:44 &  7.67 & GS-2012B-Q-29 & 09 & {\tiny R400\_G5325} & {\tiny GG455\_G0329} & 6600, 6700 & 4200 \\
   &   &  &   &  GS-2012B-Q-29 & 10 & {\tiny R400\_G5325} & {\tiny GG455\_G0329} & 6600, 6700 & 4200 \\
SPT-CL~J0245-5302 & 02:45:30.7 & -53:02:09 & ...$^{a}$ & GS-2011A-C-03 & 01 & {\tiny B600\_G5323} & {\tiny ---} & 5200, 5250 & 1500 \\
   &   &  &   &  GS-2011A-C-03 & 02 & {\tiny B600\_G5323} & {\tiny ---} & 5200, 5250 & 1500 \\
SPT-CL~J0252-4824 & 02:52:45.1 & -48:24:44 &  7.03 & GS-2013B-Q-72 & 07 & {\tiny B600\_G5323} & {\tiny ---} & 5800, 5900 & 2200 \\
   &   &  &   &  GS-2013B-Q-72 & 08 & {\tiny B600\_G5323} & {\tiny ---} & 5800, 5900 & 2200 \\
SPT-CL~J0304-4401 & 03:04:16.8 & -44:01:53 & 15.69 & GS-2012B-Q-59 & 05 & {\tiny R400\_G5325} & {\tiny GG455\_G0329} & 6100, 6200 & 3000 \\
   &   &  &   &  GS-2012B-Q-59 & 06 & {\tiny R400\_G5325} & {\tiny GG455\_G0329} & 6100, 6200 & 3000 \\
SPT-CL~J0307-6225 & 03:07:20.1 & -62:25:57 &  8.46 & GS-2012B-Q-29 & 01 & {\tiny R400\_G5325} & {\tiny GG455\_G0329} & 6200, 6300 & 3600 \\
   &   &  &   &  GS-2012B-Q-29 & 02 & {\tiny R400\_G5325} & {\tiny GG455\_G0329} & 6200, 6300 & 3600 \\
SPT-CL~J0310-4647 & 03:10:31.0 & -46:47:00 &  7.12 & GS-2013B-Q-72 & 11 & {\tiny R400\_G5325} & {\tiny GG455\_G0329} & 6700, 6800 & 3800 \\
   &   &  &   &  GS-2013B-Q-72 & 12 & {\tiny R400\_G5325} & {\tiny GG455\_G0329} & 6700, 6800 & 3800 \\
SPT-CL~J0324-6236 & 03:24:12.7 & -62:36:07 &  8.75 & GS-2013B-Q-25 & 09 & {\tiny R400\_G5325} & {\tiny GG455\_G0329} & 7000, 7100 & 5760 \\
   &   &  &   &  GS-2013B-Q-25 & 10 & {\tiny R400\_G5325} & {\tiny GG455\_G0329} & 7000, 7100 & 5760 \\
SPT-CL~J0334-4659 & 03:34:11.1 & -46:59:35 &  9.20 & GS-2013B-Q-72 & 13 & {\tiny B600\_G5323} & {\tiny ---} & 5800, 5900 & 2400 \\
   &   &  &   &  GS-2013B-Q-72 & 14 & {\tiny B600\_G5323} & {\tiny ---} & 5800, 5900 & 2400 \\
SPT-CL~J0348-4515 & 03:48:17.7 & -45:15:03 & 10.12 & GS-2012B-Q-59 & 01 & {\tiny B600\_G5323} & {\tiny ---} & 5400, 5500 & 1600 \\
   &   &  &   &  GS-2012B-Q-59 & 02 & {\tiny B600\_G5323} & {\tiny ---} & 5400, 5500 & 1600 \\
SPT-CL~J0352-5647 & 03:52:56.8 & -56:47:58 &  7.13 & GS-2013B-Q-25 & 11 & {\tiny R400\_G5325} & {\tiny GG455\_G0329} & 6800, 6900 & 4800 \\
   &   &  &   &  GS-2013B-Q-25 & 12 & {\tiny R400\_G5325} & {\tiny GG455\_G0329} & 6800, 6900 & 4800 \\
 SPT-CL~J0356-5337 & 03:56:20.5 & -53:37:59 & 6.02 & GS-2014B-Q-31 & 03 & {\tiny R400\_G5325} &  {\tiny GG455\_G0329} & 7800, 8000 & 4800 \\
   &   &  &   & GS-2014B-Q-31 & 04 & {\tiny R400\_G5325} &  {\tiny GG455\_G0329}  & 7800, 8000 & 6600 \\
SPT-CL~J0403-5719 & 04:03:52.3 & -57:19:25 &  5.86 & GS-2012B-Q-59 & 07 & {\tiny B600\_G5323} & {\tiny ---} & 5600, 5700 & 2000 \\
   &   &  &   &  GS-2012B-Q-59 & 08 & {\tiny B600\_G5323} & {\tiny ---} & 5600, 5700 & 2000 \\
SPT-CL~J0406-4805 & 04:06:54.6 & -48:05:11 &  8.13 & GS-2013B-Q-72 & 09 & {\tiny R400\_G5325} & {\tiny GG455\_G0329} & 6500, 6600 & 2500 \\
   &   &  &   &  GS-2013B-Q-72 & 10 & {\tiny R400\_G5325} & {\tiny GG455\_G0329} & 6500, 6600 & 2500 \\
SPT-CL~J0411-4819 & 04:11:15.7 & -48:19:18 & 15.26 & GS-2012B-Q-59 & 15 & {\tiny B600\_G5323} & {\tiny ---} & 5600, 5700 & 2200 \\
   &   &  &   &  GS-2012B-Q-59 & 16 & {\tiny B600\_G5323} & {\tiny ---} & 5600, 5700 & 2400 \\
SPT-CL~J0417-4748 & 04:17:22.8 & -47:48:50 & 14.24 & GS-2012B-Q-29 & 11 & {\tiny R400\_G5325} & {\tiny GG455\_G0329} & 6600, 6700 & 3600 \\
   &   &  &   &  GS-2012B-Q-29 & 12 & {\tiny R400\_G5325} & {\tiny GG455\_G0329} & 6600, 6700 & 3600 \\
SPT-CL~J0426-5455 & 04:26:04.8 & -54:55:10 &  8.85 & GS-2013B-Q-25 & 14 & {\tiny R400\_G5325} & {\tiny GG455\_G0329} & 7000, 7100 & 5760 \\
SPT-CL~J0438-5419 & 04:38:18.0 & -54:19:16 & 22.88 & GS-2011A-C-03 & 28 & {\tiny R400\_G5325} & {\tiny GG455\_G0329} & 5500, 5550 & 2700 \\
SPT-CL~J0456-5116 & 04:56:27.9 & -51:16:36 &  8.58 & GS-2013B-Q-25 & 17 & {\tiny R400\_G5325} & {\tiny GG455\_G0329} & 7000, 7100 & 4200 \\
   &   &  &   &  GS-2013B-Q-25 & 18 & {\tiny R400\_G5325} & {\tiny GG455\_G0329} & 7000, 7100 & 4200 \\
SPT-CL~J0511-5154 & 05:11:41.0 & -51:54:15 &  7.09 & GS-2011B-C-06 & 58 & {\tiny R400\_G5325} & {\tiny GG455\_G0329} & 6600, 6650 & 4800 \\
   &   &  &   &  GS-2011B-C-06 & 59 & {\tiny R400\_G5325} & {\tiny GG455\_G0329} & 6600, 6650 & 4800 \\
SPT-CL~J0539-5744 & 05:40:01.0 & -57:44:25 &  6.74 & GS-2012B-Q-29 & 17 & {\tiny R400\_G5325} & {\tiny GG455\_G0329} & 7100, 7200 & 4800 \\
   &   &  &   &  GS-2012B-Q-29 & 18 & {\tiny R400\_G5325} & {\tiny GG455\_G0329} & 7100, 7200 & 4800 \\
SPT-CL~J0542-4100 & 05:42:52.0 & -41:00:15 &  7.92 & GS-2013B-Q-25 & 19 & {\tiny R400\_G5325} & {\tiny GG455\_G0329} & 7000, 7100 & 4800 \\
   &   &  &   &  GS-2013B-Q-25 & 20 & {\tiny R400\_G5325} & {\tiny GG455\_G0329} & 7000, 7100 & 4800 \\
SPT-CL~J0549-6205 & 05:49:20.2 & -62:05:08 & 25.81 & GS-2012B-Q-59 & 09 & {\tiny B600\_G5323} & {\tiny ---} & 5500, 5600 & 1800 \\
   &   &  &   &  GS-2012B-Q-59 & 10 & {\tiny B600\_G5323} & {\tiny ---} & 5500, 5600 & 1800 \\
SPT-CL~J0555-6406 & 05:55:27.9 & -64:06:11 & 12.72 & GS-2012B-Q-59 & 11 & {\tiny B600\_G5323} & {\tiny ---} & 5600, 5700 & 2000 \\
   &   &  &   &  GS-2012B-Q-59 & 12 & {\tiny B600\_G5323} & {\tiny ---} & 5600, 5700 & 2000 \\
SPT-CL~J0655-5234 & 06:55:51.0 & -52:34:03 &  7.76 & GS-2013B-Q-72 & 15 & {\tiny R400\_G5325} & {\tiny GG455\_G0329} & 6400, 6500 & 2600 \\
   &   &  &   &  GS-2013B-Q-72 & 16 & {\tiny R400\_G5325} & {\tiny GG455\_G0329} & 6400, 6500 & 2600 \\
SPT-CL~J2017-6258 & 20:17:56.1 & -62:58:41 &  6.32 & GS-2013B-Q-72 & 01 & {\tiny R400\_G5325} & {\tiny GG455\_G0329} & 6500, 6600 & 2800 \\
   &   &  &   & GS-2013B-Q-72 & 02 & {\tiny R400\_G5325} & {\tiny GG455\_G0329} & 6500, 6600 & 2800 \\
SPT-CL~J2020-6314 & 20:20:06.6 & -63:14:36 &  5.38 & GS-2012A-Q-37 & 09 & {\tiny R400\_G5325} & {\tiny GG455\_G0329} & 6600, 6700 & 3600 \\
   &   &  &   &  GS-2012A-Q-37 & 10 & {\tiny R400\_G5325} & {\tiny GG455\_G0329} & 6600, 6700 & 3600 \\
SPT-CL~J2026-4513 & 20:26:27.5 & -45:13:36 &  5.24 & GS-2013B-Q-25 & 01 & {\tiny R400\_G5325} & {\tiny GG455\_G0329} & 7000, 7100 & 5760 \\
   &   &  &   & GS-2014B-Q-64 & 03 & {\tiny R400\_G5325} & {\tiny GG455\_G0329} & 7200, 7300 & 4800 \\
SPT-CL~J2030-5638 & 20:30:48.9 & -56:38:10 &  5.50 & GS-2013B-Q-72 & 03 & {\tiny B600\_G5323} & {\tiny ---} & 5800, 5900 & 2000 \\
   &   &  &   & GS-2013B-Q-72 & 04 & {\tiny B600\_G5323} & {\tiny ---} & 5800, 5900 & 2000 \\
   &   &  &   &  GS-2012A-Q-04 & 01 & {\tiny B600\_G5323} & {\tiny ---} & 5400, 5500 & 2000 \\
   &   &  &   &  GS-2012A-Q-04 & 02 & {\tiny B600\_G5323} & {\tiny ---} & 5400, 5500 & 2000 \\
SPT-CL~J2035-5251 & 20:35:12.3 & -52:51:06 &  9.71 & GS-2013A-Q-45 & 01 & {\tiny B600\_G5323} & {\tiny ---} & 5600, 5700 & 3900 \\
   &   &  &   &  GS-2013A-Q-45 & 02 & {\tiny B600\_G5323} & {\tiny ---} & 5600, 5700 & 2600 \\
SPT-CL~J2058-5608 & 20:58:21.1 & -56:08:43 &  5.01 & GS-2011A-C-03 & 03 & {\tiny R400\_G5325} & {\tiny GG455\_G0329} & 5500, 5550 & 3000 \\
   &   &  &   &  GS-2011A-C-03 & 04 & {\tiny R400\_G5325} & {\tiny GG455\_G0329} & 5500, 5550 & 3000 \\
SPT-CL~J2115-4659 & 21:15:12.3 & -46:59:27 &  5.18 & GS-2012A-Q-37 & 03 & {\tiny B600\_G5323} & {\tiny ---} & 5200, 5300 & 2000 \\
   &   &  &   &  GS-2012A-Q-37 & 04 & {\tiny B600\_G5323} & {\tiny ---} & 5200, 5300 & 2000 \\
SPT-CL~J2118-5055 & 21:18:55.6 & -50:55:56 &  5.54 & GS-2012A-Q-04 & 09 & {\tiny R400\_G5325} & {\tiny GG455\_G0329} & 6600, 6700 & 4800 \\
   &   &  &   &  GS-2011B-C-06 & 50 & {\tiny R400\_G5325} & {\tiny GG455\_G0329} & 6600, 6650 & 4320  \\
SPT-CL~J2136-4704 & 21:36:28.6 & -47:04:54 &  6.24 & GS-2011A-C-03 & 21 & {\tiny R400\_G5325} & {\tiny GG455\_G0329} & 5500, 5550 & 3000  \\
   &   &  &   &  GS-2011A-C-03 & 22 & {\tiny R400\_G5325} & {\tiny GG455\_G0329} & 5500, 5550 & 3000 \\
SPT-CL~J2140-5727 & 21:40:33.4 & -57:27:27 &  5.35 & GS-2012A-Q-37 & 05 & {\tiny B600\_G5323} & {\tiny ---} & 5600, 5700 & 2800 \\
   &   &  &   &  GS-2012A-Q-37 & 06 & {\tiny B600\_G5323} & {\tiny ---} & 5600, 5700 & 2800 \\
SPT-CL~J2146-4846 & 21:46:07.4 & -48:46:48 &  5.96 & GS-2011A-C-03 & 31 & {\tiny R400\_G5325} & {\tiny GG455\_G0329} & 5500, 5550 & 4200  \\
   &   &  &   &  GS-2011A-C-03 & 32 & {\tiny R400\_G5325} & {\tiny GG455\_G0329} & 5500, 5550 & 4200 \\
SPT-CL~J2146-5736 & 21:46:47.0 & -57:36:53 &  6.19 & GS-2012A-Q-04 & 03 & {\tiny R400\_G5325} & {\tiny GG455\_G0329} & 6600, 6700 & 4400 \\
   &   &  &   &  GS-2012A-Q-04 & 04 & {\tiny R400\_G5325} & {\tiny GG455\_G0329} & 6600, 6700 & 4400 \\
SPT-CL~J2155-6048 & 21:55:56.4 & -60:48:27 &  5.74 & GS-2011A-C-03 & 17 & {\tiny R400\_G5325} & {\tiny GG455\_G0329} & 5500, 5550 & 2700 \\
   &   &  &   &  GS-2011A-C-03 & 18 & {\tiny R400\_G5325} & {\tiny GG455\_G0329} & 5500, 5550 & 2700 \\
SPT-CL~J2159-6244 & 21:59:57.9 & -62:44:29 &  6.49 & GS-2012A-Q-37 & 07 & {\tiny B600\_G5323} & {\tiny ---} & 5600, 5700 & 2600 \\
    &   &  &   &  GS-2012A-Q-37 & 08 & {\tiny B600\_G5323} & {\tiny ---} & 5600, 5700 & 2600 \\
 SPT-CL~J2218-4519 &  22:19:00.0  & -45:19:11 & 5.54 & GS-2013B-Q-25 & 07 & {\tiny R400\_G5325} & {\tiny GG455\_G0329} & 6900, 7000 & 3200 \\
SPT-CL~J2222-4834 & 22:22:50.9 & -48:34:24 &  9.08 & GS-2012A-Q-04 & 12 & {\tiny R400\_G5325} & {\tiny GG455\_G0329} & 6600, 6700 & 4800 \\
   &   &  &   & GS-2012A-Q-04 & 13 & {\tiny R400\_G5325} & {\tiny GG455\_G0329} & 6600, 6700 & 4800 \\
SPT-CL~J2232-5959 & 22:32:35.7 & -59:59:25 &  8.80 & GS-2012A-Q-04 & 05 & {\tiny R400\_G5325} & {\tiny GG455\_G0329} & 6600, 6700 & 4000 \\
   &   &  &   &  GS-2012A-Q-04 & 06 & {\tiny R400\_G5325} & {\tiny GG455\_G0329} & 6600, 6700 & 4000  \\
SPT-CL~J2233-5339 & 22:33:19.1 & -53:39:00 &  8.29 & GS-2012A-Q-37 & 11 & {\tiny B600\_G5323} & {\tiny ---} & 5600, 5700 & 2800 \\
   &   &  &   &  GS-2012A-Q-37 & 12 & {\tiny B600\_G5323} & {\tiny ---} & 5600, 5700 & 2800 \\
 SPT-CL~J2245-6206 & 22:45:41.4 & -62:02:38  & 8.74 &  GS-2012A-Q-04 & 07 &  {\tiny R400\_G5325} & {\tiny GG455\_G0329} & 6700, 6800 &  3000 \\
   &   &  &   &  GS-2012A-Q-04 & 08 &  {\tiny R400\_G5325} & {\tiny GG455\_G0329} & 6700, 6800 &  3000 \\
 SPT-CL~J2258-4044 & 22:58:49.2 & -40:44:19 & 10.95 & GS-2014B-Q-31 & 05 & {\tiny R400\_G5325} & {\tiny GG515\_G0330} & 7600, 7700 & 4400 \\
   &   &  &   & GS-2014B-Q-31 & 06 & {\tiny R400\_G5325} & {\tiny GG515\_G0330} & 7600, 7700 & 4400 \\
 SPT-CL~J2301-4023 & 23:01:51.2 & -40:23:16 & 8.09 & GS-2014B-Q-64 & 04 & {\tiny R400\_G5325} & {\tiny GG455\_G0329} & 7200, 7300 & 4800 \\ 
   &   &  &   &  GS-2014B-Q-64 & 05 & {\tiny R400\_G5325} & {\tiny GG455\_G0329} & 7200, 7300 & 4800 \\ 
SPT-CL~J2306-6505 & 23:06:55.1 & -65:05:27 &  9.22 & GS-2012A-Q-37 & 13 & {\tiny R400\_G5325} & {\tiny GG455\_G0329} & 6600, 6700 & 5400  \\
   &   &  &   &  GS-2012A-Q-37 & 14 & {\tiny R400\_G5325} & {\tiny GG455\_G0329} & 6600, 6700 & 3600 \\
SPT-CL~J2325-4111 & 23:25:13.0 & -41:11:45 & 12.50 & GS-2011A-C-03 & 25 & {\tiny B600\_G5323} & {\tiny ---} & 5200, 5250 & 1800 \\
   &   &  &   & GS-2011A-C-03 & 26 & {\tiny B600\_G5323} & {\tiny ---} & 5200, 5250 & 1800 \\
SPT-CL~J2335-4544 & 23:35:08.7 & -45:44:19 & 10.37 & GS-2013B-Q-72 & 05 & {\tiny R400\_G5325} & {\tiny GG455\_G0329} & 6500, 6600 & 2800 \\
   &   &  &   &  GS-2013B-Q-72 & 06 & {\tiny R400\_G5325} & {\tiny GG455\_G0329} & 6500, 6600 & 2800 \\
SPT-CL~J2344-4243 & 23:44:44.3 & -42:43:15 & 27.44 & GS-2011A-C-03 & 29 & {\tiny R400\_G5325} & {\tiny GG455\_G0329} & 5500, 5550 & 4200 \\
   &   &  &   &  GS-2011A-C-03 & 30 & {\tiny R400\_G5325} & {\tiny GG455\_G0329} & 5500, 5550 & 4200
\enddata
\tablecomments{This table summarizes the SPT-GMOS observations for each individual spectroscopic mask 
that was observed. The columns report, from left to right, the cluster name, cluster coordinates, Gemini 
program ID, the mask number within the Gemini program, the grating used, the order-sorting filter used (if any), 
the central wavelengths used for individual spectroscopic exposures, and the total integration time for which 
the mask was exposed.}
\tablenotetext{a}{There is not a reliable SZ measurement available for this cluster due to a nearby mm-bright point source.}
\end{deluxetable*}
}

\begin{figure*}[t]
\centering
\includegraphics[scale=0.595]{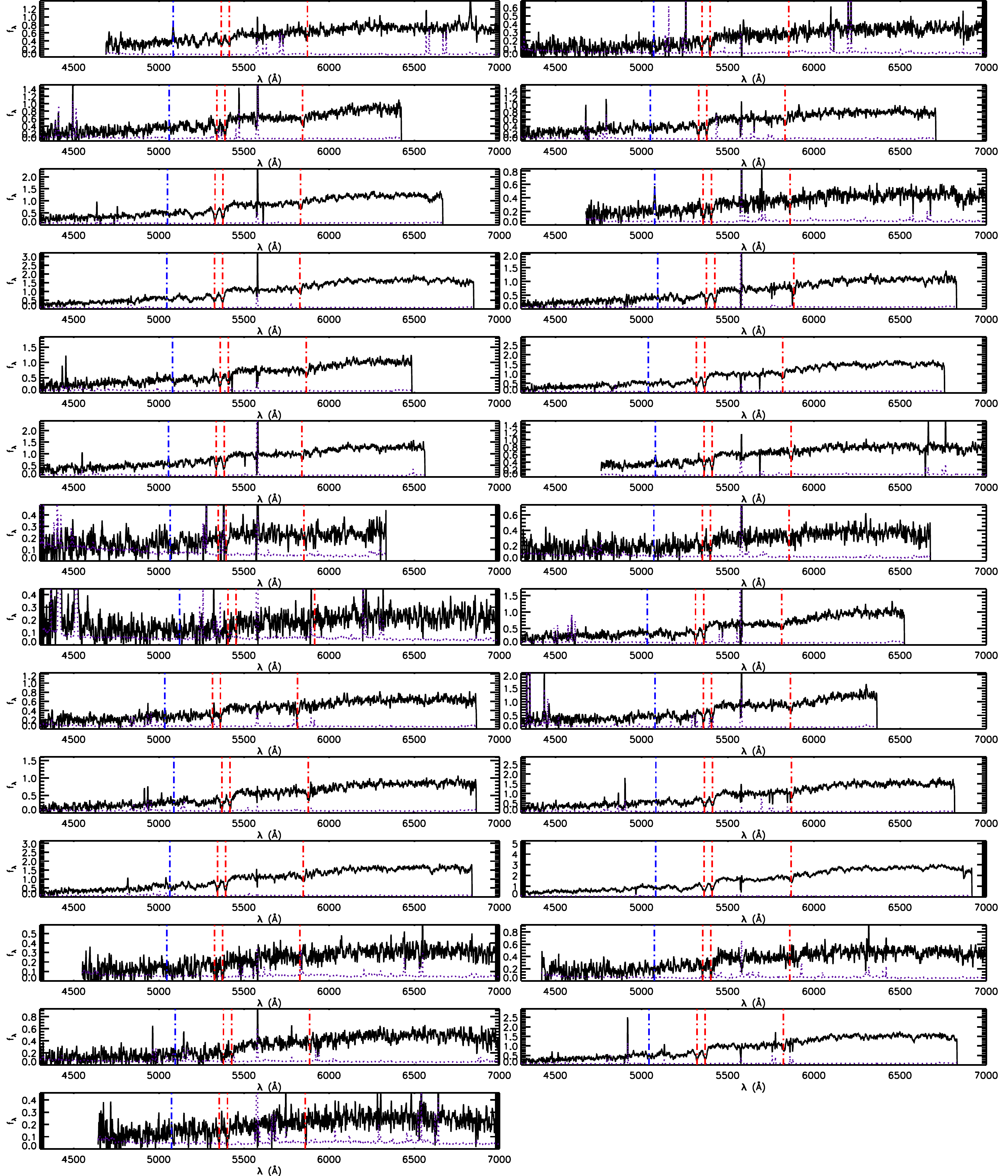}
\caption{\scriptsize{
Final 1-dimensional spectra for all 27 SPT-GMOS galaxies that we classify as cluster members 
of SPT-CL~J0348-4515 at $z=0.3592$, a galaxy cluster near the lower redshift range of clusters targeted 
in SPT-GMOS. These are example data taken in standard MOS mode. Each panel contains the 
spectrum for a single galaxy, all of 
which are plotted as a function of the observed/instrumental wavelength over a common wavelength 
interval ($\Delta \lambda = 4300-7000$), with the uncertainty per pixel over-plotted as a purple dotted 
line. In each panel we also over-plot three vertical dot-dashed lines that indicate the locations of [O {\tiny II}] 
$\lambda$3727 in emission (blue), as well as Ca {\tiny II} H \& K and G-Band in absorption (red) at the 
spectroscopic redshift measured for that galaxy. Along with Figure~\ref{fig:2222spectra}, these data 
demonstrate the typical range in S/N of SPT-GMOS cluster galaxy spectra.}}
\label{fig:0348spectra}
\end{figure*}

\begin{figure*}[t]
\centering
\includegraphics[scale=0.595]{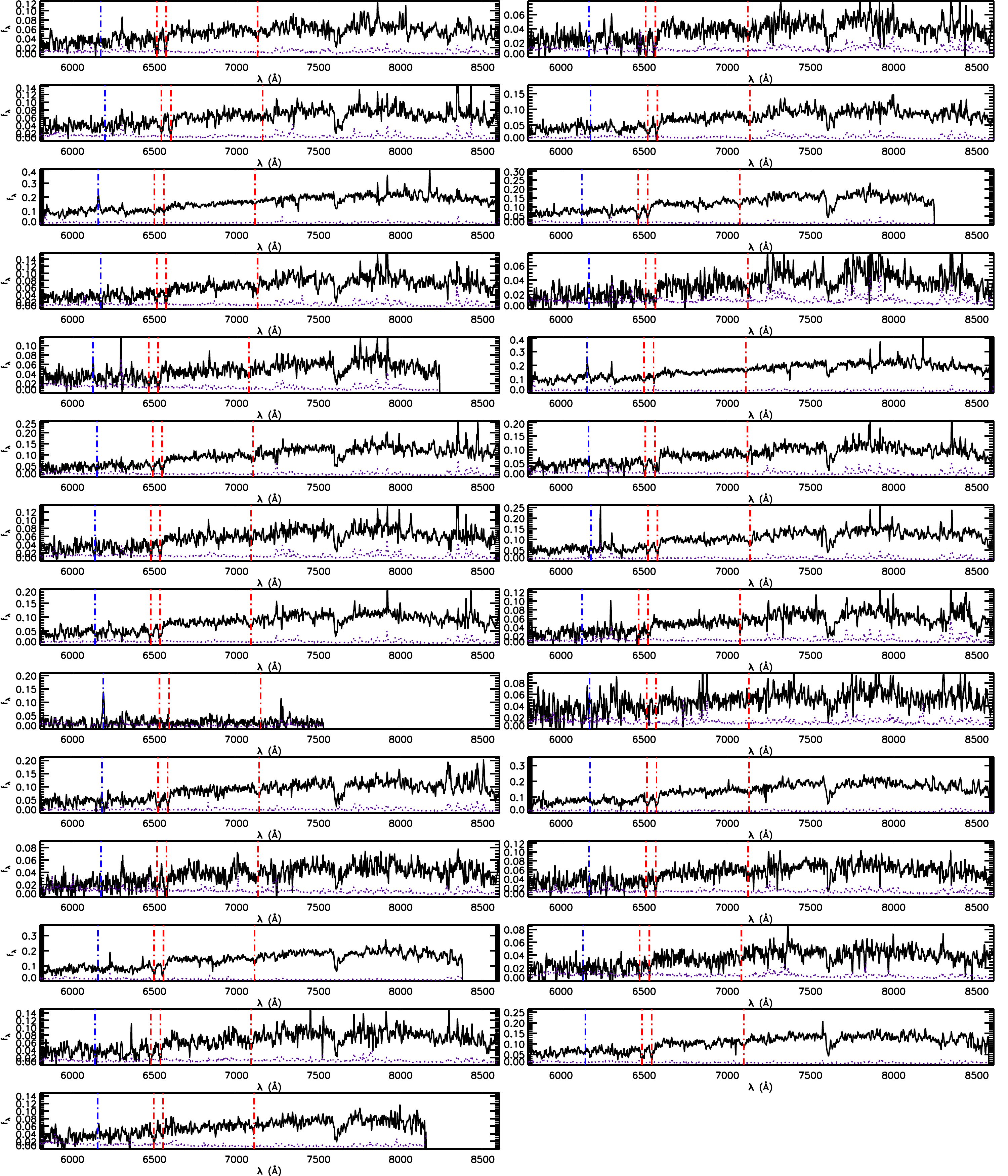}
\caption{\scriptsize{
Final 1-dimensional spectra for all 27 SPT-GMOS galaxies that we classify as cluster members of 
SPT-CL~J2222-4834 at $z=0.6519$, a galaxy cluster near the upper redshift range of clusters targeted 
in SPT-GMOS. These are example data taken in N\&S mode. Each panel contains the spectrum for a 
single galaxy, all of which are plotted as a function of the observed/instrumental wavelength over a common 
wavelength interval ($\Delta \lambda = 6200-8800$), with the uncertainty per pixel over-plotted as a 
purple dotted line. In each panel we also over-plot three vertical dot-dashed lines that locations of 
[O {\tiny II}] $\lambda$3727 in emission (blue), as well as Ca {\tiny II} H \& K and G-Band in absorption 
(red) at the spectroscopic redshift measured for that galaxy. Along with Figure~\ref{fig:0348spectra}, 
these data demonstrate the typical range in S/N of SPT-GMOS cluster galaxy spectra.}}
\label{fig:2222spectra}
\end{figure*}

\subsubsection{Instrumental Setup}
\label{subsubsec:setup}

Our observations are divided broadly into two groups: one using the B600\_G5323 grating with no 
filter, and one using the R400\_G5325 grating with the GG455\_G0329 long-pass filter. The decision 
of which grating to use was made based on the best-available photometric redshift estimate of each 
cluster \citep[e.g.,][]{song12,bleem15}, where clusters with $z_{phot} \leq 0.45$ were observed with the 
B600\_G5323 grating, and those having $z_{phot} > 0.45$ with the R400\_G5325 grating; this division 
was chosen to ensure that each cluster field was observed using a grating that has optimal throughput in 
the wavelength range where important spectral features appear at each approximate cluster redshift. 

The primary features of interest include [O {\small II}] $\lambda$3727,3729 (hereafter \oii), Ca {\small H\&K}, 
the Franhaufer G-band (a complex of Ca and Fe lines), the Balmer break, and the n $= 6 \longleftrightarrow 
2$ (hereafter \hdelta), n $= 5 \longleftrightarrow 2$ (hereafter \hgamma), and n $= 4 \longleftrightarrow 2$ (hereafter 
H-$\beta$) hydrogen Balmer lines. Central wavelengths were chosen to disperse spectra such that 
$\lambda = 4300$\AA\ would fall approximately in the middle of the detector 
for a slit placed near the middle of the GMOS-South focal plane. We binned the detector by a factor of two in 
the spectral/dispersion direction for all observations. We generally left the detector unbinned along the 
spatial direction (i.e., along the slit) to provide the best-possible sampling along the slits, though some 
early observations were binned by a factor of two along the spatial direction.

The sole exception to these standard setups were the observations of SPT-CL~J0243-4833, which  
used the R400\_G5325 grating {\it without} a long-pass order-sorting filter. This was an experimental 
setup that was used to evaluate the benefits of observing without the filter, which imposes an additional 
$\sim$5-10\% throughput loss, and relying on the throughput curve of the R400\_G5325 grating to serve 
a similar purpose to the long-pass filter. This setup was not used regularly because of the additional 
difficulties that it imposed on wavelength calibrations that result from second order images of arc lamp 
emission lines. 

Grating, filter, and central wavelength choices for all cluster observations are listed in Table~\ref{tab:obs}. 
The consistent slit widths and instrumental setups used for our observations result in spectra that all 
have similar spectral resolutions and corresponding resolving powers, $d\lambda \simeq 7 - 9$\AA\ 
and $R \simeq 600-1000$, respectively. We chose the total spectroscopic integration times 
(see Table~\ref{tab:obs}) to match the prediction from the GMOS-South integration time 
calculator\footnote{http://www.gemini.edu/sciops/instruments/integration-time-calculators/gmoss-itc} 
for the time necessary to obtain a signal-to-noise (S/N) $=$ 5 per spectral 
element immediately blueward of the 4000\AA\ break for a typical $m^{\star} +1$ passive galaxy at 
the redshift of each cluster.

\subsubsection{Micro Nod-and-shuffle Multi-object Spectroscopy}

Due to the poor GMOS-South \textit{e2v} detector performance at redder wavelengths we observed 
galaxy clusters with photometric redshift estimates $z_{phot} \geq 0.65$ in 
``microscopic'' nod-and-shuffle (N\&S) mode. This mode uses very short slitlets --- between 
3-4\arcsec\ in length in our observations --- such that the target source can be placed 
on one half of the slitlet with the other half collecting blank sky. The telescope is then 
nodded back and forth on the sky to move the target sources between the two halves of 
the slitlets, while the charge on the detector is shuffled in concert with each telescope nod. 
The resulting 2D detector image contains two separate traces (A and B) for each slitlet --- one with 
the target source at each end of the slitlet. A difference of the two traces resulting from each 
individual slitlet yields two sky-subtracted traces for the target source (one positive, one negative). 

The advantage of this mode of observation is that the nod cycle can be performed on relatively 
short timescales to match the timescale on which the intensity of sky emission varies 
--- typically a few minutes. The sky-subtraction results in nearly Poisson noise statistics. Each 
pair of slitlet traces contain complementary pairs of source$+$sky and sky-only spectra such that 
the source$+$sky spectrum from trace A was observed through the identical optical path (instrument 
and telescope) as the complementary trace B sky-only spectrum, and vice versa. We used a 
nod cycle time of 120s (i.e., one 120s interval spent integrating at each of position A and position 
B in a single nod cycle), and repeated a number of nod cycles split across two or three 
science exposures to reach the required total integration times described in 
Section~\ref{subsubsec:setup}.

\subsubsection{Mask Design}

Optical and infrared imaging observations of the 2500 deg$^{2}$ SPT cluster candidates are available 
from an extensive multi-facility campaign to identify red-sequence galaxy populations at the position 
of each candidate. These data are discussed in detail in previous SPT collaboration papers 
\citep{high10,song12,bleem15}, and we refer the reader to those publications for more information.

The pre-existing follow-up confirmation imaging is sufficient in most cases to produce photometric 
catalogs of candidate red-sequence cluster member galaxies down to at least $m^\star 
+ 1$, which is sufficient for designing masks for SPT-GMOS spectroscopy. 
For those SPT-SZ galaxy clusters that did not have follow-up imaging sufficient to reach $m^\star 
+ 1$ depths we obtained additional pre-imaging with Gemini/GMOS-South. Pre-imaging observations 
were performed in two bands, chosen to span the 4000\AA\ break (either $gr$, or $ri$), with 
integration times chosen to achieve $10\sigma$ depth for a galaxy of $m^\star + 1$ at the best-available 
estimate of the cluster photometric redshift. The GMOS pre-imaging data were reduced using the standard 
scripts from the Gemini/GMOS IRAF package\footnote{https://www.gemini.edu/node/11823}; these scripts 
subtract off the bias level for each GMOS detector, apply a flat-field correction using observations of a 
flat lamp-illuminated source within the Gemini-South dome, and map the three individual GMOS 
detectors onto a single mosaicked image using geometric transformations provided by the Gemini 
Observatory. We photometrically calibrate the pre-imaging using unsaturated stars that appear within 
the field of view of both the GMOS and pre-existing follow-up imaging.

All masks were designed with the 
Gemini MOS Mask Preparation Software (GMMPS) tool, which can use either native GMOS 
pre-imaging or ``pseudo-pre-imaging'' generated from optical imaging from other facilities. 
We designed two spectroscopic masks for each cluster using a single input catalog. GMMPS takes 
an input catalog and generates one or more spectroscopic masks with slits placed based on three 
discrete tiers of priority. For each cluster we used the highest priority to target the galaxy or galaxies 
that were identified as likely candidates to be the brightest cluster galaxy (BCG), and also occasionally 
to target other objects of special interest such as bright giant arcs resulting from strong gravitational lensing. 
The next highest priority was used to target candidate cluster member galaxies that were selected from 
the red sequence, which we identified as an overdensity in color-magnitude (e.g., $g-r$ vs $r$) and 
color-color space (e.g., $g-r$ vs $r-z$). The lowest priority tier included all galaxies that were potentially 
drawn from the ``blue cloud'' population of cluster galaxies, which we selected as all galaxies that were 
bluer than the red sequence and not obviously in the foreground (i.e., not brighter than the brightest end 
of the main sequence).

Masks were designed with slit lengths for standard multi-object masks varying between 6-8\arcsec\ in 
length depending on the typical sizes of the galaxies being observed, where the galaxies in lower redshift 
clusters have larger angular sizes than those in higher redshift clusters. For the large majority 
of our standard MOS masks we used slits with lengths between 6-6.5\arcsec. Slitlet lengths 
on the N\&S masks (targeting the higher redshift clusters) varied between 3.1-4\arcsec, with the majority 
using 3.5\arcsec\ lengths as we found this to be a good balance between optimizing the N\&S subtraction 
and maximizing the number of slitlets placed on each mask. The standard slit width for all masks was 1\arcsec, 
which is a good match to the typical size of the fainter galaxies that we targeted, and therefore strikes a 
balance between throughput and spectral resolution. Individual standard (N\&S) multi-object masks typically 
contain $\sim$30-40 (30-35) slits, with approximately one half to two thirds of those slits generally being 
placed on galaxies with a high probability of being cluster members. The slight difference in slit-packing 
between standard vs. N\&S modes simply reflects the fact that N\&S slits effectively take up twice their 
length in available detector space, and our most commonly used N\&S slit length is slightly greater than 
1/2 our most commonly used standard MOS slit length.

\begin{figure}[t]
\includegraphics[scale=0.56]{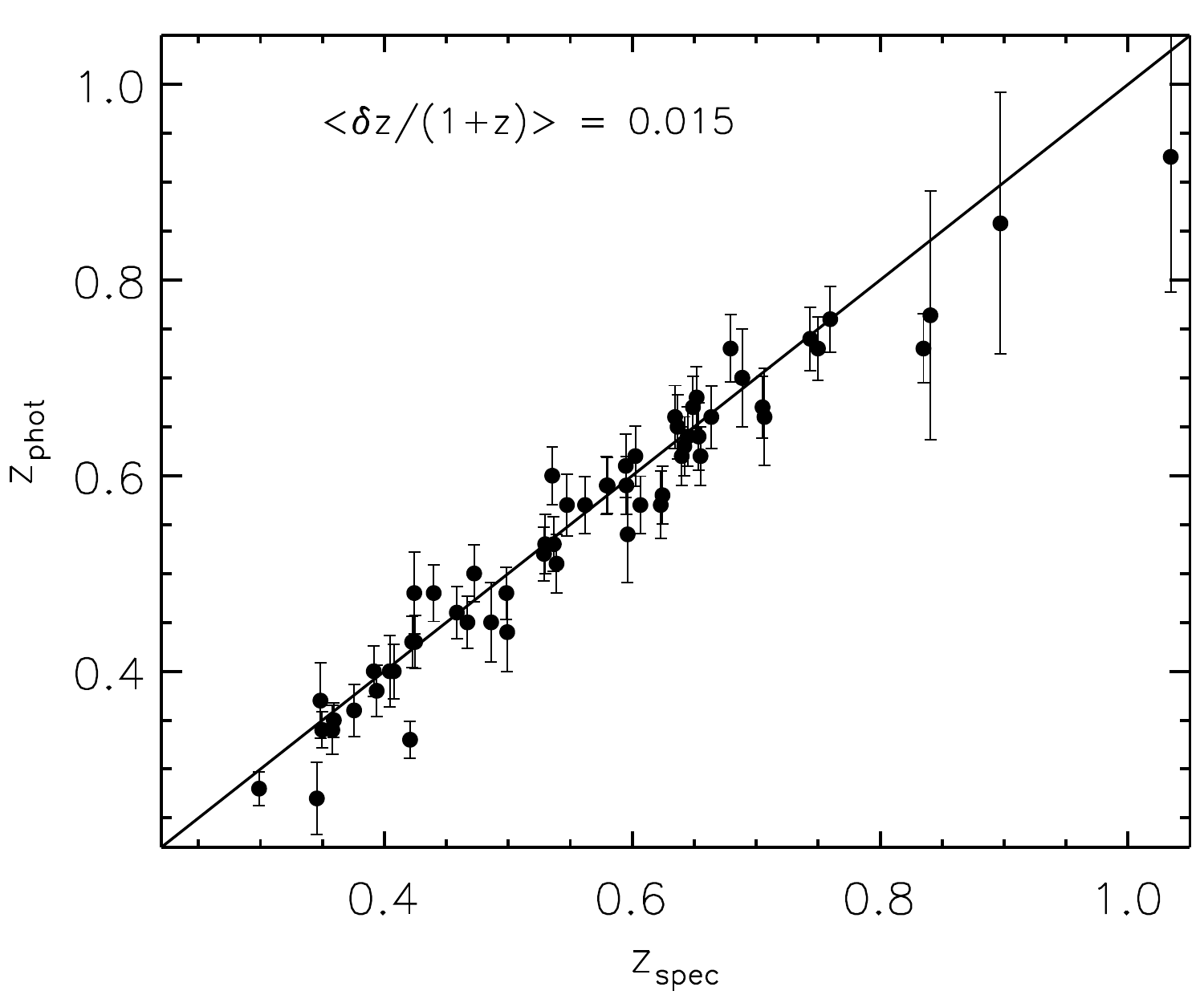}
\caption{\scriptsize{
Spectroscopic vs. photometric redshifts for the 62 SPT-GMOS clusters presented here, using photo-z's 
described in \citet{bleem15}, with the scatter between the two $\sim1.5\%$. 
SPT-GMOS spectroscopy provides a strong anchor for the photometric redshift calibration at low to 
intermediate redshift.}}
\label{fig:zcalibration}
\end{figure}

\section{Spectroscopic Data Reduction} 
 \label{sec:reduction}

We reduced all SPT-GMOS spectroscopic data uniformly using a custom pipeline. The pipeline 
relies primarily on scripts from the Gemini IRAF 
package\footnote{https://www.gemini.edu/node/10795/} developed by Gemini Observatory. We 
supplement the Gemini IRAF scripts with our own custom code --- described in more 
detail below --- that is based on the XIDL\footnote{http://www.ucolick.org/$\sim$xavier/IDL/} 
package. Some elements of the reduction pipeline vary slightly across the spectroscopic dataset, 
reflecting differences in the observing strategy --- standard multi-object spectroscopy vs 
nod-and-shuffle --- and 
differences in the GMOS-South detectors before and after the 2014B semester. Below we describe 
the reduction process for each detector and each observing mode.

\subsection{GMOS-South Spectra with e2v Detectors}

\subsubsection{Standard Multi-object Spectroscopy}

The majority of our spectroscopic observations are standard multi-object spectroscopy, and we 
use standard Gemini IRAF scripts to reduce these data. For the majority of the 
spectroscopic masks we take both arc lamp frames for wavelength calibration and quartz lamp 
flat-field frames interspersed between science frames, on sky, to ensure as little instrumental 
variation as possible between calibration and science frames. In a few cases --- primarily 
observations conducted in classical mode and already published in \citet{ruel14} --- we use 
daytime arc lamp frames and rely on sky lines in the science exposures to refine the wavelength 
calibration.

\begin{figure*}[t]
\centering
\includegraphics[scale=0.448]{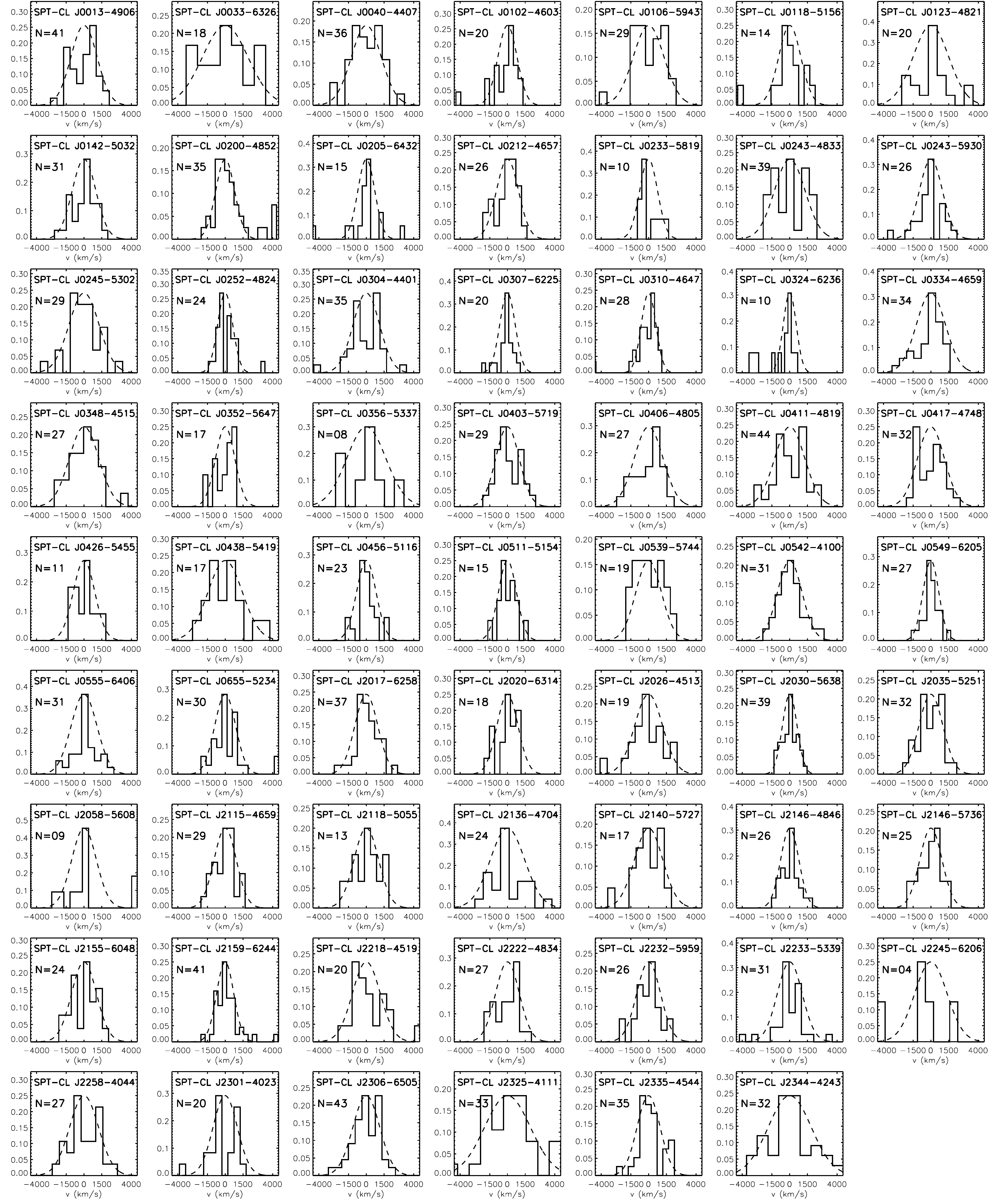}
\caption{\scriptsize{
Velocity distributions for the 62 galaxy clusters in the SPT-GMOS survey program. Over-plotted as  
dashed lines are Gaussian distributions 
with centers and widths matching the bi-weight estimates of the median and dispersion for each 
galaxy cluster. The velocities plotted here for each cluster have been converted to peculiar velocities 
relative to the bi-weight estimate of the median recession velocity for each galaxy cluster. The 
number of members used to estimate the dispersion of each cluster is indicated in each panel.}}
\label{fig:dispersions}
\end{figure*}

Our pipeline begins by subtracting a master bias frame from the science, 
arc lamp, and flat-field calibration frames. We then use the flat-field frames to identify the trace 
of each individual slit and to derive a flat-field correction. We then reduce and extract  
the 2D spectrum associated with each individual slit on the mask for both the science and arc lamp 
calibration frames. The extracted 2D arc lamp spectra are then used to fit a wavelength calibration 
for each slit using the standard line lists provided by Gemini Observatory, typically using 
20-30 arc lamp lines per slit, depending on the wavelength coverage of each slit. 
The wavelength solutions 
are applied to the science spectra, which are then used to fit a sky model that excludes the source 
trace and assumes a constant sky spectrum along the length of the slit. We subtract the sky model 
for each slit and apply cosmic ray rejection to each individual sky-subtracted 2D spectrum using 
a modified version of LACOSMIC \citep{vandokkum01}. We then use custom 
IDL\footnote{http://www.exelisvis.co.uk/ProductsServices/IDL.aspx} scripts 
to fit the source trace in each slit and extract a single 1D spectrum from each 2D science exposure. 
All of the 1D extractions for a given slit on a given mask are then combined using the XIDL 
``long\_combine'' procedure. 

We generate approximate flux calibrations in each of the instrumental configurations 
using archival ``partner'' observations of southern spectrophotometric standard stars. Specifically, 
we reduced archival data for the standards LTT1788, LTT7379 
\citep{hamuy1992,hamuy1994}, and G158-100 \citep{oke1990} using the standard Gemini IRAF 
scripts to generate average flux calibrations for 
GMOS-South in the configurations used for our science spectra. 
We apply these average flux calibrations to our science spectra, resulting in final source spectra 
that are ``flattened'' to correct for the relative throughput as a function of wavelength, but do not 
provide a reliable absolute zero point flux calibration.

\begin{deluxetable*}{lccccccc}
\tablecaption{Spectroscopy of SPT Clusters: Results\label{tab:spectable}}
\tablewidth{0pt}
\tabletypesize{\tiny}
\tablehead{
\colhead{Cluster Name} &
\colhead{N$_\mathrm{spec}$} &
\colhead{N$_\mathrm{members}$} &
\colhead{ ${\bar z_\mathrm{cluster}}$} &
\colhead{ $\sigma_\mathrm{v,BI}$ } &
\colhead{ $\sigma_\mathrm{v,G}$ }  &
\colhead{ $A^{2*}$ }  &
\colhead{ $\alpha_\mathrm{AD}$ }  \\
\colhead{  } &
\colhead{  } &
\colhead{  } &
\colhead{  } &
\colhead{ (km s$^{-1}$) } &
\colhead{ (km s$^{-1}$) } & 
\colhead{  } &
\colhead{  } }
\startdata
SPT-CL~J0013-4906 &  51 &  41 & $0.4075\pm 0.0052$ & $ 1103\pm   160$  & $ 1105\pm   160$ &  0.481 & 0.23599  \\
SPT-CL~J0033-6326 &  45 &  18 & $0.5963\pm 0.0102$ & $ 1916\pm   427$  & $ 1941\pm   433$ &  0.264 & 0.81568  \\
SPT-CL~J0040-4407\tablenotemark{a} &  44 &  36 & $0.3498\pm 0.0057$ & $ 1259\pm 195$  & $ 1276\pm 198$ &  0.238 & 0.94321  \\
SPT-CL~J0102-4603 &  48 &  20 & $0.8405\pm 0.0050$ & $  807\pm   170$  & $  803\pm   169$  &  0.533 & 0.17476 \\
SPT-CL~J0106-5943 &  50 &  29 & $0.3484\pm 0.0058$ & $ 1298\pm   225$  & $ 1304\pm   226$ &  0.390 & 0.39682  \\
SPT-CL~J0118-5156\tablenotemark{a} &  23 &  14 & $0.7051\pm 0.0053$ & $  934\pm 238$  & $  949\pm 242$ &  0.559 & 0.15117  \\
SPT-CL~J0123-4821 &  31 &  20 & $0.6550\pm 0.0083$ & $ 1505\pm   317$  & $ 1490\pm   314$  &  0.568 & 0.14325 \\
SPT-CL~J0142-5032 &  45 &  31 & $0.6793\pm 0.0056$ & $ 1000\pm   168$  & $ 1006\pm   169$ &  0.253 & 0.86765 \\
SPT-CL~J0200-4852 &  58 &  35 & $0.4991\pm 0.0040$ & $  796\pm   125$  & $  803\pm   126$ &  0.191 & 0.76500 \\
SPT-CL~J0205-6432\tablenotemark{a} &  24 &  15 & $0.7436\pm 0.0042$ & $  714\pm  175$  & $  683\pm  168$ &  0.912 & 0.02010 \\
SPT-CL~J0212-4656 &  40 &  26 & $0.6535\pm 0.0051$ & $  931\pm   171$  & $  921\pm   169$ &  0.723 & 0.05919  \\
SPT-CL~J0233-5819\tablenotemark{a} &  11 &  10 & $0.6638\pm 0.0042$ & $  754\pm  231$  & $  781\pm  239$ &  0.361 & 0.46679 \\
SPT-CL~J0243-4833 &  44 &  39 & $0.4984\pm 0.0065$ & $ 1293\pm   193$  & $ 1329\pm   198$ &  1.004 & 0.01184  \\
SPT-CL~J0243-5930 &  44 &  26 & $0.6345\pm 0.0053$ & $  975\pm   179$  & $  978\pm   180$ &  0.296 & 0.67867  \\
SPT-CL~J0245-5302\tablenotemark{a} &  38 &  29 & $0.3000\pm 0.0055$ & $ 1262\pm 219$  & $ 1260\pm 219$ &  0.267 & 0.80064  \\
SPT-CL~J0252-4824 &  42 &  24 & $0.4207\pm 0.0030$ & $  635\pm   121$  & $  656\pm   126$ &  0.441 & 0.29543   \\
SPT-CL~J0304-4401 &  48 &  35 & $0.4584\pm 0.0054$ & $ 1114\pm   175$  & $ 1116\pm   176$ &  0.479 & 0.23764  \\
SPT-CL~J0307-6225 &  36 &  20 & $0.5801\pm 0.0034$ & $  652\pm   137$  & $  618\pm   130$  & 1.071 & 0.00808  \\
SPT-CL~J0310-4647 &  45 &  28 & $0.7067\pm 0.0035$ & $  617\pm   109$  & $  628\pm   111$ &  0.515 & 0.19385 \\
SPT-CL~J0324-6236 &  33 &  10 & $0.7498\pm 0.0032$ & $  546\pm   167$  & $  520\pm   159$  &  0.904 & 0.02096 \\
SPT-CL~J0334-4659 &  51 &  34 & $0.4861\pm 0.0061$ & $ 1223\pm   195$  & $ 1203\pm   192$ &  0.907 & 0.02062 \\
SPT-CL~J0348-4515 &  41 &  27 & $0.3592\pm 0.0057$ & $ 1246\pm   224$  & $ 1255\pm   226$  &  0.277 & 0.75715 \\
SPT-CL~J0352-5647 &  33 &  17 & $0.6490\pm 0.0045$ & $  813\pm   186$  & $  812\pm   186$ &  0.509 & 0.20113  \\
SPT-CL~J0356-5337 &  36 &   8  & $1.0345\pm 0.0112$ & $ 1647\pm   572$  & $ 1691\pm   588$ &  0.279 & 0.74516  \\
SPT-CL~J0403-5719 &  52 &  29 & $0.4670\pm 0.0048$ & $  990\pm   172$  & $ 1008\pm   175$  &  0.484 & 0.23170  \\
SPT-CL~J0406-4804 &  33 &  27 & $0.7355\pm 0.0070$ & $ 1216\pm   219$  & $ 1216\pm   219$ &  0.503 & 0.20804 \\
SPT-CL~J0411-4819 &  54 &  44 & $0.4241\pm 0.0060$ & $ 1267\pm   177$  & $ 1294\pm   181$ &  0.381 & 0.41758 \\
SPT-CL~J0417-4748 &  49 &  32 & $0.5794\pm 0.0060$ & $ 1133\pm   187$  & $ 1139\pm   188$ &  0.359 & 0.47270 \\
SPT-CL~J0426-5455 &  17 &  11 & $0.6420\pm 0.0050$ & $  910\pm   265$  & $  950\pm   276$  &  0.247 & 0.89796 \\
SPT-CL~J0438-5419\tablenotemark{a} &  23 & 17 & $0.4224\pm 0.0069$ & $ 1448\pm 333$ & $ 1481\pm 340$ & 0.246 & 0.90123 \\
SPT-CL~J0456-5116 &  45 &  23 & $0.5619\pm 0.0043$ & $  821\pm 161$  & $  804\pm 157$ &  0.566 & 0.14491 \\
SPT-CL~J0511-5154\tablenotemark{a} &  23 &  15 & $0.6447\pm 0.0042$ & $  758\pm 186$  & $  779\pm 191$ & 0.356 & 0.48046 \\
SPT-CL~J0539-5744 &  44 &  19 & $0.7597\pm 0.0063$ & $ 1075\pm   233$  & $ 1118\pm   242$  &  0.316 & 0.60336 \\
SPT-CL~J0542-4100 &  44 &  31 & $0.6399\pm 0.0056$ & $ 1031\pm   173$  & $ 1036\pm   174$ &  0.229 & 0.99178 \\
SPT-CL~J0549-6205 &  47 &  27 & $0.3755\pm 0.0027$ & $   666\pm   120$  & $  669\pm   120$ &  0.197 & 0.81417 \\
SPT-CL~J0555-6406 &  53 &  31 & $0.3455\pm 0.0049$ & $ 1088\pm   182$  & $ 1073\pm   180$  &  0.617 & 0.10821 \\
SPT-CL~J0655-5234 &  50 &  30 & $0.4724\pm 0.0043$ & $  883\pm   150$  & $  902\pm   154$ &  0.258 & 0.84035 \\
SPT-CL~J2017-6258 &  54 &  37 & $0.5354\pm 0.0050$ & $  972\pm   149$  & $  961\pm   147$ &  0.315 & 0.60867 \\
SPT-CL~J2020-6314 &  43 &  18 & $0.5367\pm 0.0046$ & $  891\pm   198$  & $  890\pm   198$ &  0.506 & 0.20410 \\
SPT-CL~J2026-4513 &  47 &  19 & $0.6887\pm 0.0067$ & $ 1182\pm   256$  & $ 1227\pm   266$ &  0.232 & 0.97541  \\
SPT-CL~J2030-5638 &  67 &  39 & $0.3937\pm 0.0029$ & $  619\pm  92$  & $  631\pm  94$ &  0.180 & 0.72546 \\
SPT-CL~J2035-5251 &  61 &  32 & $0.5287\pm 0.0052$ & $ 1015\pm   167$  & $ 1022\pm   168$  &  0.271 & 0.78107 \\
SPT-CL~J2058-5608\tablenotemark{a} &  16 &   9 & $0.6065\pm 0.0056$ & $ 1038\pm 337$  & $  990\pm 322$ & 2.185 & 0.00001 \\
SPT-CL~J2115-4659 &  43 &  29 & $0.2989\pm 0.0040$ & $  934\pm 162$  & $  943\pm 164$ &  0.150 & 0.76338 \\
SPT-CL~J2118-5055\tablenotemark{a,b}  & 30 & 13 & $0.6244\pm 0.0056$ & $ 1035\pm 274$  & $ 1088\pm 289$ & 0.208 & 0.79140 \\
SPT-CL~J2136-4704\tablenotemark{a} &  28 &  24 & $0.4247\pm 0.0069$ & $ 1448\pm 277$  & $ 1461\pm 280$ &  0.364 & 0.45986 \\
SPT-CL~J2140-5727 &  47 &  17 & $0.4043\pm 0.0056$ & $ 1192\pm   274$  & $ 1176\pm   270$ &  0.375 & 0.43196 \\
SPT-CL~J2146-4846\tablenotemark{a} &  29 &  26 & $0.6230\pm 0.0042$ & $  768\pm 141$  & $  771\pm 141$ &  0.285 & 0.72234 \\
SPT-CL~J2146-5736 &  51 &  25 & $0.6025\pm 0.0050$ & $  936\pm 175$  & $  942\pm 177$ &  0.264 & 0.81491 \\
SPT-CL~J2155-6048\tablenotemark{a} &  31 &  24 & $0.5389\pm 0.0054$ & $ 1049\pm 201$  & $ 1079\pm 207$ &  0.323 & 0.58092 \\
SPT-CL~J2159-6244 &  53 &  41 & $0.3914\pm 0.0034$ & $  723\pm   105$  & $  725\pm   105$ &  0.204 & 0.75778 \\
SPT-CL~J2218-4519 &  24 &  20 & $0.6365\pm 0.0064$ & $ 1172\pm   247$  & $ 1207\pm   254$ &  0.456 & 0.27124 \\
SPT-CL~J2222-4834 &  46 &  27 & $0.6519\pm 0.0055$ & $ 1002\pm   180$  & $ 1002\pm   180$ &  0.499 & 0.21291 \\
SPT-CL~J2232-5959 &  46 &  26 & $0.5948\pm 0.0053$ & $ 1004\pm   184$  & $ 1008\pm   185$ &  0.251 & 0.87831 \\
SPT-CL~J2233-5339 &  45 &  31 & $0.4398\pm 0.0050$ & $ 1045\pm   175$  & $  975\pm   163$ &  0.839 & 0.03045 \\
SPT-CL~J2245-6206 &  46 &   4 & $0.5856\pm 0.0072$ & $ 1363\pm   724$  & $ 1406\pm   747$  &  0.550 & 0.15872 \\
SPT-CL~J2258-4044 &  44 &  27 & $0.8971\pm 0.0077$ & $ 1220\pm   220$  & $ 1248\pm   225$ &  0.322 & 0.58278 \\
SPT-CL~J2301-4023 &  55 &  20 & $0.8349\pm 0.0063$ & $ 1023\pm   216$  & $ 1045\pm   220$ &  0.911 & 0.02013 \\
SPT-CL~J2306-6505 &  57 &  43 & $0.5297\pm 0.0058$ & $ 1132\pm   160$  & $ 1138\pm   161$ &  0.188 & 0.91693 \\
SPT-CL~J2325-4111\tablenotemark{a} &  47 &  33 & $0.3579\pm 0.0088$ & $ 1932\pm 314$  & $ 1926\pm 313$ &  0.342 & 0.52026 \\
SPT-CL~J2335-4544 &  46 &  35 & $0.5473\pm 0.0050$ & $  974\pm   153$  & $  948\pm   149$ &  1.343 & 0.00171 \\
SPT-CL~J2344-4243\tablenotemark{a} &  42 &  32 & $0.5952\pm 0.0097$ & $ 1814\pm 299$  & $ 1825\pm 301$  &  0.167 & 0.79178 
\enddata
\tablecomments{A summary of the results of SPT-GMOS spectroscopy by galaxy cluster. Columns from left to right 
report the cluster name, the total number of spectra with radial velocity measurements, the number of cluster member 
galaxies, the median cluster redshift, the velocity dispersion estimates (using the bi-weight and gapper estimators) 
for each observed SPT cluster, the value of the AD test statistic, and the probability that the observed cluster velocities 
were drawn from a Gaussian velocity distribution.}
\tablenotetext{a}{~Cluster also presented in \citet{ruel14}.}
\tablenotetext{b}{~Numbers here are computed from SPT-GMOS data only, but are fully consistent with the results combining 
these data with spectra from other facilities in \citet{ruel14}.}
\end{deluxetable*}

\subsubsection{Micro N\&S Multi-object Spectroscopy}
 
Reduction of N\&S spectra differs somewhat from standard MOS data, in large part because the 
sky-subtraction step is best performed early in the process. We begin with these data by applying a 
standard bias subtraction to all frames. We then use custom IDL code based on previous work 
reducing N\&S spectra from GMOS-North \citep{bayliss11b} to create a master dark frame from 
day-time calibration exposures that reproduce the exposure times and charge shuffle patterns 
used in our science frames. The dark calibration serves primarily to identify and mask out regions 
of the detector that act as ``charge traps'' when charge is shuffled up and down along detector 
columns.

For N\&S spectra the sky subtraction step is trivially achieved by differencing each science frame 
from itself, offset along the columns of the CCD by the shuffle distance. This step results in two 2D 
sky-subtracted spectral traces for each slit, one positive and the other negative, corresponding to 
the spectra dispersed at both of the original pointed position and the nodded-to position, 
respectively. The sky-subtracted spectra are reduced using standard Gemini IRAF routines 
to apply a flat-field calibration and wavelength solution, and we apply the same custom IDL 
routines to extract a separate 1D spectrum from each 2D spectral trace --- i.e., each science 
exposure produces two 1D extractions. We combine all 1D spectra from each slit and apply 
flux calibrations identically to the standard MOS spectra described above.

\subsection{GMOS-South Spectra with Hamamatsu Detectors}

Gemini Observatory upgraded the GMOS-South detector in 2014B, replacing the old \textit{e2v} 
CCD chips with Hamamatsu chips. The SPT-GMOS survey observations were primarily 
conducted prior to 2014B, but SPT-GMOS observations performed between Dec 2014 and Dec 2015 
used the upgraded detectors. These new chips provide a tremendous improvement in the sensitivity 
of the instrument, improving the quantum efficiency by more 
than a factor of two redward of $\sim$8000\AA, and by approximately an order of magnitude 
at $\sim$10000\AA. The Hamamatsu chips are also much less prone to fringing and it is therefore 
not necessary to rely on N\&S observations to obtain high-quality spectra at redder wavelengths. 
The increased performance of these new detectors does come at the cost of a much higher 
cosmic ray hit-rate per detector pixel, such that shorter individual exposures ($t_{exp} \lesssim 
1200$s) are strongly recommended. In masks designed for the Hamamatsu detectors  
we experimented with different exposure-splitting strategies, acquiring up to six individual science 
exposures with each mask (vs. the two-exposure strategy that we used for spectra taken with the 
older detectors) with the goal of improving cosmic ray rejection for Hamamatsu spectra.

The reduction pipeline for spectra with the Hamamatsu detectors is almost 
identical to the process applied to the older \textit{e2v} detectors, with the notable exception of cosmic 
ray rejection. The shorter individual spectroscopic exposures ($\lesssim 1200$s) also result in very 
faint traces for spectra from the typical cluster member galaxies at $z \gtrsim 0.8$, which introduces 
potential problems with the extraction of 1D spectra from individual 2D spectral images. Our reduction 
pipeline for the \textit{e2v} spectra --- described above --- is not necessarily optimal for the new 
Hamamatsu spectra. Specifically, the high rate of cosmic ray 
hits and shorter individual exposure times raise concerns about the single-image cosmic ray 
rejection and 1D extraction algorithms that we apply to each individual \textit{e2v} science frame.

We have experimented with two different reduction schemes, one identical to the procedure applied 
to data taken with the old detectors, and a second in which we perform cosmic ray rejection 
simultaneously while stacking pairs of individual sky-subtracted 2D science spectra, slit by slit for 
each mask, applying the IRAF CRREJECT algorithm. This second method results in half as many 
clean 2D spectra for each mask slit --- which we trace, extract, and combine using the same custom 
IDL code described above. Careful work with the final spectra that result from each reduction 
method does not clearly favor one method as universally better; the final spectra resulting from the first method 
(the one applied to old detector data) are often high quality, and this method has the general advantage of 
producing a final S/N-weighted stack of individual 1D exposures, which allows us to make optimal use 
of different 2D spectra exposures where the average seeing (and thus the profile of the spectral trace) is 
varying between exposures. However, there are some cases where the 
second method using stacks of 2D spectra generates a final spectrum that is cleaner with better 
S/N than the first method. We generate reductions 
for all Hamamatsu spectra using both methods, and measure as many redshifts as possible from the 
two reductions. The ``best'' reduction method seems to vary across different masks and slits, and at 
this time we conclude that the best practices for reducing spectra taken with the new Hamamatsu 
detectors are still an open question.

Our spectra taken with the Hamamatsu detectors do suffer from the effects of a malfunction 
involving amplifier \#5. Some or all of the columns read out by this amplifier are occasionally and 
unpredictably ``hot'', such that the data recorded there are lost. This problem was 
diagnosed by Gemini Observatory during the 2014B semester, and announced publicly in February 
2015\footnote{https://www.gemini.edu/pio/?q=node/10004}. The impact of the defective amplifier 
is obvious in many spectra, and impacts $\sim$1/12 of any particular 2D spectrum. Unfortunately, this 
amplifier is located toward the middle of the detector array and covers wavelength ranges that include 
the 4000\AA\ break and other important features for galaxies targeted in our program. The effect of 
the faulty amplifier on our final spectra varies significantly from mask to mask, and slit to slit. Wherever 
possible we exclude spectra from exposures where the amplifier effects are severe when stacking the 
individual 1D spectra for each slit, resulting in a modest loss in S/N for the affected spectra. 
As of late 2015 the faulty amplifier is repaired and will not be a problem for future observations. 

\section{Primary Survey Data Products}
\label{sec:products}

\subsection{Galaxy Redshift Measurements}

We examine all calibrated, stacked, 1D spectra by eye, and estimate redshifts using one 
or more methods. Most redshift estimates use custom IDL code to 
cross-correlate strong, well-detected features typical of galaxy spectra such as the 4000\AA\ 
break against the same features in template galaxy spectra; \oii; Ca {\small II} H\&K$\lambda$3934, 
3969; \hdelta; \hgamma; \hbeta; G-band$\lambda$ 4305; [O {\small III}] $\lambda$$\lambda$ 4960, 
5007; and Mg {\small I}$\lambda$$\lambda$$\lambda$ 5169, 5174, 5185. In practice this means 
that we exclude regions of the spectral data that have low S/N, are contaminated by sky lines, 
or contain no notable spectral features. Independent redshift estimation was performed using 
the cross-correlation routines in the RVSAO IRAF package with the \emph{fabtemp97} template 
\citep{kurtz98}; the RVSAO routines are extremely similar, algorithmically, to our custom code, 
and results from the two methods are in excellent agreement, with typical uncertainties of 
$\delta cz \simeq 90-160$ km s$^{-1}$ from both methods after correcting the RVSAO 
uncertainties up by a factor of 1.7 to accurately reflect the true cross-correlation uncertainties 
\citep{quintana00}.

In the case of galaxy spectra with low S/N or that only have one or more emission 
lines we fit Gaussian profiles to the available lines to estimate the galaxy redshift as 
the mean redshift of all individual lines, with the standard deviation between individual 
line measurements providing the galaxy redshift uncertainty. In cases where only one line 
is detected we use the uncertainty in the centroid of the Gaussian fit to that line to estimate 
the redshift uncertainty. Single-line redshifts are only measured for spectra with a 
clear emission line that can be confidently identified based on the lack of other strong 
emission features. For example, prominent nebular emission lines (i.e., \oii, \hbeta, 
[O {\small III}] 4960, 5007, and \halpha) appear together in the spectra of 
star-forming galaxies, so that with the large wavelength coverage of our spectra we 
can often infer that a single emission line is \oii\ --- which is unresolved at our spectral 
resolution --- with the redder nebular line redshifted out of our wavelength coverage.  

Example cluster member spectra for SPT-GMOS resulting from standard 
MOS observations are shown in Figure~\ref{fig:0348spectra}, and N\&S observations in 
Figure~\ref{fig:2222spectra}. 
In total we examine 3317 individual spectra, and measure high-confidence radial velocities  
for 2595 sources ($\sim$80\% success rate). Of these sources we identify 352 as stars 
and 2243 as galaxies. These results only include high-confidence redshift 
measurements, as we generally do not report --- or include in our data release --- redshift 
interpretations that are significantly uncertain or unclear (i.e., ``best-guess'' redshifts). We 
do present a few of these best-guess redshifts in $\S~\ref{sec:giantarcs}$, where we discuss 
the spectra of bright, strongly-lensed galaxies that were observed. We make this exception 
because exceptionally bright lensed sources are particularly rare objects, and any information 
or constraints on these sources can be useful for informing follow-up efforts.

\begin{deluxetable*}{cccccccc}
\tablecaption{Sample SPT-GMOS Data Products for Candidate Brightest Cluster Galaxy Spectra\label{tab:bcgspec}} \\
\tablewidth{0pt}
\tabletypesize{\tiny}
\tablehead{
\colhead{Cluster } &
\colhead{Object } &
\colhead{ RA } &
\colhead{ Dec  } &
\colhead{ z ($\delta_{z}$)\tablenotemark{a} } &
\colhead{ $W_{0}$ \oii  }& 
\colhead{ $W_{0}$ \hdelta } &
\colhead{ D4000 } \\
\colhead{ Name} &
\colhead{ Name} &
\colhead{ (J2000.0) } &
\colhead{ (J2000.0) } &
\colhead{  } &
\colhead{ (\AA) } &
\colhead{ (\AA) } &
\colhead{  } } 
\startdata
SPT-CL~J0013-4906 & J001319.19-490638.8 & 00:13:19.19 & -49:06:38.8 & $0.40998(15)$ & $ 5.66\pm 1.50$ & $ -3.02\pm 1.32$ & $ 1.64\pm 0.03$ \\
SPT-CL~J0033-6326 & J003353.01-632641.4 & 00:33:53.01 & -63:26:41.4 & $0.59784(54)$ & $ 1.48\pm 1.31$ & $ -1.57\pm 1.02$ & $ 1.70\pm 0.02$ \\
SPT-CL~J0102-4603 & J010242.71-460416.0 & 01:02:42.71 & -46:04:16.0 & $0.84008(64)$ & $ 1.93\pm 2.61$ & $ -5.01\pm 2.08$ & $ 1.68\pm 0.03$ \\
SPT-CL~J0106-5943 & J010628.73-594313.6 & 01:06:28.73 & -59:43:13.6 & $0.35043(36)$ & $ 1.06\pm 0.53$ & $ -2.40\pm 0.97$ & $ 1.88\pm 0.01$ \\
SPT-CL~J0118-5156 & J011824.77-515628.7 & 01:18:24.77 & -51:56:28.7 & $0.70210(99)$ & $ 4.85\pm 3.07$ & $ -3.50\pm 2.54$ & $ 1.53\pm 0.04$ \\
SPT-CL~J0123-4821 & J012310.92-482122.1 & 01:23:10.92 & -48:21:22.1 & $0.65420(45)$ & $ 1.71\pm 1.18$ & $ -2.03\pm 0.76$ & $ 1.77\pm 0.02$ \\
SPT-CL~J0142-5032 & J014209.68-503231.6 & 01:42:09.68 & -50:32:31.6 & $0.67872(82)$ & $ -3.26\pm 3.93$ & $ 1.76\pm 2.14$ & $ 1.79\pm 0.06$ \\
SPT-CL~J0200-4852 & J020034.09-485215.7 & 02:00:34.09 & -48:52:15.7 & $0.49851(12)$ & $ 5.20\pm 2.57$ & $ -2.86\pm 0.80$ & $ 2.34\pm 0.23$ \\
SPT-CL~J0205-6432 & J020507.84-643226.9 & 02:05:07.84 & -64:32:26.9 & $0.74300(25)$ & $ 4.17\pm 1.69$ & $ 0.24\pm 1.75$ & $ 1.85\pm 0.03$ \\
SPT-CL~J0212-4657 & J021223.57-465713.9 & 02:12:23.57 & -46:57:13.9 & $0.65725(06)$ & $ 3.51\pm 59.08$ & $ -6.56\pm 1.52$ & $ 1.73\pm 0.02$ \\
SPT-CL~J0233-5819 & J023300.97-581937.0 & 02:33:00.97 & -58:19:37.0 & $0.66000(25)$ & $ 1.98\pm 1.36$ & $ -1.74\pm 1.38$ & $ 1.81\pm 0.03$ \\
SPT-CL~J0243-4833 & J024338.85-483339.1 & 02:43:38.85 & -48:33:39.1 & $0.49693(34)$ & $ 0.53\pm 1.25$ & $ -0.13\pm 0.56$ & $ 1.71\pm 0.02$ \\
SPT-CL~J0243-5930 & J024327.08-593100.6 & 02:43:27.08 & -59:31:00.6 & $0.63366(26)$ & $ 1.70\pm 1.50$ & $ -0.42\pm 0.84$ & $ 1.72\pm 0.02$ \\
SPT-CL~J0245-5302 & J024524.82-530145.4 & 02:45:24.82 & -53:01:45.4 & $0.30280(25)$ & $ 3.91\pm 0.90$ & $ -1.74\pm 1.04$ & $ 1.85\pm 0.02$ \\
SPT-CL~J0252-4824 & J025249.98-482458.4 & 02:52:49.98 & -48:24:58.4 & $0.42226(19)$ & $ 5.89\pm 1.51$ & $ -2.42\pm 1.18$ & $ 1.81\pm 0.04$ \\
SPT-CL~J0304-4401 & J030416.89-440131.5 & 03:04:16.89 & -44:01:31.5 & $0.45491(25)$ & $ 1.53\pm 1.70$ & $ -0.76\pm 0.87$ & $ 2.68\pm 0.28$ \\
SPT-CL~J0307-6225 & J030716.77-622647.3 & 03:07:16.77 & -62:26:47.3 & $0.57801(40)$ & $ -5.92\pm 2.63$ & $ -0.21\pm 1.21$ & $ 1.64\pm 0.02$ \\
SPT-CL~J0310-4647 & J031032.50-464708.0 & 03:10:32.50 & -46:47:08.0 & $0.70644(49)$ & $ 1.97\pm 1.39$ & $ -3.20\pm 0.89$ & $ 1.75\pm 0.02$ \\
SPT-CL~J0324-6236 & J032412.27-623555.8 & 03:24:12.27 & -62:35:55.8 & $0.74515(94)$ & $ -1.79\pm 2.60$ & $ -1.26\pm 1.67$ & $ 1.73\pm 0.03$ \\
SPT-CL~J0334-4659 & J033410.97-465945.9 & 03:34:10.97 & -46:59:45.9 & $0.48693(51)$ & $ -84.12\pm 3.71$ & $ -4.16\pm 1.18$ & $ 1.55\pm 0.02$ \\
SPT-CL~J0348-4515 & J034817.09-451500.3 & 03:48:17.09 & -45:15:00.3 & $0.36272(67)$ & $ 2.93\pm 0.74$ & $ -1.56\pm 0.67$ & $ 1.92\pm 0.02$ \\
SPT-CL~J0352-5647 & J035257.55-564751.6 & 03:52:57.55 & -56:47:51.6 & $0.64855(33)$ & $ 1.72\pm 2.12$ & $ -1.81\pm 1.16$ & $ 1.63\pm 0.03$ \\
SPT-CL~J0356-5337 & J035621.45-533752.0 & 03:56:21.45 & -53:37:52.0 & $1.03303(27)$ & $ -1.00\pm 3.38$ & $ -2.36\pm 2.46$ & $ 2.04\pm 0.06$ \\
SPT-CL~J0403-5719 & J040352.63-571946.5 & 04:03:52.63 & -57:19:46.5 & $0.45856(33)$ & $ -0.47\pm 0.89$ & $ -0.42\pm 0.98$ & $ 1.02\pm 0.01$ \\
SPT-CL~J0406-4805 & J040655.26-480457.4 & 04:06:55.26 & -48:04:57.4 & $0.73449(83)$ & $ 4.07\pm 3.31$ & $ -2.78\pm 1.98$ & $ 1.65\pm 0.04$ \\
SPT-CL~J0411-4819 & J041110.98-481939.3 & 04:11:10.98 & -48:19:39.3 & $0.41948(40)$ & $ 1.61\pm 0.84$ & $ -0.55\pm 1.12$ & $ 1.73\pm 0.02$ \\
SPT-CL~J0417-4748 & J041723.07-474848.0 & 04:17:23.07 & -47:48:48.0 & $0.58041(55)$ & $-115.46\pm 63.13$ & $ 1.79\pm 0.93$ & $ 1.70\pm 0.02$ \\
SPT-CL~J0438-5419 & J043817.63-541920.6 & 04:38:17.63 & -54:19:20.6 & $0.42170(50)$ & $ 3.04\pm 1.80$ & $ -1.09\pm 1.14$ & $ 2.05\pm 0.04$ \\
SPT-CL~J0456-5116 & J045628.11-511635.0 & 04:56:28.11 & -51:16:35.0 & $0.56270(37)$ & $ 0.54\pm 1.34$ & $ -2.64\pm 1.21$ & $ 1.59\pm 0.02$ \\
SPT-CL~J0511-5154 & J051142.95-515436.6 & 05:11:42.95 & -51:54:36.6 & $0.64880(50)$ & $ 2.34\pm 1.55$ & $ -0.67\pm 1.46$ & $ 1.68\pm 0.03$ \\
SPT-CL~J0539-5744 & J053959.92-574435.3 & 05:39:59.92 & -57:44:35.3 & $0.76873(90)$ & $ 2.53\pm 3.31$ & $ -3.29\pm 3.45$ & $ 1.65\pm 0.06$ \\
SPT-CL~J0542-4100 & J054250.05-410000.4 & 05:42:50.05 & -41:00:00.4 & $0.64176(35)$ & $ -0.07\pm 1.54$ & $ 0.25\pm 0.91$ & $ 1.74\pm 0.02$ \\
SPT-CL~J0555-6406 & J055524.99-640620.8 & 05:55:24.99 & -64:06:20.8 & $0.34496(60)$ & $ 0.58\pm 0.64$ & $ -1.04\pm 0.49$ & $ 1.97\pm 0.01$ \\
SPT-CL~J0655-5234 & J065551.98-523439.2 & 06:55:51.98 & -52:34:39.2 & $0.46816(28)$ & $ -2.39\pm 2.13$ & $ -0.93\pm 0.81$ & $ 1.55\pm 0.02$ \\
SPT-CL~J0655-5234 & J065552.75-523403.3 & 06:55:52.75 & -52:34:03.3 & $0.47286(54)$ & $ 1.43\pm 1.38$ & $ -0.23\pm 1.25$ & $ 1.79\pm 0.03$ \\
SPT-CL~J2017-6258 & J201753.08-625938.7 & 20:17:53.08 & -62:59:38.7 & $0.53624(26)$ & $ 1.09\pm 3.13$ & $ 2.49\pm 2.55$ & $ 1.65\pm 0.04$ \\
SPT-CL~J2020-6314 & J202008.39-631449.7 & 20:20:08.39 & -63:14:49.7 & $0.53761(33)$ & $ 1.19\pm 1.33$ & $ -0.51\pm 1.23$ & $ 1.61\pm 0.03$ \\
SPT-CL~J2026-4513 & J202628.26-451359.6 & 20:26:28.26 & -45:13:59.6 & $0.68694(12)$ & $ -3.35\pm 3.43$ & $ -0.54\pm 1.81$ & $ 1.64\pm 0.03$ \\
SPT-CL~J2030-5638 & J203045.25-563755.8 & 20:30:45.25 & -56:37:55.8 & $0.39321(32)$ & $ 0.68\pm 2.66$ & $ -2.56\pm 1.51$ & $ 1.88\pm 0.04$ \\
SPT-CL~J2035-5251 & J203510.69-525122.1 & 20:35:10.69 & -52:51:22.1 & $0.53465(53)$ & $ 1.65\pm 1.15$ & $ -0.29\pm 0.62$ & $ 1.66\pm 0.02$ \\
SPT-CL~J2058-5608 & J205822.28-560847.2 & 20:58:22.28 & -56:08:47.2 & $0.60610(50)$ & $ -58.69\pm 8.40$ & $ -4.12\pm 3.38$ & $ 1.21\pm 0.04$ \\
SPT-CL~J2115-4659 & J211512.78-465847.5 & 21:15:12.78 & -46:58:47.5 & $0.29587(31)$ & $ 4.57\pm 0.74$ & $ -1.81\pm 0.53$ & $ 2.06\pm 0.02$ \\
SPT-CL~J2118-5055 & J211853.67-505555.2 & 21:18:53.67 & -50:55:55.2 & $0.62380(25)$ & $ 4.63\pm 2.04$ & $ -2.24\pm 1.94$ & $ 1.86\pm 0.04$ \\
SPT-CL~J2136-4704 & J213627.85-470505.7 & 21:36:27.85 & -47:05:05.7 & $0.41710(50)$ & $ 0.70\pm 1.51$ & $ -0.39\pm 0.93$ & $ 1.74\pm 0.02$ \\
SPT-CL~J2140-5727 & J214033.43-572711.4 & 21:40:33.43 & -57:27:11.4 & $0.40778(14)$ & $ -9.73\pm 0.88$ & $ -1.35\pm 0.67$ & $ 1.77\pm 0.01$ \\
SPT-CL~J2146-4846 & J214605.93-484653.3 & 21:46:05.93 & -48:46:53.3 & $0.61770(25)$ & $ 3.91\pm 3.04$ & $ 0.48\pm 1.10$ & $ 1.81\pm 0.03$ \\
SPT-CL~J2146-5736 & J214648.41-573653.7 & 21:46:48.41 & -57:36:53.7 & $0.61060(09)$ & $ -0.95\pm 1.73$ & $ -1.15\pm 1.60$ & $ 1.51\pm 0.03$ \\
SPT-CL~J2155-6048 & J215554.65-604723.7 & 21:55:54.65 & -60:47:23.7 & $0.53480(50)$ & $ 0.90\pm 1.17$ & $ 0.41\pm 8.18$ & $ 1.77\pm 0.02$ \\
SPT-CL~J2155-6048 & J215555.47-604902.8 & 21:55:55.47 & -60:49:02.8 & $0.54190(25)$ & $ 2.41\pm 1.04$ & $ -0.53\pm 1.30$ & $ 2.00\pm 0.02$ \\
SPT-CL~J2159-6244 & J215958.67-624514.0 & 21:59:58.67 & -62:45:14.0 & $0.39186(29)$ & $ 2.47\pm 0.60$ & $ -1.99\pm 0.58$ & $ 1.84\pm 0.01$ \\
SPT-CL~J2159-6244 & J220005.53-624456.3 & 22:00:05.53 & -62:44:56.3 & $0.39108(24)$ & $ 2.74\pm 0.64$ & $ -1.69\pm 0.65$ & $ 1.89\pm 0.01$ \\
SPT-CL~J2222-4834 & J222250.73-483435.5 & 22:22:50.73 & -48:34:35.5 & $0.65122(58)$ & $ -9.84\pm 2.34$ & $ 1.21\pm 1.34$ & $ 1.41\pm 0.02$ \\
SPT-CL~J2232-5959 & J223233.83-595953.1 & 22:32:33.83 & -59:59:53.1 & $0.59564(70)$ & $ 5.37\pm 2.07$ & $ 0.26\pm 1.53$ & $ 1.62\pm 0.02$ \\
SPT-CL~J2233-5339 & J223315.62-533909.2 & 22:33:15.62 & -53:39:09.2 & $0.43847(39)$ & $ 1.27\pm 0.84$ & $ -1.32\pm 0.62$ & $ 1.90\pm 0.02$ \\
SPT-CL~J2218-4519 & J221859.22-451852.0 & 22:18:59.22 & -45:18:52.0 & $0.63555(41)$ & $ 2.42\pm0.94$  & $ -1.07\pm0.69$ &  $1.70\pm0.02$ \\
SPT-CL~J2258-4044 & J225848.27-404430.7 & 22:58:48.27 & -40:44:30.7 & $0.89652(31)$ & $ 2.72\pm4.63$ & $-1.78\pm2.02$ & $1.54\pm0.04$ \\
SPT-CL~J2301-4023 & J230151.89-402339.7 & 23:01:51.89 & -40:23:39.7 & $0.84165(41)$ & $-50.13\pm22.03$ & $4.93\pm3.28$ & $1.38\pm0.02$ \\ 
SPT-CL~J2306-6505 & J230653.57-650517.5 & 23:06:53.57 & -65:05:17.5 & $0.52850(35)$ & $ 2.93\pm 0.64$ & $ -0.33\pm 0.66$ & $ 1.79\pm 0.01$ \\
SPT-CL~J2325-4111 & J232512.01-411156.5 & 23:25:12.01 & -41:11:56.5 & $0.35390(25)$ & $ 2.33\pm 0.90$ & $ -0.23\pm 10.45$ & $ 2.01\pm 0.02$ \\
SPT-CL~J2325-4111 & J232511.71-411213.8 & 23:25:11.71 & -41:12:13.8 & $0.36240(75)$ & $ 3.31\pm 1.36$ & $ -0.98\pm 4.07$ & $ 1.91\pm 0.03$ \\
SPT-CL~J2335-4544 & J233508.51-454420.8 & 23:35:08.51 & -45:44:20.8 & $0.54592(63)$ & $ 0.57\pm 1.43$ & $ -2.31\pm 1.45$ & $ 1.69\pm 0.03$ \\
SPT-CL~J2344-4243 & J234443.90-424312.1 & 23:44:43.90 & -42:43:12.1 & $0.59810(99)$ & $ -97.05\pm 3.74$ & $ -3.73\pm 1.04$ & $ 0.90\pm 0.01$
\enddata
\tablecomments{Example of the primary SPT-GMOS data product catalog for each individual galaxy with a spectroscopic redshift measurement. The catalog includes, from left to right, the cluster name, galaxy name, galaxy coordinates, galaxy redshift, the equivalent width of [O {\tiny II}] 
$\lambda$3727, the equivalent width of \hdelta\, and the strength of the 4000\AA\ break.}
\tablenotetext{a}{The measured redshift with the uncertainty in the last two digits given in parentheses.}
\end{deluxetable*}

\subsubsection{Redshifts of Giant Arcs}

Parallel to the primary objective of the SPT-GMOS survey program, we take every opportunity to 
place spectroscopic slits on other targets of high interest, such as candidate strongly-lensed 
sources in the cores of the target SPT clusters.
Redshifts for these sources are estimated using one or both of: 1) our custom IDL 
cross-correlation code with either the \citet{shapley03} $z\sim2-3$ composite spectrum or 
the Gemini Deep Deep Survey late-type $z\sim1-2$ composite \citep[GDDS;][]{abraham04}, 
and 2) fitting Gaussian profiles to families of typical strong ultraviolet (UV) absorption lines 
(e.g., Mg {\small II}  2796, 2803, Fe {\small II} 2344, 2372, 2384, 2586, 2600, C {\small IV} 1548, 
1551, Si {\small II} 1260, 1527, and Si {\small IV} 1394, 1403) and measuring the mean and standard 
deviation in the individual line redshifts.

The spectra of some candidate strongly-lensed 
sources exhibit only weak continuum emission and do not have a clear redshift solution, but 
can have robust redshift constraints inferred from the combination of their blue colors and 
lack of emission lines \citep[e.g.,][]{bayliss11a,bayliss11b}. For example, we expect a star-forming 
galaxy at $z\gtrsim1.4$, observed over a range $\Delta \lambda \simeq5000-9000$\AA, to produce 
blue continuum emission with no strong emission lines; the most prominent features we would 
expect in such a spectrum would be absorption from low-ionization species in the inter-stellar 
medium, the strength of which can vary significantly from galaxy to galaxy and are often 
undetected in low S/N spectra. Similarly, in $\Delta \lambda \simeq5000-9000$\AA\ spectra of 
a star-forming galaxy at $z\gtrsim3.1$ we would expect to see strong features associated with 
Lyman break galaxies \citep[e.g.,][]{shapley03}, the most prominent being strong 
absorption or emission from Ly-$\alpha$. The absence of strong rest-frame optical emission 
lines and Ly-$\alpha$ in a given galaxy spectrum allows us to place lower and upper limits, 
respectively, on the redshift of that galaxy.  The redshift interpretations of giant arcs observed in 
SPT-GMOS are described in more detail in $\S~\ref{sec:giantarcs}$.

\begin{figure}[t]
\includegraphics[scale=0.425]{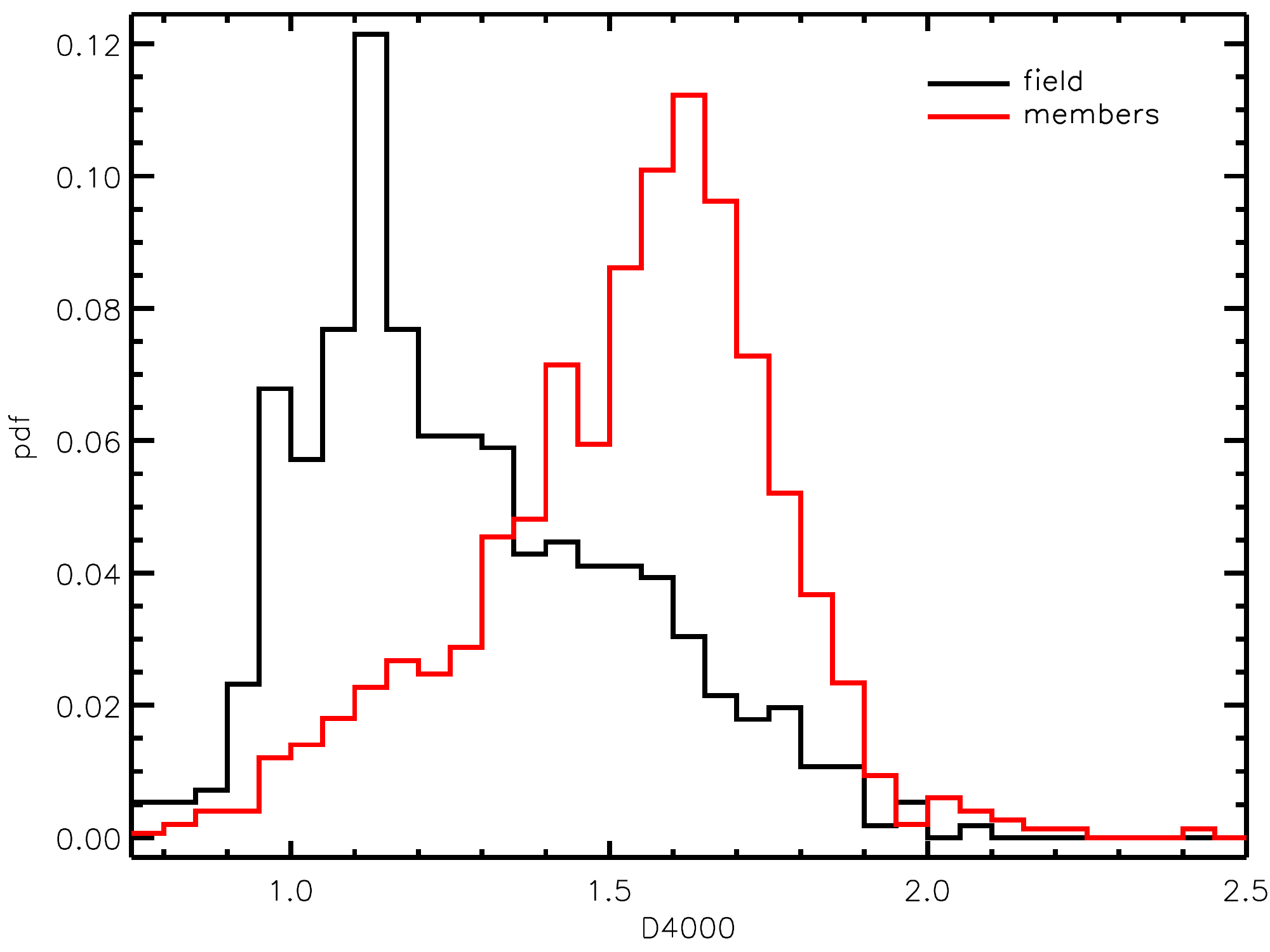}
\caption{\scriptsize{
The distribution of values for the strength of the 4000\AA\ break of both 
cluster member galaxies and non-cluster member galaxies.}}
\label{fig:gald4000}
\end{figure}

\subsection{Cluster Redshift and Velocity Dispersion Estimates}
\label{sec:zdispestimates}

The primary quantities that we want to measure for each galaxy cluster are the average cluster 
redshift, which reflects the bulk motion of the cluster in the Hubble Flow, and the velocity dispersion 
of cluster member galaxies, which scales with the depth of the cluster's gravitational potential. We 
follow the procedure described in \citet{ruel14}, which includes first computing the average cluster 
redshift, ${\bar z_\mathrm{cluster}}$, using the bi-weight location estimator as formulated by \citet{beers90}, 
and then compute the line-of-sight velocity dispersion for each cluster using two estimators: the 
square root of the bi-weight sample variance, $\sigma_{v,BI}$, and the gapper, $\sigma_{v,G}$ 
\citep{beers90}. Initial estimates of ${\bar z_\mathrm{cluster}}$, $\sigma_{v,BI}$ 
and $\sigma_{v,G}$ are generated by manually identifying each galaxy cluster as an over-density in 
velocity space and applying an initial rest-frame velocity cut of $\pm5000$ km s$^{-1}$ relative to the 
starting guess of the cluster redshift. 
The choice of 5000 km s$^{-1}$ is somewhat arbitrary, but our results are not sensitive to small 
changes in the choice of initial velocity cut.
We then iteratively compute ${\bar z_{cluster}}$, $\sigma_{v,BI}$ and 
$\sigma_{v,G}$, applying rest-frame velocity cuts of $\pm$ 3$\sigma_{v}$, where $\sigma_{v}$ is set 
equal to $\sigma_{v,BI}$ when computed from $\geq15$ spectroscopic members, and equal to 
$\sigma_{v,G}$ when computed from $<15$ members. The iterations continue until 
converging onto a single solution. The final SPT-GMOS median cluster redshift estimates inform a 
calibration of photometric redshifts measured from cluster red sequences for the full SPT-SZ cluster 
sample over a broad redshift range (Figure~\ref{fig:zcalibration}).

Table~\ref{tab:spectable} gives the final estimated values 
of ${\bar z_{cluster}}$, $\sigma_{v,BI}$ and $\sigma_{v,G}$ for the 62 SPT-GMOS clusters with 
data taken and reduced. We also show the velocity histograms with dispersion estimates over-plotted in 
Figure~\ref{fig:dispersions}. Velocity dispersion confidence intervals are computed to be 
$\pm0.92 \sigma_{v} / \sqrt{N_{Members} - 1}$, which accurately captures the total 
measurement confidence intervals including both statistical uncertainties as well as the 
systematic uncertainties from the estimators and the effects of membership selection 
\citep{ruel14}. For clusters with $<15$ members the gapper estimate of the velocity dispersion 
is generally considered more reliable, and for clusters with $\geq 15$ members the bi-weight 
estimator is likely the better choice \citep{beers90,ruel14}, though we note that the 
$\sigma_{v,BI}$ and $\sigma_{v,G}$ estimates are in excellent agreement for all 62 clusters 
measured (Table~\ref{tab:spectable}). We identify a total of 1579 cluster member galaxies 
across 62 galaxy clusters, consistent with our $N\gtrsim25$ members per cluster goal 
\citep{ruel14}.

\subsubsection{Normality of Cluster Galaxy Velocity Distributions}
\label{sec:normalitytests}

The velocity dispersion estimators that we apply to our redshift data make an implicit assumption 
about the underlying cluster velocity distributions. Specifically, we assume that they are Gaussian, 
which is not always the case. The empirical uncertainty formula that we apply does 
account, at least in part, for the additional average uncertainty in the velocity dispersion estimate 
that results from ity of cluster velocity distributions \citep{ruel14}, but it is also useful 
to test each individual cluster velocity distribution for evidence of non-Gaussianity. There are a number 
of different statistical tests that can be applied here, but the value of these tests varies strongly 
with the number of individual galaxy measurements that are available for a given galaxy cluster. 
\citet{einasto2012}, for example, restrict their analysis of velocity substructure and non-Gaussianity 
in massive clusters to systems with at least 
50 members; none of the SPT-GMOS galaxy clusters meet this member galaxy threshold. 
In a systematic study of this topic \citet{hou2009} find that some tests are ``profoundly 
unreliable'' when $<$ 30 galaxy velocities are available, which is the case for 2/3 of our sample. 
\citet{hou2009} do find, however, that the Kolmogorov and Anderson-Darling (AD) tests are robust even 
when applied to very small samples (down to at least $N=5$). It has been shown that the AD 
test is among the most statistically powerful tests for detecting departure from normality, 
whereas the Kolmogorov is among the least powerful \citep{hou2009}. We therefore compute the 
AD test statistic, in which $A^{2*}$ for the ordered data, $x_{i}$, is defined as, 

\begin{equation}
A^{2*} = A^{2}\left(1 + \frac{0.75}{n} + \frac{2.25}{n^{2}}\right), 
\end{equation}

where $A^2$ is given by,

\begin{equation}
\small
A^{2} = -n-\frac{1}{n}\sum_{i=1}^n(2i-1)(\ln\Phi(x_{i}) + \ln(1 - \Phi(x_{n+1-i}))), 
\end{equation}

\noindent
$x_{i} \leq x < x_{i+1}$, and $\Phi(x_{i})$ is the cumulative distribution function 
of the hypothetical underlying distribution. The probability that a velocity distribution tested in this 
way is non-Gaussian, $\alpha_{AD}$, is easily computed (see Equation 17 in \citet{hou2009}). The 
results of the AD test for all SPT-GMOS galaxy clusters are reported in Table~\ref{tab:spectable}, 
along with the implied probability that each cluster's velocity distribution is non-Gaussian. Two (nine) 
of the 62 SPT-GMOS clusters have velocity distributions that are discrepant from Gaussian distributions 
at the 3-$\sigma$ (2-$\sigma$) level.

\begin{figure*}
\centering
\includegraphics[scale=0.692]{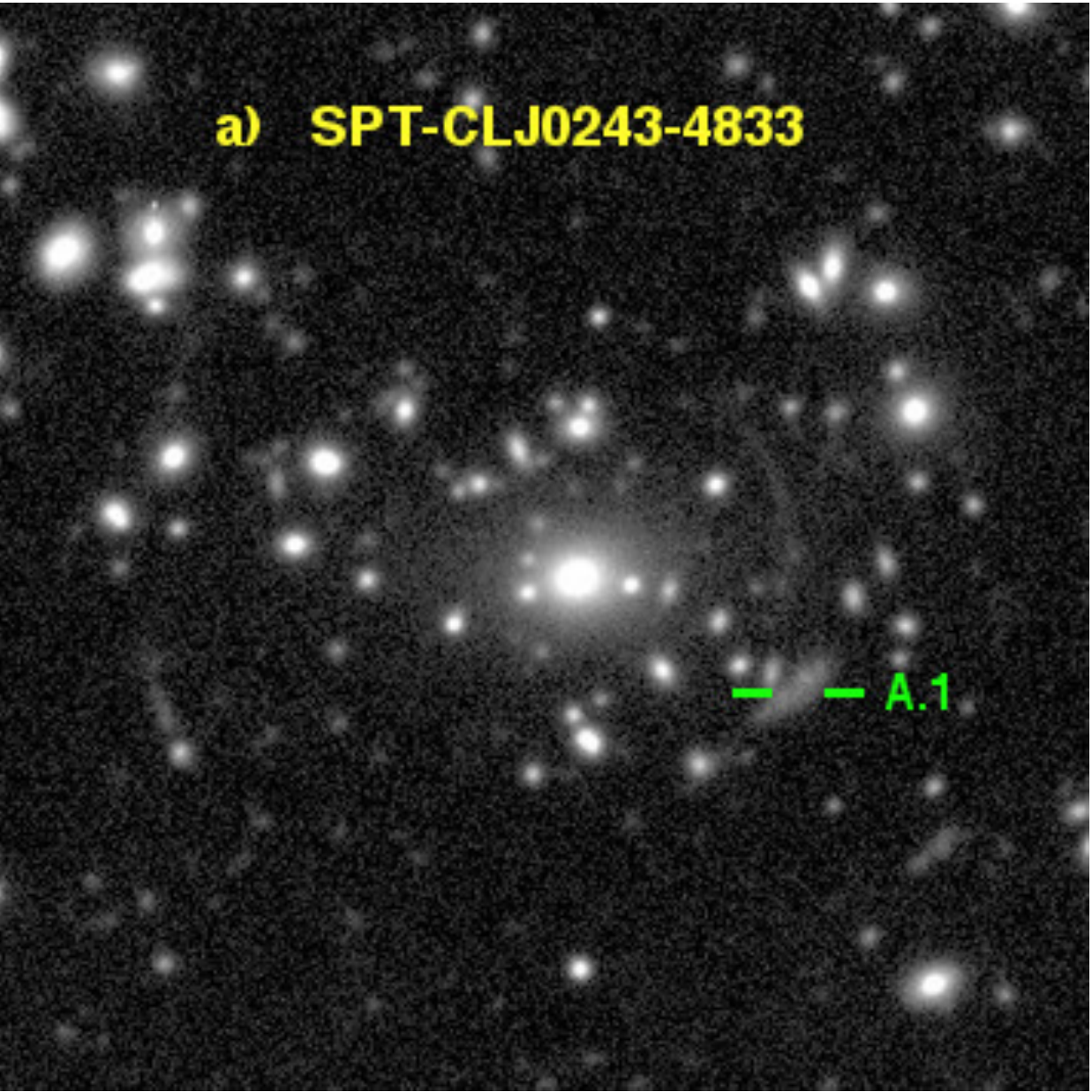}
\includegraphics[scale=0.645]{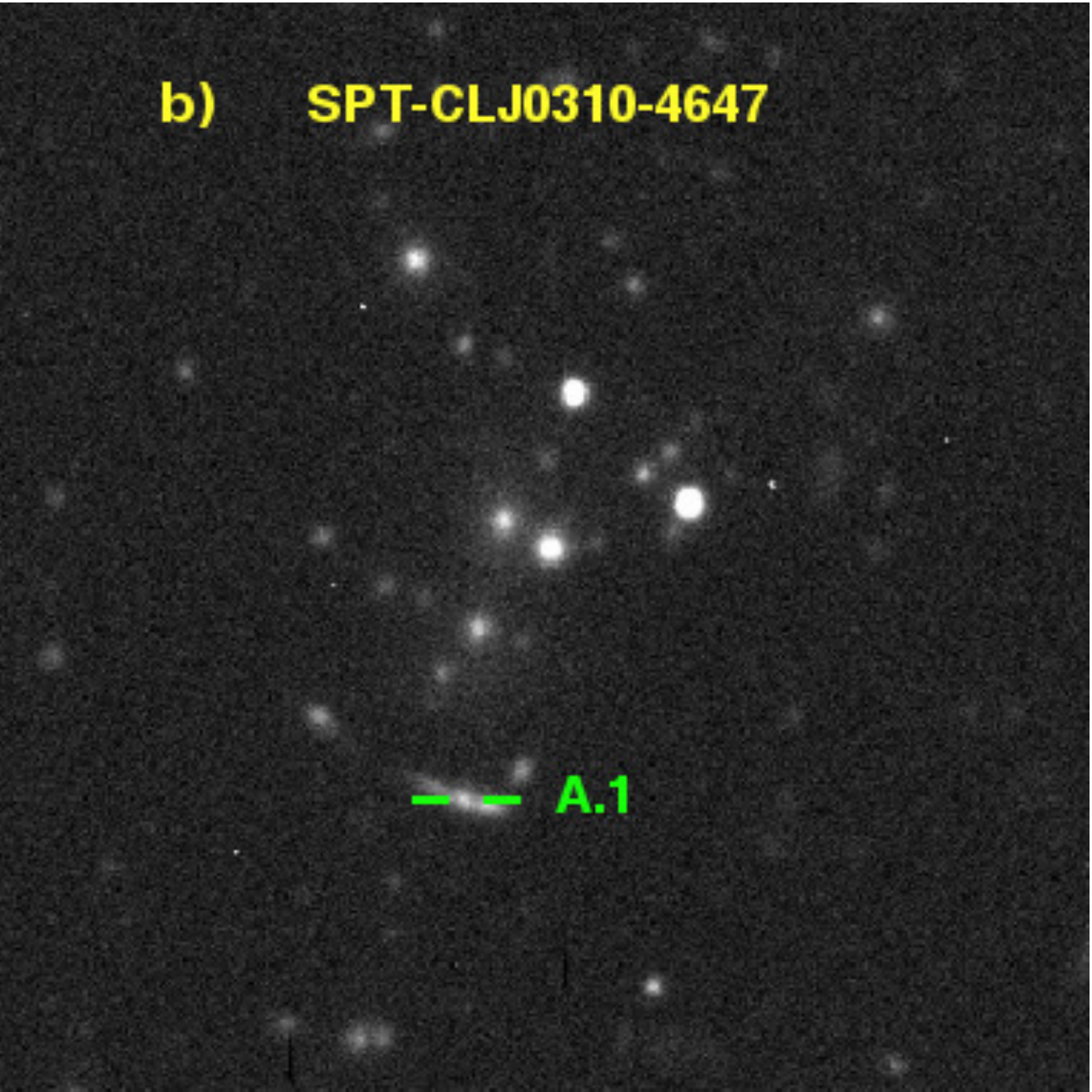}
\includegraphics[scale=0.7222]{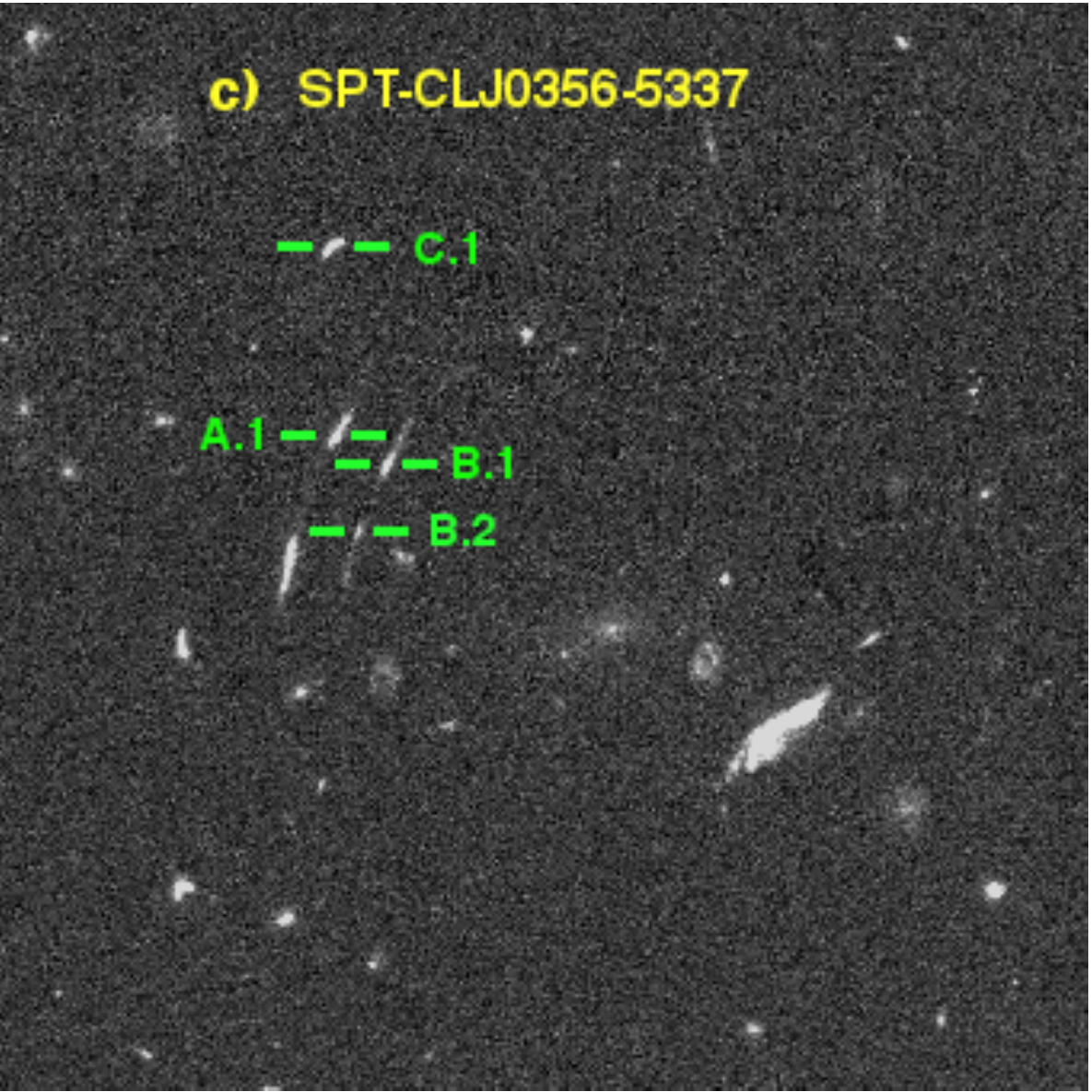}
\includegraphics[scale=0.612]{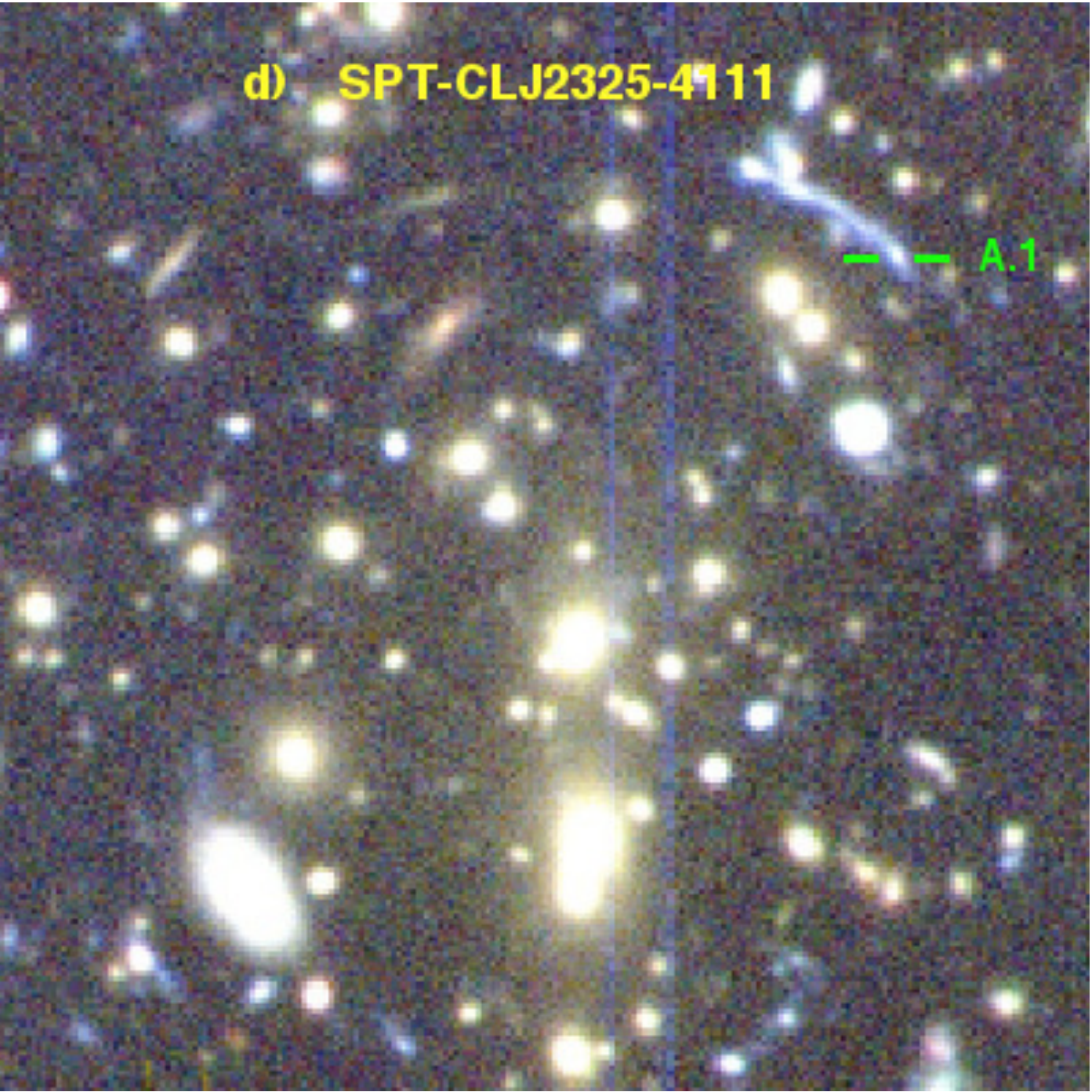}
\caption{\scriptsize{
Optical images of SPT-GMOS clusters with candidate giant arcs for which we 
measure a robust redshift of at least one arc. We show the best-available imaging for 
showing the strong lensing features, which varies depending on the data available for 
each cluster. Horizontal green dashes 
indicate individual candidate strongly-lensed sources that were targeted by a 
slit in the SPT-GMOS program. Each targeted source is labeled, and the labels 
match those referred to in Table~\ref{tab:giantarcs}. North is up and East is to the left 
in each image.
a) Magellan-II/MegaCam $r-$band image of the central 75\arcsec$\times$75\arcsec region of SPT-CL~J0243-4833.
b) Gemini/GMOS-South $r-$band image of the central 60\arcsec$\times$60\arcsec region of 
SPT-CL~J0310-4647.
c) {\it HST/ACS} F606W image of the central 40\arcsec$\times$40\arcsec region of SPT-CL~J0356-5337. 
d) Magellan-II/PISCO \citep{stalder14} $gri$-band image of the central 60\arcsec$\times$60\arcsec region of 
SPT-CL~J2325-4111.}}
\label{fig:arcs1}
\end{figure*}

\begin{figure*}
\centering
\includegraphics[scale=0.706]{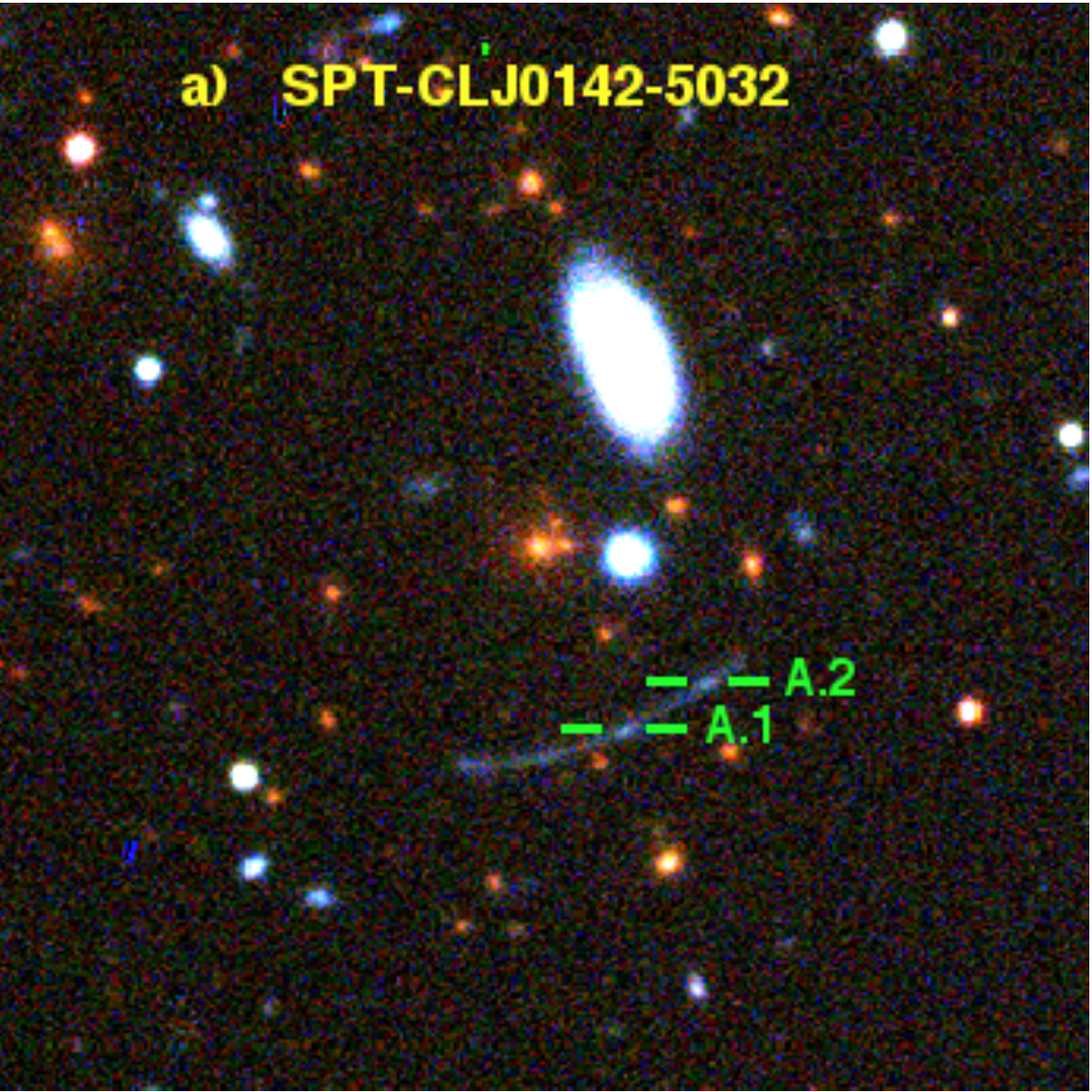}
\includegraphics[scale=0.64]{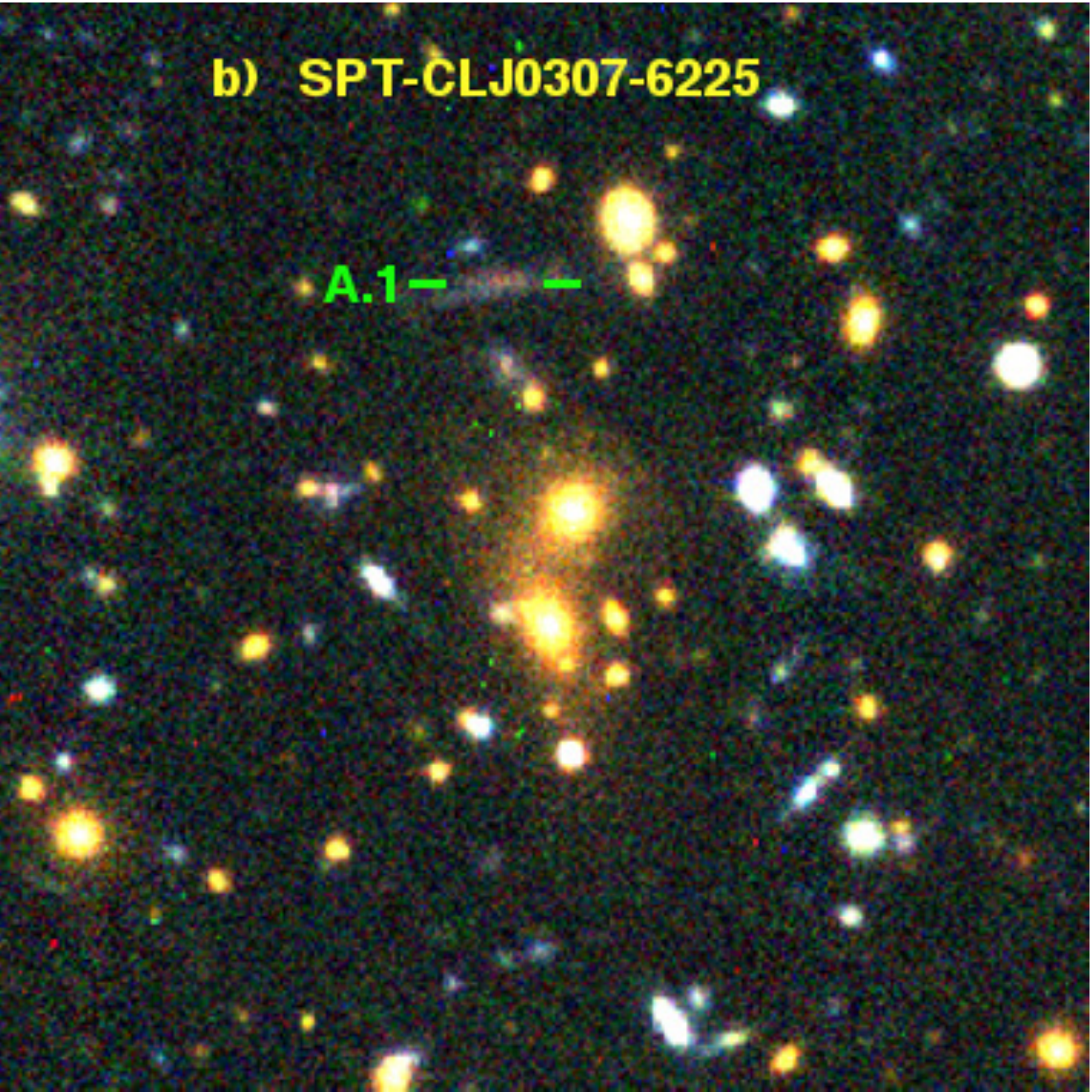}
\includegraphics[scale=0.64]{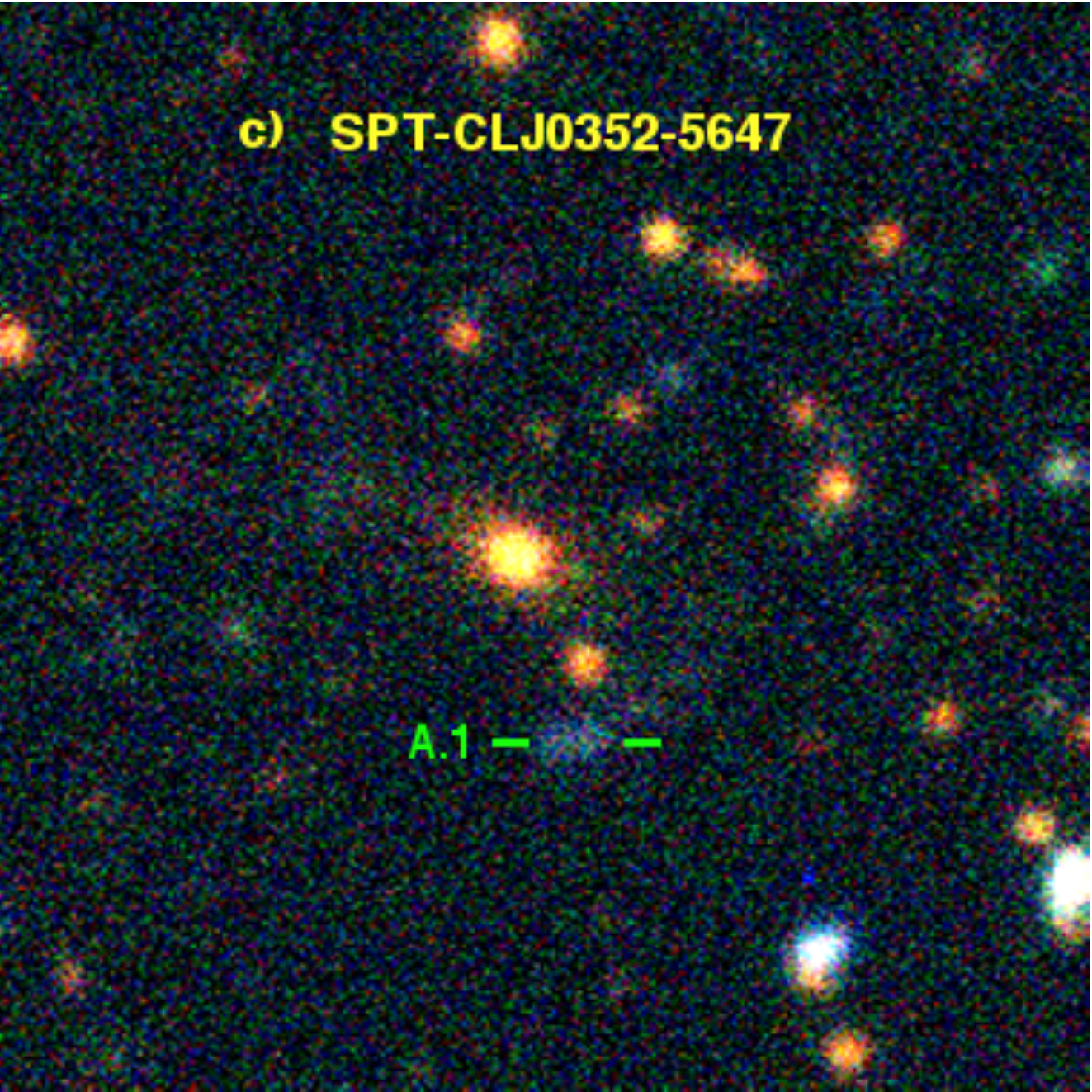}
\includegraphics[scale=0.64]{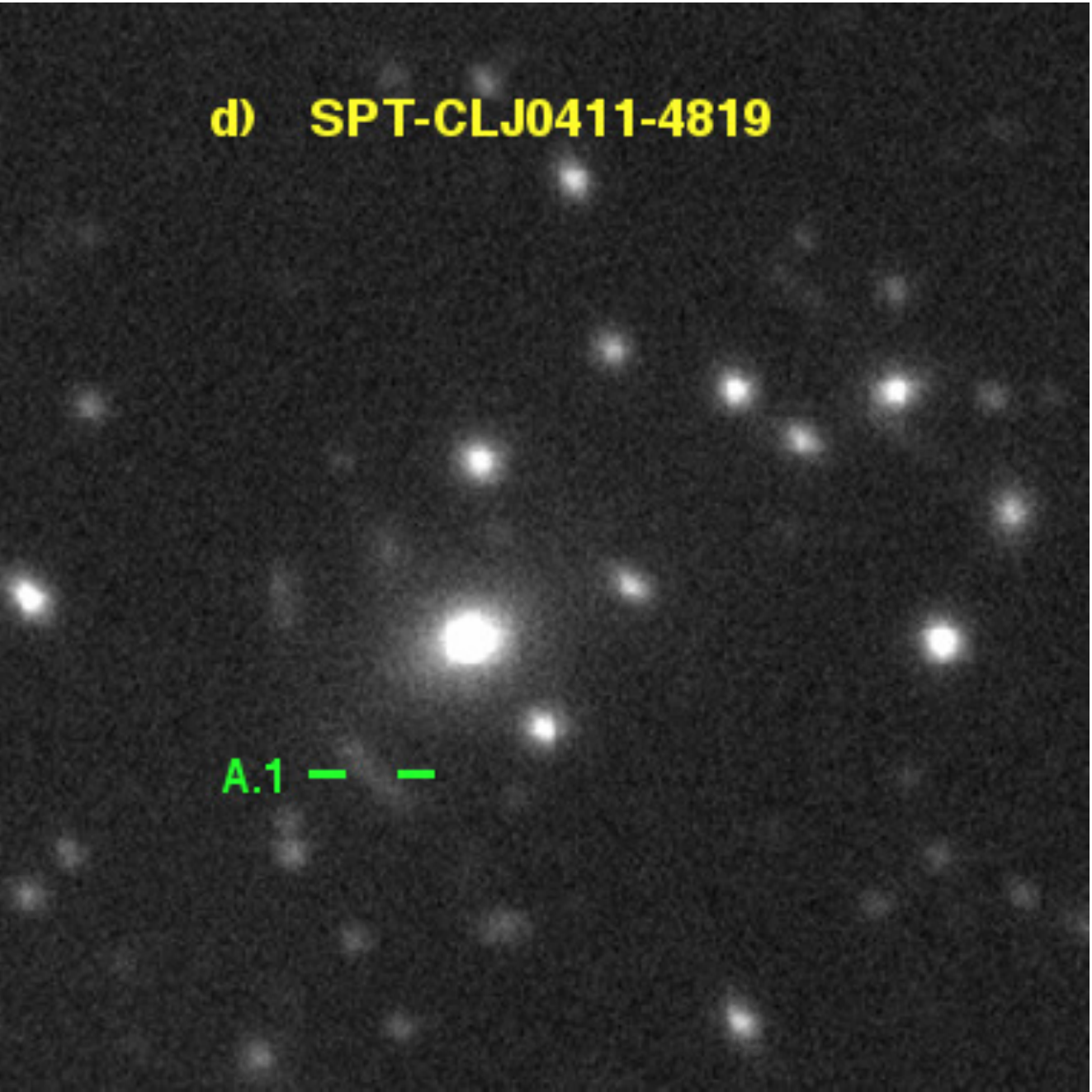}
\caption{\scriptsize{
Optical images of SPT-GMOS clusters with candidate giant arcs where the spectra provide 
possible redshift measurements. Horizontal green dashes 
indicate individual candidate strongly-lensed sources that were targeted by a 
slit in the SPT-GMOS program. Each targeted source is labeled, and the labels 
match those referred to in Table~\ref{tab:giantarcs}. North is up and East is to the left 
in each image.
a) Magellan-I/IMACS $grz$ image of the central 75\arcsec$\times$75\arcsec region of 
SPT-CL~J0142-5032.
b) Magellan-II/LDSS3 $gri$ image of the central 60\arcsec$\times$60\arcsec region of 
SPT-CL~J0307-6225.
c) Magellan-II/LDSS3 $gri$ image of the central 60\arcsec$\times$60\arcsec region of 
SPT-CL~J0352-5647.
d) Gemini/GMOS-South $r-$band image of the central 60\arcsec$\times$60\arcsec region of 
SPT-CL~J0411-4819.}}
\label{fig:arcs2}
\end{figure*}

\subsection{Galaxy Spectral Indices: \oii, \hdelta, and D4000\AA}
\label{sec:indices}

GMOS spectra taken with the B600 (R400) grating cover wavelength ranges of 
$\Delta \lambda \simeq$ 2800 (4200) \AA; this broad coverage ensures that we sample 
several well-established spectral indices for the large majority of the SPT-GMOS galaxies. 
We can, therefore, generate catalogs of galaxies with spectral index measurements of nearly 
all of the galaxies with GMOS spectra. Two important features that we focus on here are 
the \oii\ forbidden line and \hdelta. We measure rest-frame equivalent widths for each of these 
features in every galaxy where we have both a redshift measurement and where the spectra 
cover the appropriate rest-frame wavelengths. The equivalent width of a transition with 
$\lambda$ is defined by the equation, 

\begin{equation}
W_{\lambda} = \int (1 - F_{\lambda}/F_\mathrm{cont}) d\lambda
\end{equation}

\noindent
where all quantities have been converted into the rest-frame. We compute W$_{0}$ for 
\oii\ and \hdelta\ using the well-established intervals that define the flux density per pixel in the 
spectral line, $F_{\lambda}$, and the flux density per pixel in the continuum, $F_{cont}$, from 
\citet{balogh99}. We also compute the strength of the 4000\AA\ break (D4000) using wavelength 
intervals also defined by \citet{balogh99}. The distribution of D4000 values for cluster members 
--- as defined during our iterative velocity dispersion estimation described above in 
\S~\ref{sec:zdispestimates} --- and non-members is shown in Figure~\ref{fig:gald4000}. Using these 
standard index definitions ensures uniform measurements, independent of both spectral resolution 
and the S/N of the spectra. Sample measurements of redshift and spectral indices are presented in 
Table~\ref{tab:bcgspec}; this table is an abridged version of the complete dataset where here we only 
show one or two candidate BCG(s) for each cluster. For SPT-SZ galaxy clusters that appear in 
\citet{mcdonald16} we use the same BCG candidates, and we perform the same visual selection 
of candidate BCGs for the rest of the SPT-GMOS sample. Not all clusters have a single clear BCG, 
so that in some cases we flag multiple bright galaxies as possible BCGs; we also note that a few 
of the clusters have BCGs that were not given a slit during the mask design step.

\section{Secondary Survey Products}
\label{sec:discussion}

In this section we describe several additional ancillary and derived data products beyond the primary 
measurements described above. This includes reporting redshift measurements and redshift constraints 
(for spectra where there is no clear redshift solution) for a sample of bright strongly lensed sources, 
matching the SPT-GMOS spectroscopic catalogs against photometric catalogs from previously 
published SPT-SZ follow-up, and derived properties of SPT-GMOS galaxies 

\begin{figure*}
\centering
\includegraphics[scale=0.502]{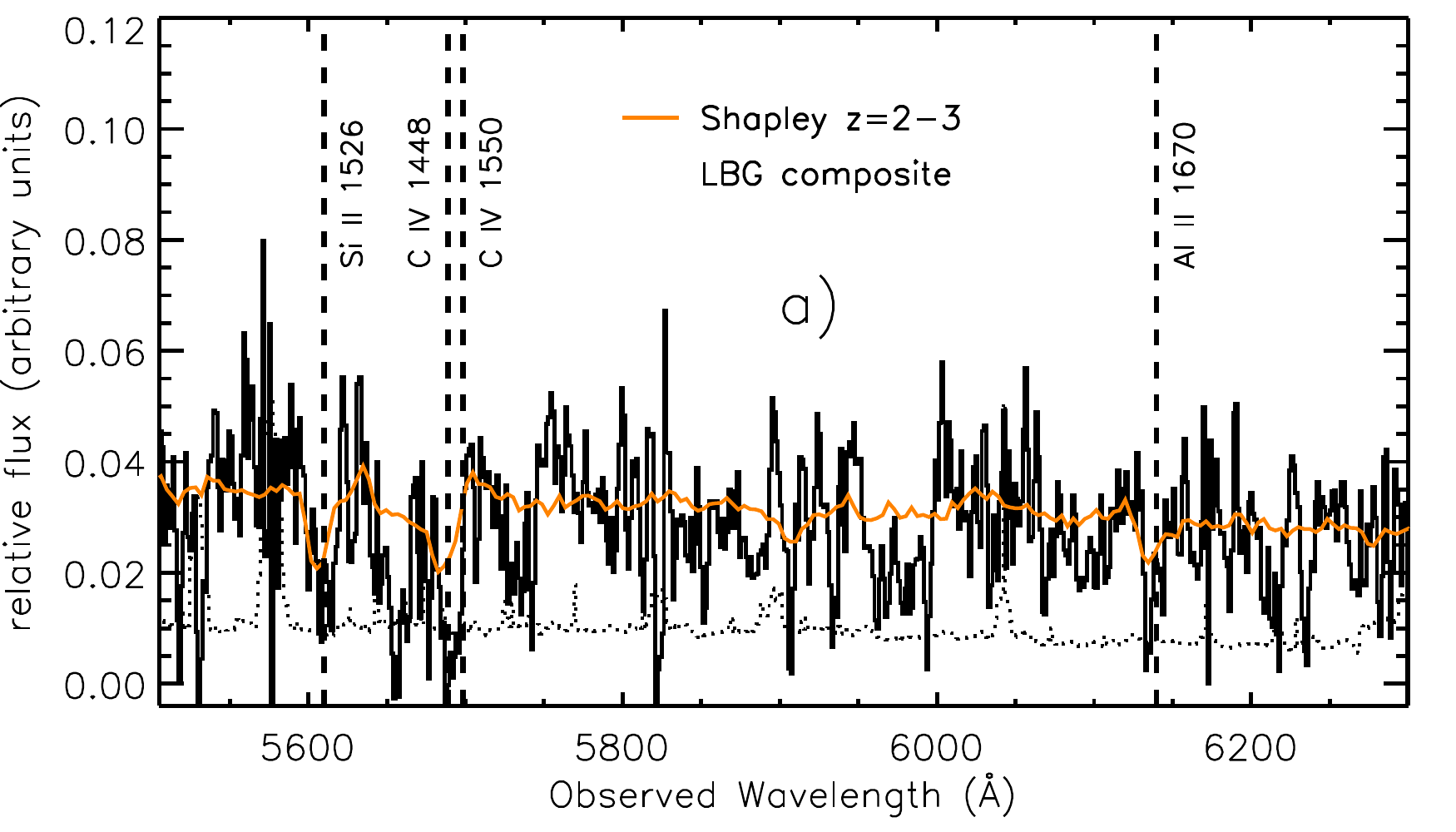}
\includegraphics[scale=0.502]{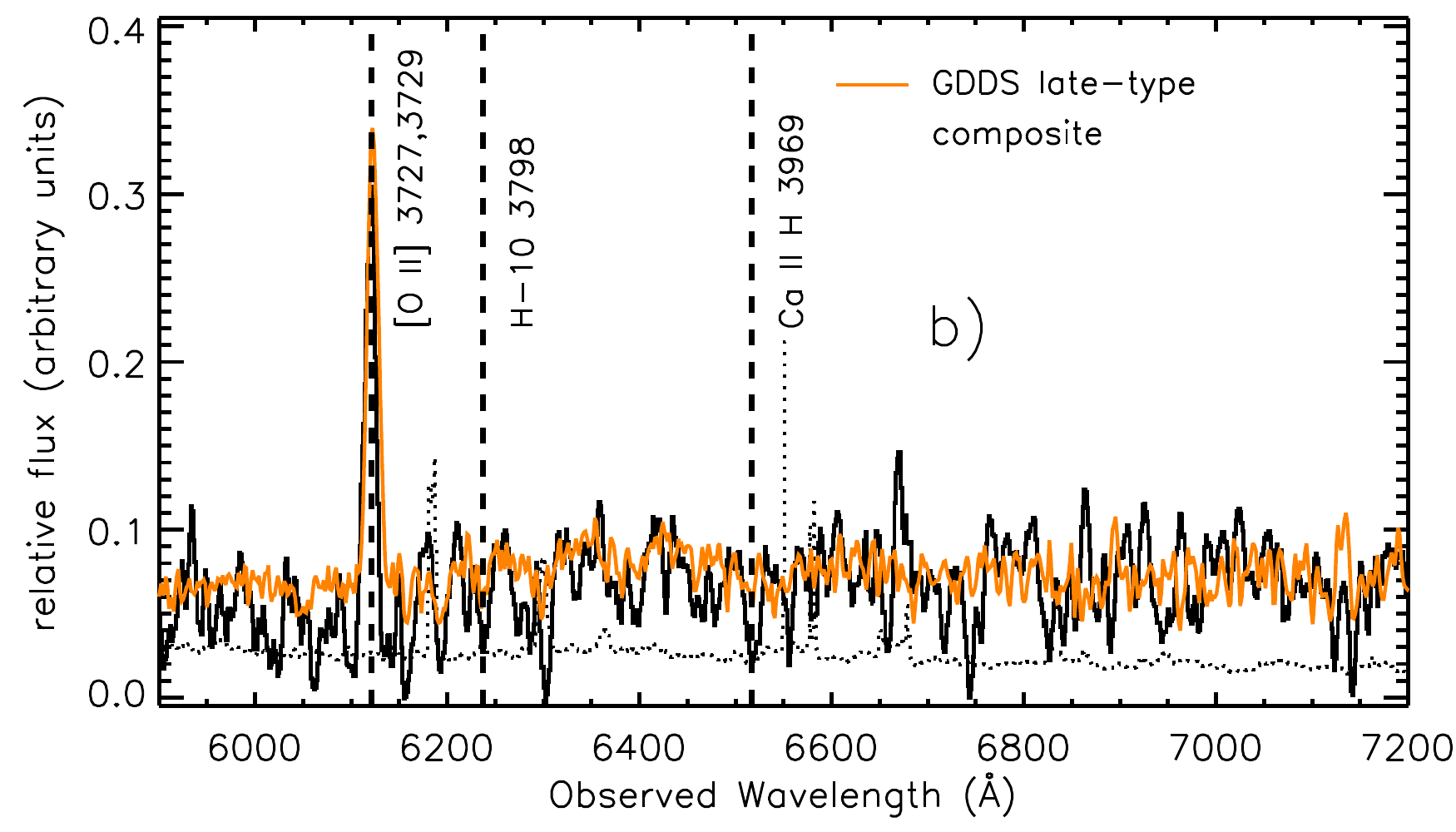}
\includegraphics[scale=0.502]{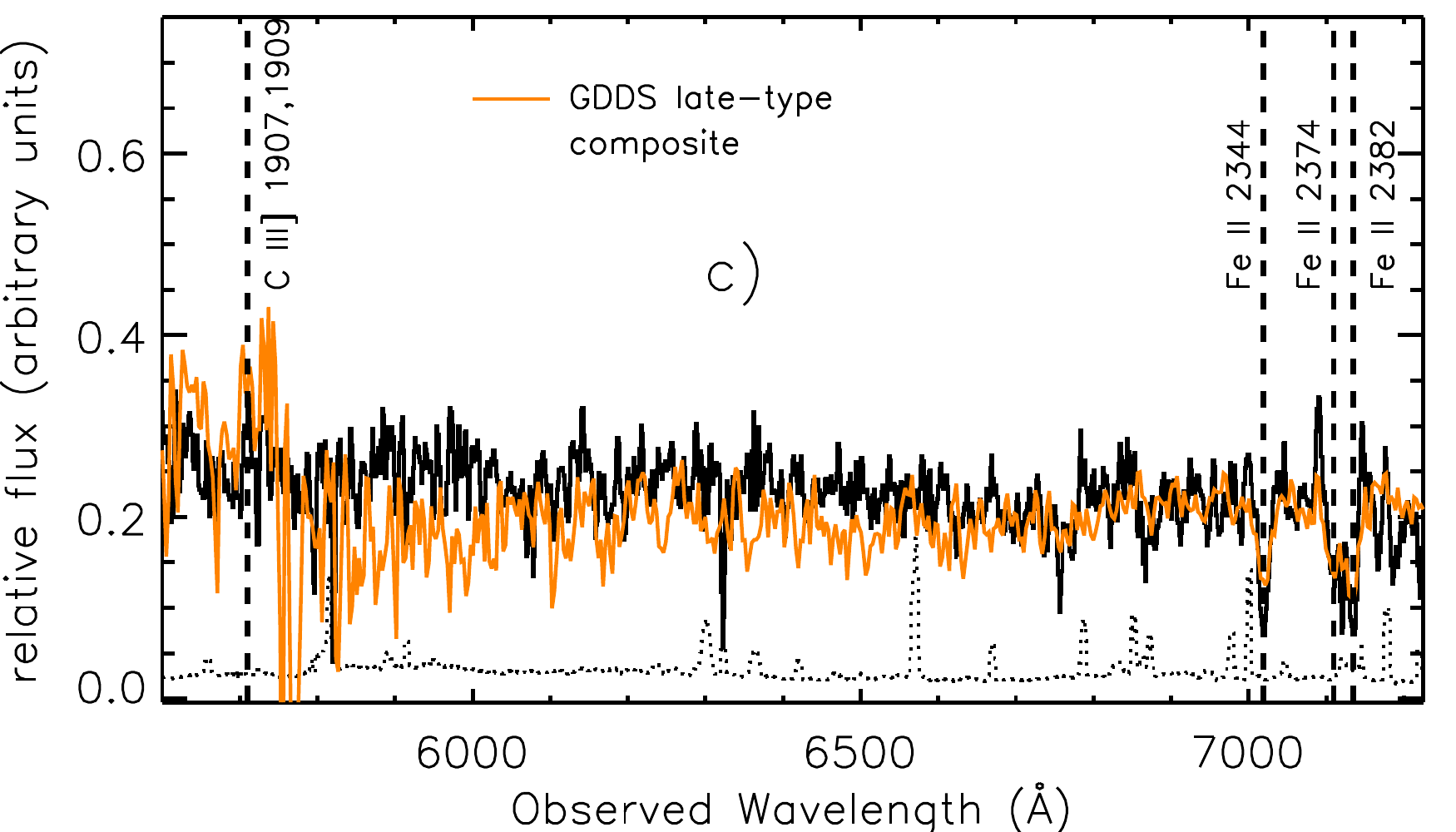}
\includegraphics[scale=0.502]{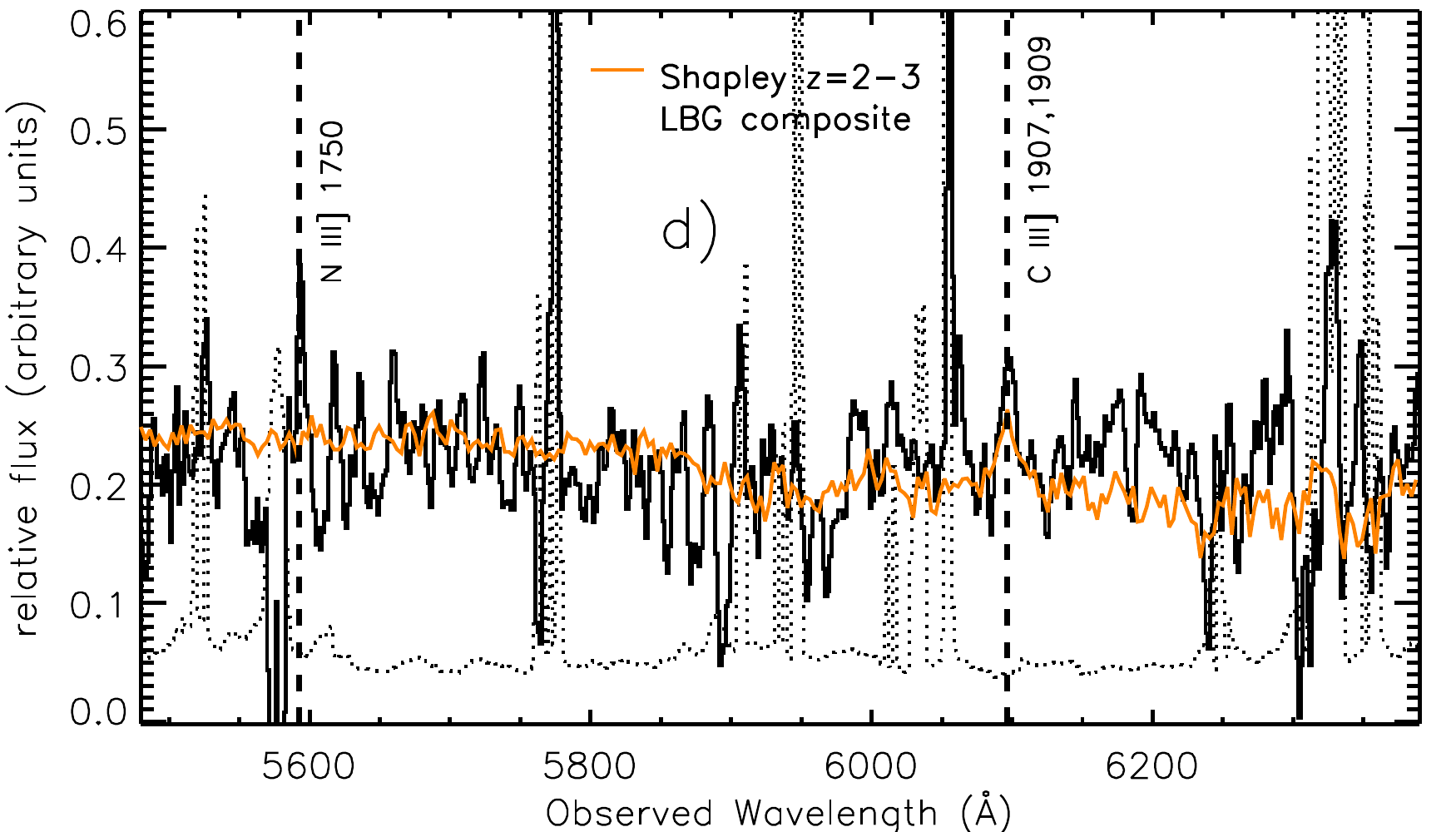}
\includegraphics[scale=0.502]{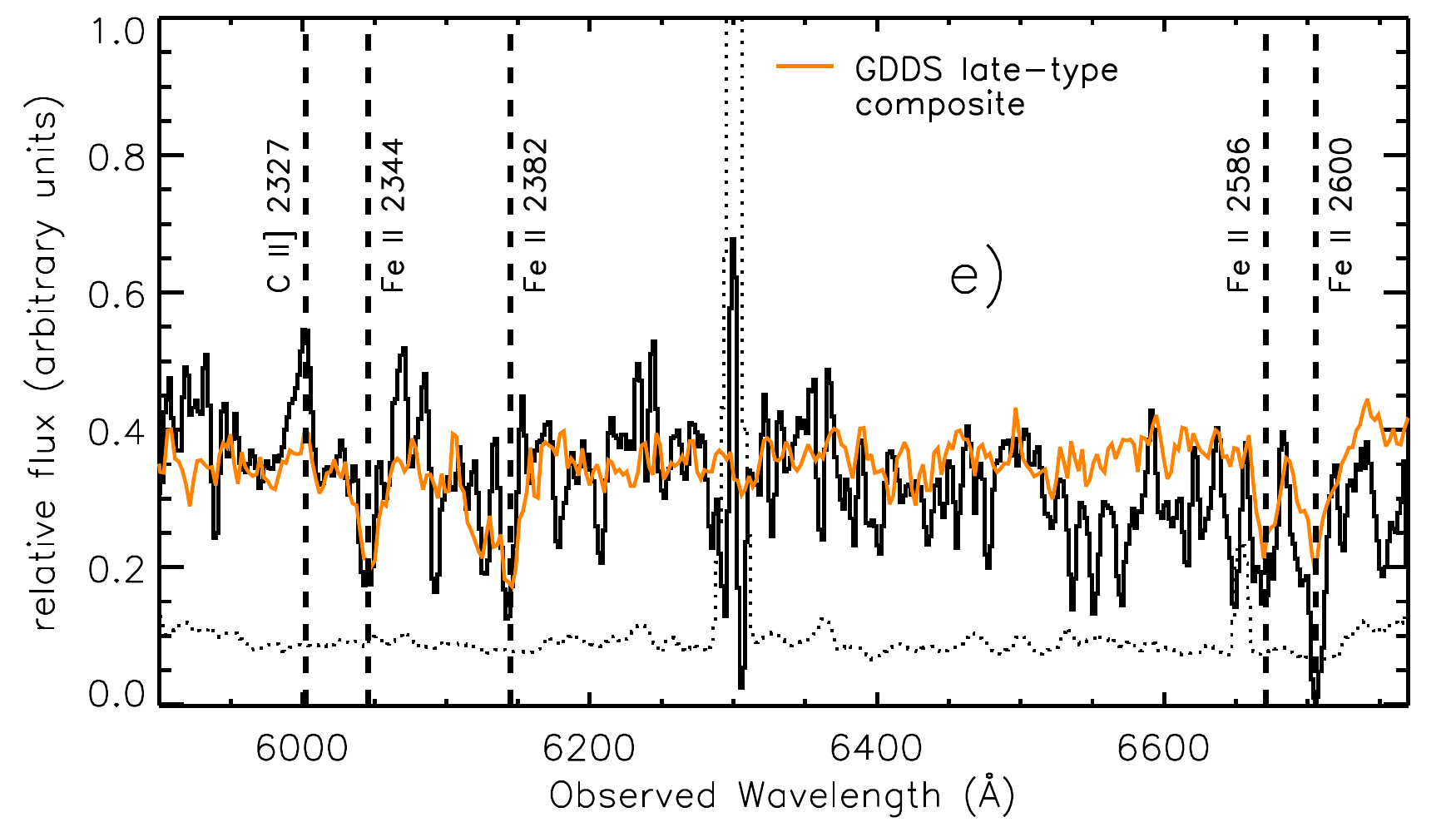}
\caption{\scriptsize{
GMOS spectra (black solid line) of four giant arcs from which we measure redshifts, with the pixel-by-pixel 
uncertainties plotted as dotted lines. In each panel we also over-plot in orange one of either the Shapley 
et al.~(2003) composite $z\simeq 2-3$ LBG composite spectrum, or the Abraham et al.~(2004) GDDS 
$z \simeq 1-2$ late-type composite spectrum. 
a) Spectrum of A.1$+$A.2 in SPT-CL~J0142-5032 with the Shapley et al. (2003) LBG composite redshifted to 
match that of the arc. 
b) Spectrum of A.1 in SPT-CL~J0243-4833 with the GDDS late-type composite redshifted to match 
that of the arc. 
c) Spectrum of A.1 in SPT-CL~J0356-5337 with the Shapley et al. (2003) LBG composite redshifted to match 
that of the arc. 
d) Spectrum of A.1 in SPT-CL~J0310-4647 with the GDDS late-type composite redshifted to match 
that of the arc. 
e) Spectrum of A.1 in SPT-CL~J2325-4111 with the GDDS late-type composite redshifted to match 
that of the arc.}}
\label{fig:arcspec}
\end{figure*}

\subsection{Giant Arc Redshifts}
\label{sec:giantarcs}

We report new redshift measurements for several strongly-lensed background galaxies 
that appear near the cores of SPT galaxy clusters. New redshift constraints for giant arcs are given 
in Table~\ref{tab:giantarcs}, and images of each arc appear in Figures~\ref{fig:arcs1} \& \ref{fig:arcs2}. 
The new SPT-GMOS giant arcs redshift measurements are consistent with the 
redshift distributions observed in other well-defined giant arc samples \citep{bayliss11a,bayliss12}, and 
generally follow the peak era of unobscured star formation as 
traced by high surface brightness star-forming galaxies. 

Bright strongly-lensed galaxies offer rare opportunities to study distant galaxies at a level of detail that is 
impossible in the field 
\citep{pettini00,bayliss10,bian10,koester10,wuyts12a,bayliss14b,james14}, 
and it is inevitable that the brightest high-redshift, strongly-lensed galaxies will be the most thoroughly studied and 
best-understood galaxies at their respective redshifts. Recent efforts have had great success finding the brightest 
strongly-lensed galaxies in wide-area surveys, primarily in the north. The new giant arcs presented here 
are a preliminary step toward extending the search for highly magnified distant galaxies into the south 
\citep[e.g.,][]{buckleygeer11,nord16}, and will support future strong lensing analyses of SPT strong lensing 
clusters, as well as tests of giant arc statistics within the SPT cluster sample. 

\begin{deluxetable*}{lccccl}
\tablecaption{Giant Arc Redshift Constraints \label{tab:giantarcs}}
\tablewidth{0pt}
\tabletypesize{\tiny}
\tablehead{
\colhead{SPT Cluster Lens} &
\colhead{Arc ID} &
\colhead{ Slit RA } &
\colhead{ Slit Dec } &
\colhead{Redshift}  &
\colhead{ Spectral Features and Comments} \\
\colhead{ } &
\colhead{} &
\colhead{ (J2000.0) } &
\colhead{ (J2000.0) } &
\colhead{  }  &
\colhead{  } 
}
\startdata
Solid Redshifts: &    &    &    &      \\
\hline 
SPT-CL~J0243-4833 & A.1 & 02:43:37.36 & -48:33:46.9  & $0.6418 \pm  0.0003$  & \oii, H-10 $\lambda$3798, Ca {\small II} $\lambda$3969 ; Fig.~\ref{fig:arcspec}  \\
SPT-CL~J0310-4647 & A.1 & 03:10:32.27 & -46:46:52.8  & $1.9942 \pm 0.0002$ & C {\small III}] $\lambda\lambda$1907,1909; Fe {\small II} $\lambda$2344, 2374, 2382; Fig.~\ref{fig:arcspec}   \\
SPT-CL~J2325-4111& A.1 & 23:25:10.20 & -41:11:20.0 &  $1.5790\pm0.0010$ & C {\small III}] $\lambda\lambda$1907,1909; Fe {\small II} $\lambda$2344, 2382, 2586, 2600; Fig.~\ref{fig:arcspec} \\
\hline \\
Possible Redshifts: &   &   &    &  Best Guess z:  &      \\
\hline 
SPT-CL~J0142-5032 & A.1 & 01:42:09.08 & -50:32:42.0 & $2.6740\pm0.0010$ & possible Si {\small II} $\lambda$1526, C {\small IV} $\lambda$1550, Al {\small II} $\lambda$1670; Fig.~\ref{fig:arcspec} \\
  & A.2 & 01:42:08.52 & -50:32:38.9  & $2.6740\pm0.0010$ & possible Si {\small II} $\lambda$1526, C {\small IV} $\lambda$1550, Al {\small II} $\lambda$1670; Fig.~\ref{fig:arcspec} \\
SPT-CL~J0356-5337 & A.1 & 03:56:20.23 & -53:37:53.6  & $2.1955 \pm 0.0007$ & possible C {\small III}] $\lambda$1909, N {\small III}] $\lambda$1750; See Fig.~\ref{fig:arcspec} \\
\hline  \\
Redshift Limits: &   &    &    &  z Constraints:  &      \\
\hline 
SPT-CL~J0142-5032 & A.1 & 01:42:09.08 & -50:32:42.0 &   $1.44<z<3.1$  & $\Delta \lambda = 4900-9100$\AA; see also best guess z above \\  
  & A.2 & 01:42:08.52 & -50:32:38.9  &  $1.44<z<3.1$   & $\Delta \lambda = 4900-9100$\AA; see also best guess z above  \\  
SPT-CL~J0307-6225 & A.1 & 03:07:17.17 & -62:26:28.9  & $1.30<z<2.8$ & $\Delta \lambda = 4430-8590$\AA; weak continuum only \\
SPT-CL~J0352-5647 & A.1 & 03:52:57.13 & -56:48:02.0  & $1.40<z<3.0$ & $\Delta \lambda = 4680-8980$\AA; weak continuum only \\
SPT-CL~J0356-5337 & A.1 & 03:56:20.23 & -53:37:53.6  & $1.78<z<3.9$ & $\Delta \lambda = 5920-10350$\AA; see also best guess z above \\
 & B.1  & 03:56:20.46 & -53:37:55.2 & $1.78<z<3.9$ & $\Delta \lambda = 5920-10350$\AA; weak continuum only \\
  & B.2  & 03:56:20.46 & -53:37:55.2 & $1.78<z<3.9$ & $\Delta \lambda = 5920-10350$\AA; weak continuum only \\
  & C.1 & 03:56:19.88 & -53:37:58.9 & $1.78<z<3.9$ & $\Delta \lambda = 5920-10350$\AA; weak continuum only  \\
SPT-CL~J0411-4819 & A.1 & 04:11:10.59 & -48:19:44.3 & $0.84<z<2.3$ & $\Delta \lambda = 3900-6850$\AA; weak continuum only  
\enddata
\tablecomments{Details of giant arcs observed in SPT-GMOS. Results are sorted into three categories. ``Solid redshifts'' are high-confidence 
measurements that appear in the SPT-GMOS spectroscopic catalogs. ``Possible redshifts'' are best-guess measurements and represent the 
most likely redshift for these sources based on a small number of very weak spectroscopic features, but could be mis-interpretations of the data. 
``Redshift limits'' are constraints that we place on the redshifts of sources based on the \emph{lack} of identifiable spectroscopic features in the 
available data.}
\end{deluxetable*}

\subsubsection{Giant Arc Redshift Measurements}

There are five giant arcs with SPT-GMOS spectra for which 
we report new spectroscopic redshift measurements. Three of these redshifts are 
unambiguous, but we consider the other two to be more speculative. The redshift interpretations of these 
five individual lensed galaxy spectra are briefly described below. We designate candidate strongly-lensed 
sources for each cluster, ordered arbitrarily, as A, B, C, etc., and indicate individual images as, for 
example, A.1 specifying the first image of the A lensed system, A.2 the second image, and so on 
(see Figure~\ref{fig:arcs1}a-d, Figure~\ref{fig:arcs2}a).

\vspace*{0.2cm}

\noindent
{\bf SPT-CL~J0142-5032: } This is a very low-confidence and speculative redshift solution owing to the low S/N 
($\sim$ 1-1.5 per pixel) of the spectrum. The possible redshift solution is based on weak absorption features 
that coincide with strong lines common in Lyman break galaxies at $z=2.674$. As shown in Figure~\ref{fig:arcs2}a 
we targeted this arc with slits at two different positions and we see these very weak absorption features 
in the low S/N spectra from both slits. In Figure~\ref{fig:arcspec} we show a stack of the spectra from 
the two slit positions. Even if we are mis-interpreting these weak features then we can still confidently conclude 
that this source falls within the redshift range, $1.44<z<3.1$, based on our wavelength coverage and the lack of 
strong spectral features.

\vspace*{0.2cm}

\noindent
{\bf SPT-CL~J0243-4833: } The strong \oii\ emission line and 
accompanying weak absorption features (Ca {\small II} H $\lambda$3969 and H-10 Balmer 
line) provide a robust redshift solution.

\vspace*{0.2cm}

\noindent
{\bf SPT-CL~J0310-4647: } We note strong C {\small III}] $\lambda$1909 emission and a family 
of corroborating Fe {\small II} absorption lines; these features inform a clear redshift for the 
giant arc.

\vspace*{0.2cm}

\noindent
{\bf SPT-CL~J0356-5337: } The two possible emission lines identified here are 
very low-confidence, so we consider this the most likely redshift for this source, but not an 
unambiguous redshift measurement. The possible N {\small III}] $\lambda$1750 emission line, 
in particular, is at least as likely to be a noise spike as a real feature, because this line is not 
generally observed even in high S/N spectra of distant star-forming galaxies \citep[e.g.,][]{pettini00,
shapley03,quider09,quider10,bayliss14b}. We only highlight this potential line in the figure because 
it does coincide quite well with the putative C {\small III}] $\lambda$1909 emission feature. In the 
case where we are mis-interpreting these weak features we can still confidently conclude that this 
source falls within the redshift range, $1.78<z<3.9$, based on our wavelength coverage and the 
lack of strong spectral features.

\vspace*{0.2cm}

\noindent
{\bf SPT-CL~J2325-4111: } The spectrum exhibits strong C {\small II}] $\lambda$2326 emission 
along with four strong Fe {\small II}  absorption lines. These lines inform a clear redshift 
solution. 

\subsubsection{Giant Arc Redshift Limits/Constraints}

There are also five giant arc candidates identified in four SPT-GMOS clusters that received 
spectroscopic slits but did not result in a precise redshift estimate. Instead we place redshift 
constraints on each of these sources based on the lack of strong spectral features in the GMOS 
spectra. Specifically, the arc candidates that we identify are blue, which would guarantee the 
presence of strong emission lines in their spectra, and if we detect continuum 
emission but fail to observe one or more of the rest-frame optical emission lines that are typical of 
blue star-forming galaxies in the spectrum of a given giant arc then we can conclude that it is at a 
sufficiently high redshift to move the bluest of those strong lines --- \oii\ --- 
beyond the red end of our spectra. In the case of spectra that extend blueward of $\sim5000$\AA\  
we can also place an upper limit on the redshift based on the lack of Ly-$\alpha$ observed either 
in emission or absorption (apparent as a strong spectral break). Giant arc candidates discussed 
below are shown in Figure~\ref{fig:arcs1}c, and Figure~\ref{fig:arcs2}b-d.

\vspace*{0.2cm}

\noindent
{\bf SPT-CL~J0307-6225: } We identify a blue, extended arc running tangentially relative to the 
center of the cluster. From the presence in its spectrum of low S/N continuum emission, the lack of 
emission lines, and the wavelength coverage ($\Delta \lambda = 4430-8590$\AA) we infer 
that the arc has a redshift $1.30<z<2.8$.

\vspace*{0.2cm}

\noindent
{\bf SPT-CL~J0352-5647: } This cluster exhibits a blue extended source with a long but faint extended tail. 
In the spectrum of this source we again see weak continuum with no emission features, and from the 
wavelength coverage ($\Delta \lambda = 4680-8980$\AA) we infer a redshift in the range $1.4<z<3$ for 
this source.

\vspace*{0.2cm}

\noindent
{\bf SPT-CL~J0356-5337: } We identify two strongly-lensed systems in this cluster --- B and C, 
both having clear multiple images visible in {\it HST} imaging obtained as part of the SPT-SZ ACS 
Snapshot Survey ({\it HST }Program ID 13412, PI: Schrabback) --- that received slit coverage on our 
spectroscopic masks but did not result in redshifts (Figure~\ref{fig:arcspec}). The spectra of both the B and 
C sources exhibit low S/N blue continuum emission with no emission lines, and given the wavelength 
coverage ($\Delta \lambda = 5920-10350$\AA) of the GMOS spectra we can infer that both sources have 
redshifts in the range, $1.78<z<3.9$.

\vspace*{0.2cm}

\noindent
{\bf SPT-CL~J0411-4819: } There is a clear giant arc candidate, consisting of at least two segments, 
near the core of this cluster. The spectrum of this source contains low S/N blue continuum emission with 
no emission lines, from which we infer a redshift in the range $0.84<z<2.3$ for the wavelength coverage 
($\Delta \lambda = 5920-10350$\AA) of the SPT-GMOS data.

\begin{deluxetable}{cccc}
\tablecaption{Galaxy Spectral Type Classification\label{tab:galtypes}}
\tablewidth{0pt}
\tabletypesize{\tiny}
\tablehead{
\colhead{Spectral Type} &
\colhead{W$_{0,3727}$ (\AA)} &
\colhead{W$_{0,H\delta}$ (\AA)} &
\colhead{Classification} }
\startdata
k          & none  &  $<$ 3  & passive \\
k$+$a &  none  &   $\geq$ 3, $\leq$ 8  & post-starburst \\
a$+$k &  none  &  $>$ 8  & post-starburst \\
e(c)     &  $>$ -40  & $<$ 4  & star-forming \\
e(b)     &  $\leq$ -40  & any & star-forming \\
e(a)     & yes  & $\geq$ 4  & star-forming 
\enddata
\tablecomments{This table lists the criteria used to classify galaxy type based on the strength of spectral 
features. The columns listed, from left to right, are the name of the specific galaxy spectral type, the 
criterion for the equivalent width of the [O II]$\lambda$ 3727 feature, the criterion for the equivalent width 
of the \hdelta\ feature, and the general classification (i.e., passive, post-starburst, or star-forming).}
\end{deluxetable}

\subsection{Magnitude Distribution of Cluster Members With Spectra}
\label{subsec:luminosities}

In addition to exploring their spectral types, it is also interesting to explore the distribution of 
galaxy magnitudes for which we measure spectra. Here we describe how the SPT-GMOS dataset 
can be combined with existing imaging catalogs of SPT-SZ galaxy clusters. This would enable, for 
example, systematic tests of velocity dispersion measurements such as investigating how velocity 
dispersion estimates change as a function of the fraction of brighter vs. fainter cluster member galaxies 
included. There is a large pool of optical imaging data available for the SPT-SZ galaxy cluster catalog 
\citep{high10,song12,bleem15}, and we combine our spectra with these photometric measurements 
to determine the luminosities of those cluster member galaxies for which we now have spectra 
relative to the characteristic luminosity, 
$L^\star$ (or more precisely, its magnitude equivalent $M^\star$). The combination of broadband 
magnitudes and spectral line equivalent widths 
($\S$~\ref{sec:indices}) also provides a straightforward way to estimate spectral line fluxes, which 
can be used, for example, to estimate the instantaneous star formation rate from \oii.

Most of the existing photometry is already in the Sloan Digital Sky Survey \citep[SDSS;][]{york2000} 
photometric system, but there is also a significant amount of imaging data that were taken 
using the older Johnson-Cousins filter sets (\emph{BVRcIc}). Additionally, imaging acquired with the 
Swope 1m telescope Las Campanas is in a photometric system that is close to, but not precisely the 
same as Johnson-Cousins \citep{bleem15}. We use transformations determined 
empirically from standard star fields of the older photometric system that also appear in the SDSS 
\citep{jordi2006}, which are optimized for stars rather than galaxies but are the only established 
transformations that use the photometric bands available to us. These transformations rely on applying 
an offset to a band that is closest to the SDSS band of interest, and then applying a color term. We 
are able to compute transformations into the SDSS $r-$band for all galaxies in our photometric 
catalogs, and are often but not always able to compute transformations into the SDSS $i-$band. 
Systematic uncertainties are unavoidable when applying these kinds of empirical transformations 
between photometric systems; \citet{jordi2006} report systematic uncertainties of 0.06 magnitudes 
in their transformations between $VRI$ and $gri$ magnitudes, for instance. The transformed $r-$ and 
$i-$band magnitudes that we recover therefore have typical total uncertainties of $\sim$0.1 magnitudes. 

\begin{figure}[t]
\includegraphics[scale=0.482]{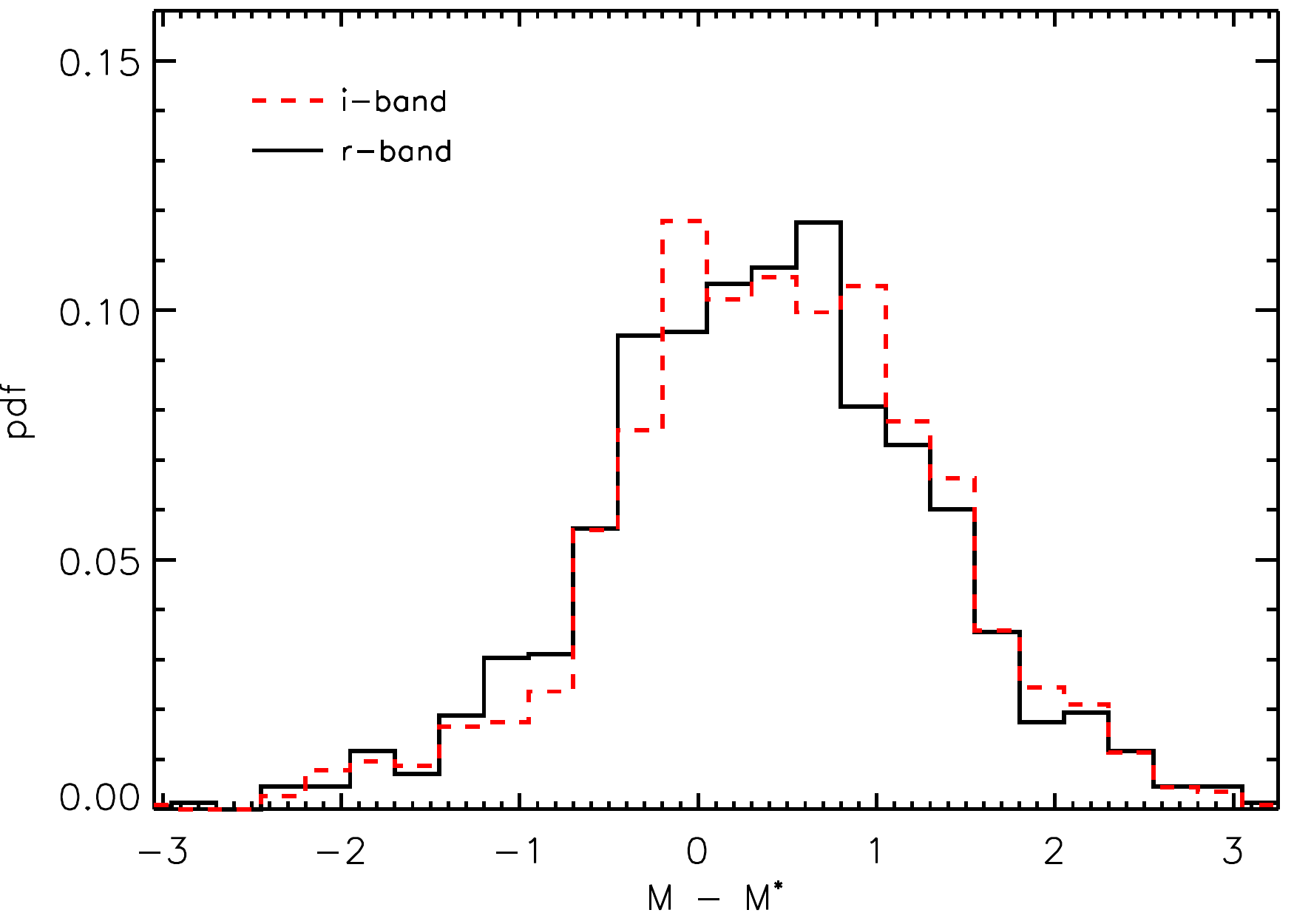}
\caption{\scriptsize{
The distribution of magnitudes of cluster member galaxies with spectra measured in this work 
in the $r-$ and $i-$bands. Individual galaxies are plotted in units of magnitude minus 
the characteristic magnitude, $M^{*}$, at their redshift. The $r-$ and $i-$band magnitudes for 
individual galaxies are estimated using the procedure described in Section~\ref{subsec:luminosities}, 
such that data in various different optical bandpasses are all transformed into the SDSS $r-$ and 
$i-$bands.}}
\label{fig:gals_mstar}
\end{figure}

The distribution of cluster member galaxy magnitudes, relative to $m^{*}$ at each cluster redshift, 
is shown in Figure~\ref{fig:gals_mstar} for each of the $r-$ and $i-$bands. Recall that our mask 
design strategy focused on acquiring 
spectra for cluster members down to $\sim m^{*}+1$, and that strategy is reflected in the magnitude 
distribution of spectroscopically-observed cluster galaxies, though we also find that we were able to 
measure redshifts for some cluster member galaxies as faint as $m^{*}+2.5$. Having magnitudes in 
units of $m^{\star}$ in-hand for galaxies with spectroscopic data could be a valuable piece of 
information to fold into future analyses of galaxy cluster scaling relations involving velocity 
dispersions, as there is evidence from simulations that velocity dispersions can be biased by 
using only, for example, the very brightest cluster member galaxies \citep{saro13,gifford13a}. 
This bias could potentially be removed from velocity dispersion estimates if we know the 
relative luminosities of the galaxies used to estimate the dispersion. We do point out, however, that 
while these magnitude-relative-to-$m^\star$ measurements inform us about the relative brightness 
of cluster member galaxies for which we have spectra, they \emph{do not} fully characterize the 
\emph{selection} function with respect to the luminosity of cluster galaxies with redshift measurements. 
Matched magnitudes for SPT-GMOS galaxies are available in the full SPT-GMOS data release 
(see \S~\ref{sec:motivation}).

\begin{figure}[h]
\centering
\includegraphics[scale=0.53]{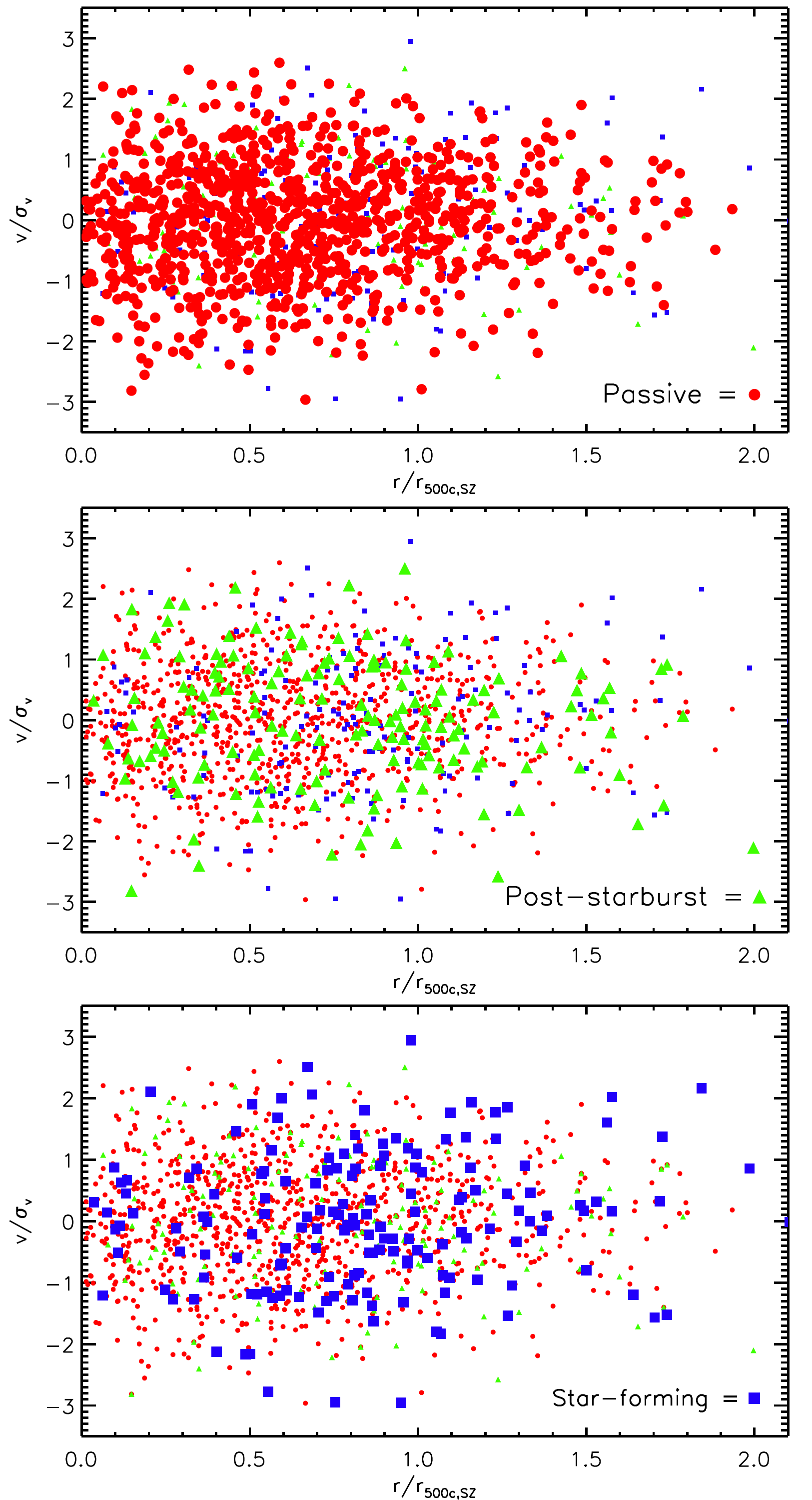}
\caption{\scriptsize{
These three panels all show the full ensemble of member galaxies from all clusters within SPT-GMOS 
with each galaxy plotted according to its peculiar velocity relative to the median redshift of its host cluster. 
Individual galaxies are plotted in one of three colors indicating whether they were classified as 
passive (red), post-starburst (green), or actively star-forming (blue). In each panel we emphasize one 
galaxy type above the others: \emph{Top:} passive galaxies, \emph{Middle:} post-starburst galaxies,
and \emph{Bottom:} star-forming galaxies. The median projected radial distances of post-starburst 
and star-forming cluster members in the SPT-GMOS spectroscopic sample are both 
$\sim 0.9R_{500c,SZ}$, while for passive cluster members it is $\sim 0.67R_{500c,SZ}$.}}
\label{fig:phasespace}
\end{figure}

\subsection{Galaxy Spectral Type Classification}

Using the \oii\ doublet and the H-$\delta$ equivalent width data products described above it is 
straightforward to classify each galaxy with a GMOS spectrum and a spectroscopic redshift using the 
spectral index criteria described in Table 6 of \citet{dressler99}. Table~\ref{tab:galtypes} lists the 
specific criteria that we use to classify SPT-GMOS galaxies as one of six types of galaxies; we 
apply these criteria exclusively to galaxy spectra that have reliable redshift measurements 
(2243 in total). Briefly, these six spectral galaxy types identify a given galaxy as either passive 
(dominated by old stars), actively star-forming, or post-starburst (i.e., transitioning between 
star-forming and passive, with star formation likely recently quenched). Where the galaxy 
classification of \oii\ emission criteria reads ``None'', this refers to an absence of 
the \oii\ emission feature, or emission detected at $\leq 2 \sigma$ significance, while ``yes'' refers 
to a detection of emission with W$_{0,3727} < 0$ at a significance of $>2\sigma$. We identify no 
broad line emission objects --- i.e., no strong active galactic nuclei (AGN) --- in our spectra. Technically, 
because we target galaxies over a wide range of redshifts using masks that were observed with a wide 
range of different exposure times, our sensitivity to \oii\ emission varies somewhat 
across the SPT-GMOS sample, and no special effort was made to achieve a uniform effective detection 
limit in units of star formation (e.g., M$_{\Sun}$ yr$^{-1}$). However, the SPT-GMOS program was 
designed to measure absorption line redshifts, and slits were deliberately placed on galaxies 
bright enough to produce decent S/N continuum spectra (S/N $\gtrsim5$ per spectral element). 
This results in spectra that allow us to place useful limits on the presence of 
\oii\ emission in the vast majority of SPT-GMOS galaxy spectra. Exact depths vary from mask to 
mask due to variable observing conditions and differences between the photometric redshift 
used to plan the observations vs. the true redshifts of each galaxy cluster, but our spectra are 
typically sensitive down to $L_{\mathrm[O II]} \simeq 5-10 \times 10^{40}$ erg s$^{-1}$, 
corresponding to a star formation rate, 
SFR$_{\mathrm [O II]} \sim 1$ M$_{\odot}$ yr$^{-1}$ \citep{kennicutt98}.

Galaxy spectral type information is useful for exploring astrophysical trends such as the relationship 
between galaxy evolution with environment \citep[e.g.,][]{muzzin13,hennig16,zenteno16}. For example, 
Figure~\ref{fig:gald4000} demonstrates the 
difference in the distribution of D4000 values for SPT-GMOS galaxies in clusters vs. those in the 
field (i.e., non-cluster members as defined in \S~\ref{sec:zdispestimates}). The galaxy type 
information can also be used to investigate how cluster member galaxies of different  
spectral types occupy the phase space of line-of-sight velocity and projected radial distance from 
the galaxy cluster center. To demonstrate this we plot the ensemble of all SPT-GMOS galaxy 
cluster members, where each cluster member galaxy recession velocity is converted into a normalized 
peculiar velocity relative to the mean recession velocity/redshift of its host galaxy cluster, scaled into 
units of $\pm \sigma_{v}$. We also compute the projected physical radial distance 
between each cluster member and the SZ cluster centroids, normalized by $R_{500,SZ}$. With these 
quantities in-hand we can generate a sort of ensemble phase space for all SPT-GMOS galaxy clusters, 
which we show in Figure~\ref{fig:phasespace}. Plotted in this way the SPT-GMOS 
sample of cluster members exhibits the same qualitative trends that have been observed in other studies 
\citep{mohr96a,lewis02c,rines05,pimbblet06,dressler13}, such as post-starburst and star-forming cluster 
member galaxies residing preferentially at larger cluster-centric radii. It is worth emphasizing the 
\emph{qualitative} nature of this agreement, but we caution we note that the field of view of GMOS imposes an 
upper limit on the projected radial separation within which we have spectra for a given galaxy cluster. The 
precise limit varies with cluster mass and redshift, but generally prevented us from targeting galaxies with 
projected radial separations greater than $\sim$2$R_{500,SZ}$. We note here the different \emph{relative} 
median projected radii of different types of galaxies, but the specific median projected radii that we measure 
--- $\sim 0.9R_{500c,SZ}$ for post-starburst and star-forming cluster members vs. $\sim 0.67R_{500c,SZ}$ 
for passive cluster members --- do not necessarily represent the true median radial distributions of all 
cluster member galaxies.

These plots serve as an example 
of the kinds of analyses that the SPT-GMOS/ data products will enable, but we leave more rigorous 
analyses to future work, as the goal of this paper is to present the survey dataset and data products. 
The precise projected physical radii of SPT-GMOS galaxies are sensitive to the exact cosmological 
parameter values used to compute the angular diameter distance, and so we do not provide these 
values in the data release, but they are straightforward to compute from the galaxy positions 
(e.g., Table~\ref{tab:bcgspec}) and SPT-SZ galaxy cluster centroids provided in Table~\ref{tab:obs}.

\section{Conclusions and Summary}

We present the full spectroscopic data release of the SPT-GMOS survey of 62 SPT-SZ galaxy clusters, 
which includes 2595 spectra with radial velocity measurements, 2243 of which are galaxies (1579 cluster 
members). Some of the 
SPT-SZ galaxy clusters are identified as strong-lensing systems in the available imaging, and we 
measure spectroscopic redshifts (or redshift constraints/limits) for candidate strongly-lensed 
background sources where possible. In addition to redshifts, we also measure standard 
spectral index measurements of the strength of the \oii\ doublet, \hdelta, 
and the 4000\AA\ break. These indices are useful for spectrally classifying galaxies, and introduce 
the potential to investigate the properties of SPT-SZ member galaxies as a function of galaxy type.

The SPT-GMOS survey can be combined with previously published results from other spectroscopic programs 
\citep{sifon13,ruel14,sifon16} to provide $>$100 SPT-SZ galaxy clusters with spectroscopic follow-up 
(longslit or MOS), and more than 90 clusters with $N \geq 15$ member velocity dispersion measurements. 
These data contribute to a broad effort to obtain multi-wavelength follow-up of SPT-SZ galaxy clusters --- 
including extensive X-ray observations \citep{williamson11,mcdonald13,mcdonald14} 
and ongoing weak lensing measurements \citep[e.g.,][]{high12} --- that will inform future multi-wavelength 
efforts to cross-calibrate the SZ mass-observable relation, and thereby enable future cosmological studies 
using galaxy clusters.

\acknowledgments{We thank the anonymous referee for helpful and thoughtful feedback that improved this 
paper. This work is supported by the National Science Foundation through Grant AST-1009012. 
The South Pole Telescope is supported by the National Science Foundation through grant PLR-1248097. 
Partial support is also provided by the NSF Physics Frontier Center grant PHY-1125897 to the Kavli Institute 
of Cosmological Physics at the University of Chicago, the Kavli Foundation, and the Gordon and Betty Moore 
Foundation grant GBMF 947. Galaxy cluster research at SAO is supported in part by NSF grants AST-1009649 
and MRI-0723073. R.J.F.\ gratefully acknowledges support from the Alfred P.\ Sloan Foundation.
Argonne National Laboratory work was supported under U.S. Department of Energy contract 
DE-AC02-06CH11357. BB is supported by the Fermi Research Alliance, LLC under Contract No. 
DE-AC02-07CH11359 with the United States Department of Energy.  Support for program 
\#HST-GO-13412.004-A was provided by NASA through a grant from the Space Telescope Science 
Institute, which is operated by the Association of Universities for Research in Astronomy, Inc., under 
NASA contract NAS 5-26555.

The data presented here were taken with the Gemini Observatory, which is operated by the Association of 
Universities for Research in Astronomy, Inc., under a cooperative agreement with the NSF on behalf of the 
Gemini partnership: The United States, Canada, Chile, Australia, Brazil, and Argentina. Gemini data used in 
this work was taken as a part of the following Gemini programs: GS-2011A-C-03, GS-2011A-C-04, GS-2011B-C-06, 
GS-2011B-C-33, GS-2012A-Q-04, GS-2012A-Q-37, GS-2012B-Q-29, GS-2012B-Q-59, GS-2013A-Q-05, 
GS-2013A-Q-45, GS-2013B-Q-25, GS-2013B-Q-72, GS-2014B-Q-31, GS-2014B-Q-64.
Additional supporting data were obtained with the 6.5 m Magellan Telescopes, which 
are located at the Las Campanas Observatory in Chile. This work is also partly based on observations made with 
the NASA/ESA \emph{Hubble Space Telescope}, obtained from the Data Archive at the Space Telescope Science 
Institute, which is operated by the Association of Universities for Research in Astronomy, Inc., under NASA contract 
NAS 5-26555; these observations are associated with program \#13412.}

{\it Facilities:} \facility{Gemini:South}, \facility{SPT}, \facility{Magellan:Baade (IMACS)}, 
\facility{Magellan:Clay (LDSS3, PISCO, Megacam)}, \facility{{\it HST} (ACS)}

\bibliographystyle{fapj}
\bibliography{sptgmos}

\end{document}